\documentclass[twocolumn,aps,superscriptaddress,multicol,amsmath,amssymb]{revtex4-2}

\makeatletter
\newcommand*{\rom}[1]{\expandafter\@slowromancap\romannumeral #1@}
\makeatother
\usepackage{xcolor}
\usepackage{color}
\usepackage{graphicx}
\usepackage{amsmath,amssymb}
\usepackage[utf8]{inputenc}
\usepackage{epstopdf}
\usepackage[font=small,labelfont=bf,justification=justified]{caption}
\usepackage[font=small,labelfont=bf,justification=justified]{subcaption}
\usepackage[colorlinks=true, linkcolor=blue, urlcolor=blue, citecolor=blue]{hyperref}
\usepackage{amssymb}
\usepackage{amsmath}
\usepackage{graphicx}
\usepackage{epstopdf}
\usepackage{bm}% bold math
\usepackage{gensymb}
\usepackage{soul}
\usepackage{sidecap}
\usepackage[normalem]{ulem}
\usepackage{times}
\usepackage{booktabs}

\usepackage{float}
\usepackage{enumerate}
\usepackage{multirow}
\usepackage{tabularx}
\usepackage{array}
\usepackage{url}
\usepackage{slantsc}
\usepackage{lmodern}
\usepackage{cancel}
\usepackage{extarrows}
\usepackage{xcolor}
\usepackage{ragged2e} 
\usepackage{tikz}
\usepackage{feynmp}
\usepackage{tikz-feynman}
\tikzfeynmanset{compat=1.1.0}
\usepackage[normalem]{ulem}
\usepackage{xparse}
\usepackage{soul}
\usepackage{todonotes}
\usepackage{amsmath}
\usepackage{mathtools}
\makeatletter
\NewDocumentCommand{\sotwo}{O{red}O{black}+m}
{%
	\begingroup
	\setulcolor{#1}%
	\setul{-.5ex}{.4pt}%
	\def\SOUL@uleverysyllable{%
		\rlap{%
			\color{#2}\the\SOUL@syllable
			\SOUL@setkern\SOUL@charkern}%
		\SOUL@ulunderline{%
			\phantom{\the\SOUL@syllable}}%
	}%
	\ul{#3}%
	\endgroup
}
\makeatother
\usepackage{mathtools}
\newcommand{\overparent}[1]{\overbracket[0.8pt][0.8ex]{#1}}
\graphicspath{{figs/}}
\def\beq{\begin{equation}}
	\def\eeq{\end{equation}}
\def\bea{\begin{eqnarray}}
	\def\eea{\end{eqnarray}}

\begin{document}
	\title{Foundations of Many-Body Theory of Quantum Unified Statistics:
		Green functions and Linear Response Theory}
	\author {M. Beyrami}
	\affiliation{Department of Physics, University of Mohaghegh Ardabili, P.O. Box 179, Ardabil, Iran}
	\author {M. N. Najafi}
	\email{morteza.nattagh@gmail.com }
	\affiliation{Department of Physics, University of Mohaghegh Ardabili, P.O. Box 179, Ardabil, Iran}
	\author {H. Esmaili}
	\affiliation{Department of Physics, University of Mohaghegh Ardabili, P.O. Box 179, Ardabil, Iran}
	\author {M. Dournama}
	\affiliation{Department of Physics, University of Mohaghegh Ardabili, P.O. Box 179, Ardabil, Iran}
	\author {H. Mohammadzadeh}
	\affiliation{Department of Physics, University of Mohaghegh Ardabili, P.O. Box 179, Ardabil, Iran}


\begin{abstract}
We develop a comprehensive many-body theory for systems of particles obeying quantum unified statistics, or quons. After exploring the properties of the Fock space of this system, we formulate a systematic $S$-Matrix expansion and a Generalized Wick's theorem. A consistent Green-function formalism is constructed at both zero and finite temperatures, accompanied by a generalized Wick’s theorem appropriate for infinite-statistics operator algebras. Within this framework, we establish diagrammatic rules for interacting quon systems. Employing the random phase approximation, we derive the dielectric function and reveal the emergence of anomalous plasmon modes that have no direct counterpart in conventional Bose or Fermi systems. We further analyze the ground-state energy, energy-loss function, generalized Thomas–Fermi screening wave vectors, and Friedel oscillations, elucidating how infinite statistics qualitatively modifies collective behavior and screening properties. 
\end{abstract}
	
\maketitle
	
\tableofcontents

\section{Introduction}\label{1}
%===================================================
Understanding collective quantum configurations and the macroscopic properties of matter is fundamentally tied to the statistical laws governing elementary particles. The conventional Bose-Einstein and Fermi-Dirac frameworks \cite{einstein1924sitzber,dirac1926theory} have long formed the cornerstone of monumental advances in condensed matter physics, successfully describing the behavior of countless bosonic and fermionic systems. However, as theoretical investigation has expanded into low-dimensional, topological, and strongly correlated systems, this conventional paradigm has revealed inherent limitations \cite{RevModPhys.80.1083}, since a growing number of physical phenomena resist a complete description in terms of purely bosonic or purely fermionic statistics alone. These limitations have, in turn, driven a sustained search for generalized statistical structures capable of capturing entirely new physical regimes \cite{wilczek1982quantum,haldane1991fractional,macfarlane1989q,lavagno2002generalized,tsallis1988possible,yan2021statistical}. Among the resulting structures, infinite quantum statistics stands out as a fundamental algebraic limit: by completely removing the symmetry constraints ordinarily imposed by particle permutation, it opens up a unique and remarkably vast theoretical space for the discovery of novel quantum phenomena \cite{greenberg1991particles}.

The systematic expansion of quantum statistical frameworks beyond the conventional Bose-Fermi dichotomy began with Wilczek's introduction of anyons, which predicted the emergence of fractional statistics in two-dimensional systems \cite{wilczek1982quantum}. This concept subsequently provided the theoretical foundation for the fractional quantum Hall effect, in which quasiparticles carrying fractional statistics play a central role \cite{PhysRevLett.50.1395,PhysRevLett.53.722,PhysRevLett.52.1583}. Around the same time, Haldane introduced a complementary notion, fractional exclusion statistics, which generalizes the Pauli exclusion principle through a continuous exclusion parameter \cite{PhysRevLett.67.937}. Although this generalization is formally defined in arbitrary spatial dimension, it is exactly realized in certain one-dimensional systems, where it directly governs the fractional statistics of quasiparticle excitations \cite{haldane1991fractional,PhysRevLett.73.1574,PhysRevLett.73.3331}. Building on this broader effort to move beyond conventional statistics, Greenberg introduced the concept of infinite statistics, describing particles now commonly referred to as quons, whose creation and annihilation operators obey the generalized commutation relation $a^{}_k a^\dagger_l-qa^\dagger_la^{}_k=\delta^{}_{kl}$ \cite{greenberg1991particles}. In the special limiting case $q=0$, this infinite-statistics framework becomes maximally permissive, admitting any representation of the permutation group, a property that endows it with a uniquely rich algebraic structure \cite{PhysRevLett.65.3361}. This flexibility, in turn, enables infinite statistics to model systems in which the occupation of a single quantum state is, in principle, unlimited \cite{altherr1993thermal}. More recently, emerging applications of such generalized statistics in describing topological order, spin liquids, and strongly correlated systems have opened new horizons for the study of exotic quantum phases \cite{wen1995topological}.

Despite these considerable advances in the theory of fractional statistics and anyons, developing a coherent many-body framework for infinite statistics remains an open and largely unresolved challenge. Early work in this direction focused primarily on algebraic aspects and on the structure of the associated Fock spaces \cite{greenberg1991particles,mishra1995generalized,MISHRA1994210}, laying important mathematical groundwork but stopping short of a full dynamical theory. Later attempts to construct a finite-temperature quantum field theory for quons largely remained at the level of basic formulations, and did not lead to the development of an operational perturbation theory or the computational tools that are indispensable in condensed matter physics \cite{altherr1993thermal}. Crucially, several key concepts and methods central to conventional many-body theory-including finite-temperature Green's functions, a generalized Wick's theorem, systematic diagrammatic expansions, and linear response formalism-have not been systematically defined and extended to systems obeying infinite statistics \cite{mahan2013many}. This fundamental gap has, in practice, precluded the investigation of concrete physical questions: for instance, whether a superconducting phase can emerge in such systems, what the structure of their collective response functions is, or whether quasi-correlated behaviors characteristic of ordinary matter can be simulated simply by tuning the underlying statistical parameters. It is precisely this gap that the present work aims to address-namely, the absence of an applicable, computationally equipped many-body theory for infinite statistics, one capable of bridging the abstract algebraic structure of these particles with concrete, observable physical phenomena \cite{zagier1992realizability}.

Within the narrower domain of fractional statistics, some attempts have already been made to define single-particle Green's functions in one-dimensional systems \cite{PhysRevB.83.085426}. These efforts, however, remain largely confined to non-interacting calculations or simple lattice models, and they lack a systematic field-theoretic framework capable of deriving diagrammatic rules or computing response functions from first principles. In a related but distinct direction, other studies on deformed algebras have examined their formal algebraic structure and associated coherent states \cite{A:Macfarlane_1989}, further underscoring the need for a unified, dynamically complete many-body formalism-the central objective pursued in the remainder of this work.
%===================================================
\section{Summary of the Paper and Achievements}
This study presents a quantum many-body framework tailored for deformed particles-sometimes referred to as unified-statistics particles, or quons-governed by unified quantum statistics, thereby establishing a direct link between particle statistics and the resulting effective interaction strength. By treating the deformation parameter $\delta$ as an adjustable variable, we show that a diverse range of effective interactions can be realized simply through modifications of the underlying statistics, without introducing any additional interaction terms by hand.

The work is organized into two principal parts: the first develops the fundamental theoretical framework for systems exhibiting unified quantum statistics, while the second applies these theoretical tools to derive key physical quantities relevant to practical applications.

In the first part, we construct a generalized $S$-matrix expansion (Sec.~\ref{SEC:SMatrix}) together with a corresponding extension of Wick's theorem (Sec.~\ref{SEC:WICKsTHEOREM}), formulated specifically for particles obeying unified quantum statistics. Together, these two developments provide the foundation for an advanced Feynman diagrammatic technique (Sec.~\ref{SEC:DiagrammaticParticles}), which is essential for the systematic computation of Green's functions in this generalized statistical framework.

The second part of the paper begins by defining the single-particle Green's function for the unified statistics gas. We derive explicit expressions for this Green's function in both direct (real-space) and momentum space (Sec.~\ref{SEC:FiniteTemperatureGreenFunction}), and subsequently obtain the corresponding spectral density and density of states (Sec.~\ref{SpectralDensity}), both of which follow directly from the Green's function derived earlier. Building on these results, we then analyze key physical properties of the interacting system, in particular the dielectric response function within the random phase approximation (RPA) for the deformed fermion gas, thereby establishing a connection to fundamental concepts in collective phenomena (Sec.~\ref{AverageEnergy}). Finally, in Sec.~\ref{DielectricFunction}, we employ this dielectric function to discuss Friedel oscillations, plasmonic excitations, and energy loss functions (Sec.~\ref{Friedel-Oscillations}) within the RPA, illustrating how the deformation parameter $\delta$ manifests itself in these observable collective phenomena.
%%%%%%%%%%%%%%%%%%%%%%%%%%%%%%%%%%%%%%%%%
\section{Construction of Fock Space}\label{SEC:FockSpace}
This section outlines the Fock space structure for $N$ identical particles governed by unified quantum statistics proposed in~\cite{yan2021statistical,PhysRevE.109.044104} based on exchange‑statistics factor operators. Representing the coordinate of $i$th particle by $\xi_i$, and the whole wave function by $\hat{\Psi}(\xi_1,\ldots,\xi_j,\ldots,\xi_i,\ldots,\xi_N)$, the effect of particle exchange is governed by: 
\begin{align}\label{Eq:wave-function}
	\hat{\Psi}&(\xi_1,\ldots,\xi_j,\ldots,\xi_i,\ldots,\xi_N)\nonumber\\
	=
	&\hat{q}_{ij}
	\hat{\Psi}(\xi_1,\ldots,\xi_i,\ldots,\xi_j,\ldots,\xi_N),
\end{align}
where $\hat{q}_{ij}$ is an exchange‑statistics factor operator. As noted in~\cite{yan2021statistical} this operator is independent of the particles' indices $i$ and $j$, leading us to show it by $\hat{q}$. Given that two successive exchanges of the same particles has a trivial effect, we have $\hat{q}^2=\textbf{1}$, where $\textbf{1}$ is an identity operator. This leads the eigenvalues of $\hat{q}$ to be $+1$ and $-1$: 
\begin{equation}
	\hat{q}|\pm ,k\rangle
	=
	\pm|\pm ,k\rangle,
\end{equation}
where $|\pm ,k\rangle$ are eigenvectors that form a complete orthonormal set for the Hilbert space, with $k$ denoting the internal degrees of freedom. A general single-particle state can then be expanded as follows:
\begin{equation}\label{Eq:Internal-State}
	|\mathcal{\phi}\rangle=\sum^{}_{s=\pm}\sum^{}_\nu c^{(s)}_\nu|s,\nu\rangle,
\end{equation}
where $c^{(\pm)}_\nu$ are the expansion coefficients.
\subsection{The Fock Space}
As a first step toward developing a field theory for the deformed particles, we introduce the Fock space in this subsection. The many-body Hilbert space is spanned by the following basis states (for the details of the derivation of this representation refer to SEC.~\ref{SEC:FockSpaceApp}):
\begin{widetext}
    \begin{equation}
	\begin{split}
|\hat{\Psi}^{\hat{q}}_{}\rangle\equiv |\nu_{P_1}&,\nu_{P_2},...,\nu_{P_N}\rangle^{\hat{q}} =\sqrt{\frac{\prod_\nu n_\nu !}{N!}}\textbf{1}^{\hat{q}}\sum_{P_E}\hat{q}^{[P_E]}\left| \nu_1\right\rangle \otimes \left| \nu_2\right\rangle\otimes...\otimes\left| \nu_N\right\rangle,
	\end{split}
	\end{equation}
\end{widetext}
where $N$ is the total number of particles, $\left\lbrace  \left| \nu_i\right\rangle\right\rbrace$ is an orthonormal basis of the Hilbert space of single particles, and $n_\nu$ is the number of deformed particles in the state $\nu$. In this relation $P_E$ is a members of the permutation group that exchange particles between \textit{different} quantum states, and $[P_E]$ is the number of permutations needed for that permutation. The generalized identity operator is defined as
\begin{equation}
    \mathbf{1}^{\hat{q}}_{} =
    \begin{cases}
        \mathbf{1}, & \text{if all } n_{\nu_i} = 0 \text{ or } 1, \\
        \frac{1}{2}(\mathbf{1} + \hat{q}), & \text{otherwise},
    \end{cases}
\end{equation}
Noting that $\left(\mathbf{1}^{\hat{q}}_{}\right)^2=\mathbf{1}^{\hat{q}}_{}$, one can readily confirm by inspection that $\langle\hat{\Psi}^{\hat{q}}_{}|\hat{\Psi}^{\hat{q}}_{}\rangle= \textbf{1}^{\hat{q}}$. Note that this operator is configuration-dependent.\\
It is more convenient to use the particle number representation $ |\{n_\nu\}\rangle^{\hat{q}}\equiv\left| n_{\nu_1},n_{\nu_2},...\right\rangle^{\hat{q}}_{}$, which shows explicitly a state with $n_{\nu_1}$ deformed particles is the state $\nu_1$, etc. 
Then, the above calculations result in the following completeness relation:
\begin{equation}
	\sum_{\{n_\nu\}}
	|\{n_\nu\}\rangle^{\hat{q}}\;{}^{\hat{q}}\langle \{n_\nu\}|
	=\mathbf{1}^{\hat{q}},
\end{equation}
where the symbol $\{n_\nu\}$ in the summation shows that the summation runs over all $n_{\nu}$ configurations. The creation and annihilation operators, $\hat{a}_\nu^\dagger$ and $\hat{a}_\nu$ of deformed particles are defined in the same way as the ordinary particles, and are defined according to their operation on the states:
\begin{equation}
   \begin{split}
    &\hat{a}_{\nu_i}\left| n_{\nu_1},...,n_{\nu_i},...\right\rangle^{\hat{q}}_{}=C_{\hat{q}}^-\left| n_{\nu_1},...,n_{\nu_i}-1,...\right\rangle^{\hat{q}}_{}\\
    &\hat{a}_{\nu_i}^\dagger\left| n_{\nu_1},...,n_{\nu_i},...\right\rangle^{\hat{q}}_{}=C_{\hat{q}}^+\left| n_{\nu_1},...,n_{\nu_i}+1,...\right\rangle^{\hat{q}}_{}
   \end{split}
\end{equation}
where $C_{\hat{q}}^-\equiv\hat{q}^{\sum_{l=1}^{i-1}n_{\nu_l}}\textbf{1}^{\hat{q}}\sqrt{n_{\nu_i}}$ and $C_{\hat{q}}^+\equiv\hat{q}^{\sum_{l=1}^{i-1}n_{\nu_l}}\textbf{1}^{\hat{q}}\sqrt{n_{\nu_i}+1}$. By using these equations
it is easy to show that the Fock-like space is spanned by
\begin{equation}
\left| n_{\nu_1},...,n_{\nu_i},...\right\rangle^{\hat{q}}_{}=\prod_{i}\frac{\hat{a}_{\nu_i}^\dagger}{\sqrt{n_{\nu_i}!}}\left| 0\right\rangle^{\hat{q}}_{},
\label{Eq:states}
\end{equation}
where $\left| 0\right\rangle^{\hat{q}}_{}$ is a vacuum state. These operators obey the generalized commutation relations~\cite{yan2021statistical}: 
\begin{align}\label{Eq:algebraQhat1}
	[\hat{a}^{}_{\nu},\hat{a}^{\dagger}_{\nu'}]^{}_{\hat{q}}
	&=\mathbf{1}^{\hat{q}}\delta_{\nu\nu'},  \\
	\label{Eq:algebraQhat2}
	[\hat{a}^{}_{\nu},\hat{a}^{}_{\nu'}]_{\hat{q}}
	&=[\hat{a}^{\dagger}_{\nu},\hat{a}^{\dagger}_{\nu'}]_{\hat{q}}=0,
\end{align}
where $[A,B]_{\hat{q}}\equiv AB-\hat{q}BA$. The appearance of the operator $\textbf{1}^{\hat{q}}$ on the right-hand side of Eq.~\eqref{Eq:algebraQhat1} indicates that these commutation relations depend on the occupation numbers $n_\nu$ of the entire state, a feature absent for ordinary particles.\\  
Then we define the number operator as $\hat{n}_\nu\equiv \hat{a}_\nu^\dagger\hat{a}_\nu$, so that
\begin{equation}
\hat{n}_{\nu_i}\left| n_{\nu_1},...,n_{\nu_i},...\right\rangle^{\hat{q}}_{}=\textbf{1}^{\hat{q}}n_{\nu_i}\left| n_{\nu_1},...,n_{\nu_i},...\right\rangle^{\hat{q}}_{}.
\end{equation}
This leads to the following commutation relations (see \eqref{Eq:commutator1} and \eqref{Eq:commutator2}):
\begin{align}\label{Eq:commutators1}
	\left[\hat{n}^{}_{\mu},\hat{a}^{}_\nu\right]&=-\mathbf{1}^{\hat{q}}\delta^{}_{\nu\mu}\hat{a}^{}_\mu,\\
	\label{Eq:commutators2}
	\left[\hat{n}^{}_{\mu},\hat{a}^{\dagger}_\nu\right]&=\mathbf{1}^{\hat{q}}\delta^{}_{\nu\mu}\hat{a}^{\dagger}_\mu,
\end{align}
which will be very useful in constructing the Fock space and in the remainder of the paper.

\subsection{The Occupation Number and Fermi Energy}
We consider an ideal gas of deformed
particles with the non-interacting Hamiltonian
\begin{equation}
\hat{H}^0_{\hat{q}}=\sum_\nu\varepsilon_\nu\hat{n}_\nu, 
\end{equation}
where $\varepsilon_\nu$ are the single-particle energy levels. The thermal average of an observable $\hat{O}$ is shown by
\begin{equation}
    \langle\hat{O}\rangle\equiv \mathcal{Z}_{\hat{q}}^{-1}\text{Tr}_{\hat{q}}\left[\hat{\rho}_{\hat{q}}\hat{O}\right]  
    \label{Eq:rhohat}
\end{equation} 
where $\hat{\rho}_{\hat{q}}\equiv e^{-\beta (H^0_{\hat{q}}-\mu\hat{N})}$ is the equilibrium density matrix, and  $\beta=1/kT$ with $T$ being the temperature of the gas, $\mu$ is the chemical potential, $\hat{N}\equiv \sum_\nu \hat{n}_\nu$ is the operator of the total number of particles, and the grand partition function is defined as
\begin{equation}\label{Eq:Partition-Function1}
	\mathcal{Z}(\hat{q})={\rm Tr}^{}_{\hat{q}}
	\exp\left[\hat{\rho}_{\hat{q}}\right],
\end{equation}
The details of taking trace in the Fock space is represented in the APPENDIX~\ref{AppendixUSpf}, where we show that:
\begin{equation}\label{Equation:gpfunction}
	\mathcal{Z}(\hat{q}) = \prod_\nu \frac{1 - \frac{1 - \hat{q}}{2} z^{2}e^{-2\beta\epsilon_{\nu}}}{1-ze^{-\beta\epsilon_{\nu}}},
\end{equation}
and $z=e^{\beta\mu}_{}$ is the fugacity. In this stage, $\hat{q}$ can be replaced by its expectation value. Showing the statistical status of the system in a linear combination of $\left|+ \right\rangle $ and $\left|- \right\rangle $, i.e. $\left|\text{system} \right\rangle =c_+\left|+ \right\rangle +c_-\left|- \right\rangle$, we find
\begin{equation}
    \left\langle \hat{q}\right\rangle=|c_+|^2-|c_-|^2=1-2\delta,   
\end{equation}
where we used the fact that $|c_+|^2+|c_-|^2=1$, and defined $\delta=|c_-|^2$. Using these relations we end up with:
\begin{equation}\label{Eq:Partition-Function2}
	\mathcal{Z}(\delta)=\prod_{\nu}\frac{1-\delta z^{2}e^{-2\beta\epsilon_{\nu}}}{1-ze^{-\beta\epsilon_{\nu}}}.
\end{equation}
The standard thermodynamic relation for the number of particles in the $\nu$ state for the grand canonical ensemble is		
\begin{equation}\label{Equation:OCCnumber}
	\langle n_\nu \rangle = \frac{1}{\beta} \frac{\partial \ln \mathcal{Z}_{\hat{q}}^{(\nu)}}{\partial \mu}.
\end{equation}
where $\mathcal{Z}_{\hat{q}}^{(\nu)}$ is the single-mode partition function corresponding to the $\nu$th energy level. The mean occupation number of a single-particle state in energy level $\varepsilon_\nu$ is obtained:
\begin{equation}\label{Eq:DFunction}
	\bar{n}_\delta(\beta,\epsilon_\nu)\equiv\langle n_\nu\rangle=\frac{1}{e^{\beta(\epsilon_\nu-\mu)}-1}-\frac{2\delta}{e^{2\beta(\epsilon_\nu-\mu)}-\delta}.
\end{equation}
In the limits $\delta\!=\!0$ and $1$, this distribution reduces to the Bose-Einstein and Fermi-Dirac distributions, respectively.
For this distribution to be physically admissible (i.e., non‑negative for all single‑particle energies $\epsilon_\nu$), we should have either 
\begin{equation}
    \epsilon<\mu+\frac{1}{2\beta}\ln\delta \ \ \ \ \text{or}\ \ \ \ \epsilon>\mu .
\end{equation}
This tells us that there is a \textit{forbidden region} in the spectrum of deformed particles $\mu-\frac{1}{2\beta}|\ln\delta|\le\epsilon\le\mu$, which is absent in Bosonic and Fermionic limits where the forbidden region shrinks to zero. The extreme case is Bosonic for which the forbidden region is $-\infty<\epsilon<\mu$ (at $\epsilon=\mu$ Bose-Einestein condensation takes place), while for the Fermionic case there is no forbidden region, and the average occupasion number is unity (zero) for $\epsilon<\mu$ ($\epsilon>\mu$), identifying the Fermi level. For Fermion-like region ($0.5<\delta\le 1$), in the limit $\beta\to\infty$ we have
\begin{equation}
    \bar{n}_\delta(\beta\to\infty,\epsilon_\nu)=\begin{cases}
    0 & \epsilon_\nu>\mu\\
    1 & \epsilon_\nu<\mu
\end{cases}.
\label{Eq:FermionAccuNumZero}
\end{equation}
The generalized Fermi surface is defined for fermion-like particles ($0.5<\delta\le 1$) via the relation
\begin{equation}
    \int_0^{p_F^{(\delta)}}n_\delta(T=0,\epsilon_p)\text{d}^dp=N
\end{equation}
We know that in this limit Eq.~\eqref{Eq:FermionAccuNumZero} applies, where $\epsilon_F^{(\delta)}\equiv \epsilon_{p_F^{(\delta)}}$ is the generalized Fermi energy. Using the fact that $\sum_p \bar{n}_\delta(\beta\to\infty,\epsilon_p)=N$, $\sum_p \bar{n}_\delta(\beta\to\infty,\epsilon_p)\epsilon_p=E_F^{(\delta)}$, for a parabolic dispersion $\epsilon_p=\frac{p^2}{2m}$ one finds
\begin{equation}
   p_F^{(\delta)}=\frac{2\pi}{(2\delta-1)^{1/d}} \left(C_dn_d\right)^{1/d},
   \label{Eq:FermiMomentum}
\end{equation}
where $C_1=2$, $C_2=\pi$, $C_3=\frac{4\pi}{3}$, and $n_d=N/V_d$ is the density of $d$-dimensional system with volume $V_d$. One also can find the generalized Fermi energy
\begin{equation}
    E_F^{(\delta)}=\frac{NC_d^2}{d+2}\epsilon_{p_F^{(\delta)}}
\end{equation}
This formulation provides the algebraic and statistical foundation required for developing the Green’s functions and diagrammatic techniques for unified statistics in the subsequent sections.
% \subsection{mode expansion}
\section{$S$-MATRIX THEORY }\label{SEC:SMatrix}
%%%%%%%%%%%%%%%%%%%%%%%%%%%%%%%%%%%%%%%%%%
For a quantum field theory description, one typically start with an $S$-Matrix theory. We consider $d+1$ dimensions with the space‑time coordinates $x=(t,\textbf{x})$, where $\textbf{x}$ is $d$-dimensional Euclidean space, and the corresponding momentum is represented by $p$, the scalar product of which is represented by $p.x\equiv\epsilon_\textbf{p}t-\textbf{p}.\textbf{x}$, $\epsilon_\textbf{p}$ representing the energy of the particle with momentum $\textbf{p}$. For a translational invariant system, the field operators can be expanded in terms of plane waves as follows:
\begin{equation} \label{Eq:mode-expantion-qParticle}
	\hat{\psi}(x)=
	\frac{1}{\sqrt{V}}\sum_\textbf{p} e^{ip.x} a_\textbf{p} 
	\ , \ \hat{\psi}^\dagger(x)=
	\frac{1}{\sqrt{V}}\sum_\textbf{p} e^{-ip.x} a^\dagger_\textbf{p},
\end{equation}
where $V$ is the system volume, and $a_p$ and $a_p^\dagger$ are the annihilation and creation operators in the momentum space satisfying the unified algebra given in Eqs.~\eqref{Eq:algebraQhat1} and \eqref{Eq:algebraQhat2}. Note that, all the fields and operators we deal with in the rest of the paper correspond to deformed particles, and need a subscript $\hat{q}$, but we drop it to avoid unnecessary complications in notation. The fields $\hat{\psi}(x)$ and $\hat{\psi}^\dagger(x)$ are characterized and recognized by their commutation relations
\begin{equation}\label{Eq:modeexpantion}
	\begin{aligned}
		\left[\hat{\psi}(x_1),\hat{\psi}^\dagger(x_2)\right]_{\hat{q}}
		=&\frac{1}{V}\sum_{\textbf{p}^{}_1 \textbf{p}^{}_2}e^{ip^{}_1.x^{}_1-ip^{}_2.x^{}_2}\left[ a_{\textbf{p}^{}_1}, a^\dagger_{\textbf{p}^{}_2}\right]_{\hat{q}}\\&=\frac{1}{V}\sum_{\textbf{p}^{}_1}e^{ip^{}_1.(x^{}_1-x^{}_2)}\mathbf{1}^{\hat{q}},
	\end{aligned}
\end{equation}
Note that, for the ordinary particles, $\textbf{1}^{\hat{q}}$ is replaced by the ordinary identity operator, resulting in a Dirac delta function in case $t_1=t_2$, i.e. the equal-time commutation.

\subsection{Preliminary Definitions: Generalized Time Ordering and Normal Ordering}
We now develop the perturbative framework for unified statistics, which requires generalizing the time‑ordering, normal‑ordering, the $S$‑matrix expansion, and Wick’s theorem adapted with $\hat{q}$. The time-ordering operator of two operators $\hat{O}_1(t),\hat{O}_2(t')$ (composed of creation and annihilation operators) is defined as
\begin{equation}
	\begin{split}			T_{\hat{q}}\left[\hat{O}_1(t)\hat{O}_2(t')\right]&\equiv\Theta(t-t')\hat{O}_1(t)\hat{O}_2(t')\\
		&+\hat{q}^{m}\Theta(t'-t)\hat{O}_2(t')\hat{O}_1(t)+\ldots,
	\end{split}
\end{equation}
where $m$ is the number of particle‑index exchanges needed to swap the two operators.
One can show readily
\begin{equation}\label{Eq:Time-ordering}
	\begin{aligned}
		&T_{\hat{q}}\left[\hat{O}_1(t)\hat{O}_2(t')\right]
		\\&=\Big(\Theta(t-t')+\hat{q}\Theta(t'-t)\Big)T\left[\hat{O}_1(t)\hat{O}_2(t')\right],
	\end{aligned}
\end{equation}
where $T$ is the ordinary time ordering.\\
For the definition of normal ordering, we note that any general field operator $\hat{\psi}(x)$ is decomposed into two positive- and negative-frequency parts:	
\begin{equation}\label{Eq:field_operator}
    \hat{\psi}(x)\equiv A\hat{\psi}^+(x)+B\hat{\psi}^-(x),
\end{equation}
$A$ and $B$ are constant coefficient and $\hat{\psi}^+(x)$ and $\hat{\psi}^-(x)$ denote the creation and annihilation components of the field operator, respectively. These operators are typically defined by their action on the non-interacting ground state in quantum many-body systems, or equivalently to the non-interacting vacuum state in quantum field theories (see Eq.~\eqref{Eq:states} for the definition). Denoting this state by $\left| \Phi_0\right\rangle^{\hat{q}}_{}$ we have:
\begin{equation}
	\hat{\psi}^+(x)\left| \Phi_0\right\rangle^{\hat{q}}=0 ,\ \  \left(\hat{\psi}^-(x)\right)^\dagger\left| \Phi_0\right\rangle^{\hat{q}}=0.
\end{equation}
From the plane‑wave mode expansion in Eq.~\eqref{Eq:mode-expantion-qParticle}, the creation and annihilation parts of the field operator are
\begin{equation}\label{Eq:psi_plus_minus}
	\begin{split}
		&\hat{\psi}^-(x)=\frac{1}{\sqrt{V}}\sum_p e^{-ip.x} a_p^{\dagger}\\
		&\hat{\psi}^+(x)=\frac{1}{\sqrt{V}}\sum_p e^{ip.x} a_p.
	\end{split}
\end{equation}
The normal ordering moves all negative frequency operators 
$\hat{\psi}^-(x)$ to the left of positive frequency one $\hat{\psi}^+(x)$, inserting a factor of $\hat{q}$ for each transposition
\begin{equation}\label{Eq:normal-order}
	\begin{aligned}
		&N_{\hat{q}}\{\hat{\psi}_1(x_1)\hat{\psi}_2(x_2)...\hat{\psi}_n(x_n)\}\\&=(\hat{q})^\mathcal{P}\{\hat{\psi}^-(x_{i_1})...\hat{\psi}^-(x_{i_m})\hat{\psi}^+(x_{j_1})...\hat{\psi}^+(x_{j_{n-m}})\},
	\end{aligned}
\end{equation}
where $\mathcal{P}$ represents the number of permutations and the subscript sets $(i_1,...,i_m)$, and $(j_1,...,j_{n-m})$ represent the positions of the positive and negative frequency operators, respectively.\\

Specifically, for two fields case, one has
\begin{equation}\label{Eq:Tq_definition}
	T_{\hat{q}}\{\hat{\psi}^+(x_1)\hat{\psi}^-(x_2)\}=
	\begin{cases}
		\hat{\psi}^+(x_1)\hat{\psi}^-(x_2) & ;~t_1>t_2\\
		\hat{q}\hat{\psi}^-(x_2)\hat{\psi}^+(x_1) & ;~t_2>t_1
	\end{cases}
\end{equation}
and
\begin{equation}\label{Eq:Nq_definition}
	N_{\hat{q}}\{\hat{\psi}^+(x_1)\hat{\psi}^-(x_2)\}=\hat{q}\hat{\psi}^-(x_2)\hat{\psi}^+(x_1),
\end{equation}
using of which we define the contraction of two fields $\hat {\psi}(x_1)$ and $\hat {\psi}(x_2)$ by
\begin{equation}\label{Eq:contract-q}
	\overparent{\hat {\Psi}(x_1) \hat {\Psi}}(x_2)\equiv T_{\hat{q}}\{\hat {\Psi}(x_1) \hat {\Psi}(x_2)\} - N_{\hat{q}}\{\hat {\Psi}(x_1) \hat {\Psi}(x_2)\}. 
\end{equation}
This contraction reduces to the free‑particle Green’s function in the absence of interactions (Appendix~\ref{SEC:wick-nonzero}). For the finite temperatures we define the contraction as 
\begin{equation}
\begin{split}
\overparent{\hat {\Psi}(x_1) \hat {\Psi}}(x_2)&\equiv\left\langle T_{\hat{q}}\left[\hat {\Psi}(x_1) \hat {\Psi}(x_2)\right]\right\rangle_0\\
&=\text{Tr}\left\{\hat{\rho}_{\hat{q}}T_{\hat{q}}\left[\hat {\Psi}(x_1) \hat {\Psi}(x_2)\right]\right\},
\end{split}
\end{equation}
Specifically
\begin{equation}
    \begin{split}
    \overparent{\hat {a}_\nu(\tau) \hat {a}^\dagger_{\nu}}=\frac{[\hat {a}_\nu(\tau), \hat {a}^\dagger_{\nu'}]}{1-\hat{q}e^{\lambda\tau\zeta_\nu}},
    \end{split}
\end{equation}
where $\lambda=\mp1$ for $\psi$ ($\psi^\dagger$) respectively (Eq.~\eqref{Eq:contr}), and $\hat{\rho}_{\hat{q}}$ was already defined in Eq.~\eqref{Eq:rhohat}. This expression is processed in Appendix~\ref{SEC:wick-nonzero}. We notice that the subscript 0 denotes the fact that the averages are taken with respect to $H_0$. An important example of contraction is the thermal average of the time-ordered product of an annihilation operator and a creation operator (Matsubara Green's function), which is of especial importance in this study.

\subsection{Green's Functions} 	
We have an option to choose the definition of the zero temperature Green function based on the different definitions of time ordering:
\begin{equation}
	iG_{T}(p;t,t')=\frac{\langle\Psi_0|T_{}\left[a_p(t)a^{\dagger}_p(t')\right]|\Psi_0\rangle}{\langle\Psi_0|\Psi_0\rangle},
	\label{Eq:Green0}
\end{equation}
and
\begin{equation}
	iG_{\hat{q}}(p;t,t')=\frac{\langle\Psi_0|T_{\hat{q}}\left[a_p(t)a^{\dagger}_p(t')\right]|\Psi_0\rangle}{\langle\Psi_0|\Psi_0\rangle},
	\label{Eq:Green0}
\end{equation}
where $|\Psi_0\rangle$ is the interacting ground state of the system and $a_p(t)$, $a^{\dagger}_{p}(t')$  are the annihilation and creation operators in the Heisenberg picture. The denominator ${\langle\Psi_0|\Psi_0\rangle}$ ensures proper normalization of the expectation value. These two definitions are related using Eq.~\eqref{Eq:Time-ordering}:
\begin{equation}
	\begin{split}
		iG_{\hat{q}}(p;t,t')&
		=\left[\Theta(t-t')+\hat{q}\Theta(t'-t)\right]iG_T(p;t,t').
	\end{split}
\end{equation}
In \textit{finite temperature} systems, we employ the Matsubara formalism, and define the single-particle Matsubara Green's function in terms of $\tau\equiv it$ as the imaginary‑time. To make distinction between imaginary and real times, we use the abbreviation $\mathcal{X} \equiv(\tau,\textbf{x})$, showing a ($d+1$)-Euclidean coordinate.  representation for unified quantum statistics:
\begin{equation}
	\begin{aligned}
		G^{\mathcal{M}}_{\hat{q}}(\mathcal{X},\mathcal{X}')&=-\langle T_{\hat{q}}[\hat{\psi}_I^{}(\mathcal{X})\hat{\psi}_I^{\dagger}(\mathcal{X}'\tau')]\rangle
		\\&=-\frac{1}{\mathcal{Z}_{\hat{q}}}\text{Tr}_{\hat{q}}\left(\hat{\rho}_{\hat{q}}T_{\hat{q}}[\hat{\psi}_I^{}(\mathcal{X})\hat{\psi}_I^{\dagger}(\mathcal{X}')]\right)
	\end{aligned}
\end{equation}
or, in the $\nu$-space:
\begin{equation}
	\begin{aligned}
		G^{\mathcal{M}}_{\hat{q}}(\nu;\tau)
		&=-\left\langle T_{\hat{q}}a_\nu(\tau)a^{\dagger}_{\nu}(0)\right\rangle,
	\end{aligned}
\end{equation}
where $\mathcal{Z}_{\hat{q}}$ is the grand partition function. An important property of the Matsubara Green's function is its special kind of periodicity for $\tau<0$ (see Eq.~\eqref{Eq:period} in Appendix~\ref{GF-EOM})
\begin{equation}
    G^{\mathcal{M}}_{\hat{q}}(\tau+\beta)=\hat{q}G^{\mathcal{M}}_{\hat{q}}(\tau).
\end{equation}
Note that the imaginary time dependent operators for non-interacting system read~\eqref{Eq:anonInt} and~\eqref{Eq:adagger}:
\begin{equation}
    (a_\nu(\tau),a^\dagger_\nu(\tau))=\mathbf{1}^{\hat{q}}(e^{-\tau\zeta_\nu}a_\nu,e^{\tau\zeta_\nu}a^\dagger_\nu),
    \label{Eq:aadagger}
\end{equation}
which will be used in the following sections.
\subsection{$S$-Matrix Expansion}
Consider a system with the Hamiltonian
\begin{equation}
    H=H_0+H_{\text{int}},
\end{equation}
where $H_0$ is the quadratic part of the Hamiltonian for free deformed particles, and $H_{\text{int}}$ is the interaction part. The question to be processed in this section is how the correlations for the full interacting system is obtained perturbatively in terms of the correlations of the free system, which is called $S$-matrix expansion. This expansion is expressed in terms of the evolution operator, which is expanded according to the standard Dyson series~\cite{PhysRevA.111.022223}:
\begin{equation}
	\begin{aligned}
		S(t,t')=&\sum^\infty_{n=0}\frac{(-i)^n}{n!}\int^t_{t'}dt_1\int^t_{t'}dt_2...\int^t_{t'}dt_n
		\\&\times T\left[H^{}_{\text{int}}(t_1)H^{}_{\text{int}}(t_2)...H^{}_{\text{int}}(t_n)\right].
	\end{aligned}
\end{equation} 
Let $\hat{O}_H^{(1)}(t)$ and $\hat{O}_H^{(2)}(t')$ be two observables in the Heisenberg picture so that
\begin{equation}\label{representation}
	\hat{O}_H\equiv e^{iHt}\hat{O}(0)e^{-iHt}
\end{equation}
Using the Gell‑Mann–Low theorem, the expectation value in the interacting ground state $|\Psi_0\rangle$ is expressed in terms of the non‑interacting (free) ground state 
$|\Phi_0\rangle$ and operators in the interaction picture. This expansion can be shown to be:
\begin{equation}
	\begin{aligned}
		&\frac{\langle\Psi_0|T_{\hat{q}}\left[\hat{O}_H^{(1)}(t)\hat{O}_H^{(2)}(t')\right]|\Psi_0\rangle}{\langle\Psi_0|\Psi_0\rangle}
		\\&
		~=\frac{\Theta(t-t')}{\langle\Phi_0|\hat{S}|\Phi_0\rangle}\langle\Phi_0|\sum^\infty_{\nu=0}\left(-i\right)^\nu\frac{1}{\nu!}\int^\infty_{-\infty}dt_1...\int^\infty_{-\infty}dt_\nu \\&~~~~\times
        % \mathbf{1}^{\hat{q}}
        T\left[\hat{H}_{\text{int}}(t_1)...\hat{H}_{\text{int}}(t_\nu)\hat{O}_{\text{int}}^{(1)}(t)\hat{O}_{\text{int}}^{(2)}(t')\right]|\Phi_0\rangle\\&
    	~+\frac{(\hat{q})^m\Theta(t'-t)}{\langle\Phi_0|\hat{S}|\Phi_0\rangle}\langle\Phi_0|\sum^\infty_{\nu=0}\left(-i\right)^\nu\frac{1}{\nu!}\int^\infty_{-\infty}dt_1...\int^\infty_{-\infty}dt_\nu
		\\&~~~~\times
        % \mathbf{1}^{\hat{q}} 
        T\left[\hat{H}_{\text{int}}(t_1)...\hat{H}_{\text{int}}(t_\nu)\hat{O}_{\text{int}}^{(1)}(t')\hat{O}_{\text{int}}^{(2)}(t)\right]|\Phi_0\rangle,
	\end{aligned}
	\label{Eq:genSMatrix}
\end{equation}
% {\color{blue}Please double check this: I made a change} 
where $m$ is the number of permutations needed to interchange $O_1$ and $O_2$. This equation furnishes a systematic and complete framework for computing arbitrary correlation functions including all single‑particle and many‑particle Green’s functions to any order in perturbation theory for systems governed by intermediate quantum statistics.
All effects stemming from the intermediate statistics are encapsulated solely in the factor $\hat{q}$ 
% and $1^{\hat{q}}$
, which appear at each interaction vertex and in the time-ordering prescription. The remaining structure of the expansion-namely, the sum over interaction orders, time integrals, and expectation values in the free theory-remains entirely analogous to that of conventional quantum field theory.
%======================
% \subsubsection{zero temperature}
\section{GENERALIZED WICK'S THEOREM }\label{SEC:WICKsTHEOREM}
Having established the generalized $S$‑matrix theory, we now derive the Wick’s theorem appropriate for unified statistics. This theorem decomposes the time-ordered product of $n$ field operators into a sum over all possible contractions, multiplied by the normal-ordered product of the remaining operators. For systems obeying unified statistics, this decomposition must correctly incorporate the statistical factors $\hat{q}$ corresponding to each specific configuration of contractions. In the following, we prove both the zero temperature and finite temperature Wick's theorems.
 
\subsection{Wick's Theorem in Zero Temperature}\label{SEC:Wick0}
Using the definitions of time ordering, normal ordering (Eqs.~\eqref{Eq:Time-ordering},~\eqref{Eq:normal-order}) and contraction (Eq.~\eqref{Eq:contract-q}), we can derive the generalized Wick' theorem for unified statistics (see Appendix \ref{wick-zero}).
We now present the generalized Wick theorem applicable to unified quantum statistics, derived as follows:
\begin{equation}
	\begin{aligned}
		T_{\hat{q}}&\left\lbrace\hat{\Psi}_1(x_1)\hat{\Psi}_2(x_2)...\hat{\Psi}_n(x_n)\right\rbrace
		=N_{\hat{q}}\Big\{\hat{\Psi}_1(x_1)\hat{\Psi}_2(x_2)...\hat{\Psi}_n(x_n)\Big\}
		\\
		&~+\bigg(\overparent{\hat{\Psi}_1(x_1)\hat{\Psi}_2}(x_2)
		N_{\hat{q}}\Big\{\hat{\Psi}_3(x_3)...\hat{\Psi}_n(x_{n})\Big\}+\cdots+\\&~\text{all possible contractions(for even permutations)}\bigg)
		\\
		&~+\hat{q}\bigg(\overparent{\hat{\Psi}_1(x_1)\hat{\Psi}_3}(x_3)
		N_{\hat{q}}\Big\{\hat{\Psi}_2(x_2)...\hat{\Psi}_n(x_{n})\Big\}+\cdots\\&~+\text{all possible contractions(for odd permutations)}\bigg).
	\end{aligned} 
\end{equation}
Here, each term in the second and third lines represents a specific configuration of pairwise contractions. The ellipses indicate the sum over all possible ways to pair the operators, while respecting the rules of even or odd permutations.
First term is the fully normal-ordered product of all $n$ operators, with no contractions. Second term (without pre-factor $\hat{q}$ includes all contractions where the operators are paired in their natural order (i.e.,$x_1$ with $x_2$, then $x_1$ with $x_4$, etc.). These configurations correspond to even permutations of the operators from the original order and thus acquire no additional statistical phase (i.e., pre-factor $\hat{q}^0$=1).
Third term (with pre-factor $\hat{q}$)
includes all contractions whose formation requires an odd permutation of the operators relative to the original order. The simplest example is contracting $\hat{\Psi}_1(x_1)$ with $\hat{\Psi}_3(x_3)$, which necessitates skipping over $\hat{\Psi}_2(x_2)$.
Each odd permutation introduces an additional statistical factor $\hat{q}$.
According to the definition of a contraction for zero temperature Eq.~\eqref{Eq:contract-q} one ends up with
\begin{equation}
	\overparent{\hat{\psi}_I^{}(x)\hat{\psi}_I^{\dagger}}(x')=-G^0_{\hat{q}}(x,x'),
\end{equation}
where $G^0_{\hat{q}}(x,x')$ is the free Green's function.
% ({\color{blue}How did we prove this Habib}).
Then, by substituting each contraction in its definition, it is found that:
\begin{widetext}
	\begin{equation}
		\begin{aligned}
			T_{\hat{q}}&\left\lbrace\hat{\Psi}_1(x_1)\hat{\Psi}_2(x_2)...\hat{\Psi}_n(x_n)\right\rbrace
			=N_{\hat{q}}\Big\{\hat{\Psi}_1(x_1)\hat{\Psi}_2(x_2)...\hat{\Psi}_n(x_n)\Big\}
			\\
			&+\bigg(G_{\hat{q}}^0(x_1,x_2)
			N_{\hat{q}}\Big\{\hat{\Psi}_3(x_3)...\hat{\Psi}_n(x_{n})\Big\}+\cdots+G_{\hat{q}}^0(x_1,x_2)G_{\hat{q}}^0(x_1,x_4)\cdots G_{\hat{q}}^0(x_{n-1},x_n)\bigg)
			\\
			&+\hat{q}\bigg(G_{\hat{q}}^0(x_1,x_3)
			N_{\hat{q}}\Big\{\hat{\Psi}_2(x_2)...\hat{\Psi}_n(x_{n})\Big\}+\cdots+G_{\hat{q}}^0(x_1,x_3)G_{\hat{q}}^0(x_2,x_4)\cdots G_{\hat{q}}^0(x_{n-2},x_n)\bigg).
		\end{aligned} 
	\end{equation}
\end{widetext}
In this equation, the second line corresponds to even permutation of fields, while in the third line, a $\hat{q}$ appears showing that the number of permutations is odd. Note also that $\hat{q}$ has remained in this formula, showing that the average over the statistics has not been yet taken. We will use this theorem in
the following sections, especially in designing the Feynman diagrammatics. In the next section, we process the bare Green's function.
%=======================================
\subsection{Wick's Theorem for Finite Temperatures}\label{SEC:Wick1}
Here we consider finite temperaures, for which it is customary to consider imaginary time $\tau\equiv it$ (Matsubara formalism). The fields then take the following form in the interaction picture:
\begin{equation}\label{Eq:InteractionPicture}
	\hat{\psi}_I(\mathcal{X})~~\text{or}~~ \hat{\psi}^\dagger_I(\mathcal{X})=\sum_{j}\chi_j(\mathcal{X})\alpha_j,
\end{equation}
where $\alpha_j$, denotes $a_j$, or $a^\dagger_j$ and $\chi_j(\text{x}\tau)$ are basis functions of $x$ and $\tau$ (more explicitly $\phi^{0}_j(\text{x}) e^{-\tau\zeta_j}$ or $\phi^{0}_j(\text{x})^* e^{\tau\zeta_j}$ for each case respectively).	 
	
The generalized Wick's theorem then asserts that $\langle T_{\hat{q}}[\hat{A}\hat{B}\hat{C}\cdots\hat{F}]\rangle_0$ is equal to the sum over all possible fully contracted terms. The proof of this theorem is given in Appendix~\ref{SEC:wick-nonzero}. This theorem states (for $\tau_A>\tau_B>\tau_C>\tau_D\cdots>\tau_F$):
\begin{equation}
	\begin{aligned}
		&\langle \hat{A}\hat{B}\hat{C}\hat{D}\hat{E}\cdots\hat{F}\rangle_0
		\\&=\left([\overparent{\hat{A}\hat{B}}\hat{C}\hat{D}\hat{E}\cdots\hat{F}]+[\overparent{\hat{A}\hat{D}}\hat{B}\hat{C}\hat{E}\cdots\hat{F}]+\cdots\right)
		\\&+\hat{q}\left([\overparent{\hat{A}\hat{C}}\hat{B}\hat{D}\hat{E}\cdots\hat{F}]+[\overparent{\hat{A}\hat{E}}\hat{B}\hat{C}\hat{D}\cdots\hat{F}]+\cdots\right).
	\end{aligned}
\end{equation}
Therefore, the generalized Wick theorem at finite temperature has a similar formalism to the usual case, with the difference that when the operators are moved, a generalized algebra is used. Note that, like by definition
\begin{equation}
    \overparent{\hat{\psi}(\mathcal{X})\hat{\psi}^\dagger}(\mathcal{X}')=G_{\hat{q}}^{\mathcal{M}}(\mathcal{X},\mathcal{X}').
\end{equation}
Which allows us to break up the higher order terms in terms of the non-interacting Green's functions.
%=======================================
\section{FINITE TEMPERATURE GREEN'S FUNCTION}\label{SEC:FiniteTemperatureGreenFunction}
To describe the particles obeying unified quantum statistics at non-zero temperature, we develop the Matsubara formalism and finite-temperature Green's functions. This development requires a proper definition of the Green's function in the complex (imaginary-time) representation. 
\subsection{Evaluation of Non-Interacting Matsubara Green's Function}
Substituting $a_{\nu}(\tau)$ and using the orthonormal occupation‑number basis, the Green's function can be computed explicitly: For non-interacting case, using the relation~\eqref{Eq:aadagger}, after some straightforward calculations (Appendix~\ref{AppendixB}):
\begin{equation}
	\begin{aligned}
	G^{\mathcal{M}0}_{\hat{q}}(\nu;\tau)&=-\frac{e^{-\zeta_\nu\tau}}{\mathcal{Z}_{\hat{q}}}
	\sum_{\{n_i\}}\langle\{n_i\}|e^{-\beta(\hat{H}_0-\mu \hat{N})}\mathbf{1}^{\hat{q}}\hat{a}_\nu \hat{a}^{\dagger}_{\nu})|\{n_i\}\rangle\\
    &=-\frac{e^{-\tau\zeta_\nu}}{\mathcal{Z}_{\hat{q}}}\Bigg(\mathcal{Z}_{\hat{q}}+\hat{q}\sum_{\{n_i\}}e^{-\beta\sum_{i}\zeta_i n_i}\mathbf{1}^{\hat{q}}n_\nu\Bigg)
	\end{aligned}
\end{equation}
where the superscript $0$ shows that it is a bare Green's function, and $\zeta_\nu=\epsilon_\nu-\mu$.
Using the properties of $\textbf{1}^{\hat{q}}$ we finally find (Appendix~\ref{AppendixB}):
\begin{equation}
	\begin{aligned}
	G^{\mathcal{M}0}_{\hat{q}}(\nu;\tau)&=-e^{-\tau\zeta_\nu}\Bigg(1-\frac{\hat{q}}{\zeta_\nu}\frac{\partial}{\partial\beta}\ln{\mathcal{Z}^{(\nu)}_{\hat{q}}}\Bigg),
	\end{aligned}
\end{equation}
where $\mathcal{Z}^{(\nu)}_{\hat{q}}$ is the contribution of state $\nu$ to the grand partition function.
By taking the derivative and substituting the expectation value of the operator $\hat{q}$, ($1 - 2\delta\equiv\vartheta$), the Green’s function reduces to a purely numerical form that depends on the parameter $\delta$:
\begin{equation}\label{Eq:GF-main}
	\begin{aligned}
		G^{\mathcal{M}0}_{\delta}(\nu,\tau)&=-e^{-\zeta_\nu  \tau } \left[1+\vartheta\left(\frac{1}{e^{\beta\zeta_\nu }-1}-\frac{2 \delta }{e^{2 \beta  \zeta_\nu }-\delta}\right)\right]
		\\&=-e^{-\zeta_\nu  \tau } \left(1+\vartheta n_\nu\right)
	\end{aligned}
\end{equation}
where $\vartheta=(1-2 \delta)$. The	Fourier transformation of $G^{\mathcal{M}0}_{\delta}(\nu,\tau)$ over $[0,\beta]$ gives an analytic formula, and gives the explicit form of the generalized Matsubara frequencies as follows:
\begin{equation}
   \begin{aligned}
   \label{Eq:GFomega}
		G^{\mathcal{M}0}_{\delta}(\nu,i\omega_n^{(j)})&=\int_0^\beta d\tau e^{i\omega_n^{(j)}\tau}_{}G^{\mathcal{M}0}_{\delta}(\nu,\tau),  
       \\&=\dfrac{\mathcal{A}_\delta(\zeta_\nu,i\omega_n^{(j)})}{i\omega_n^{(j)}-\zeta_\nu},
   \end{aligned}
\end{equation}
which is served as the spectral representation of the single‑particle Green’s function. The generalized $\delta$-dependent Matsubara frequencies read ($j=1,2$ , see Appendix~\ref{AppendixB})
\begin{equation}
	i\beta\omega_n^{(j)}=(2-j)2in\pi+(j-1)\frac{2in\pi+\ln{\delta}}{2}.
\end{equation}
Note that $\omega_n$ now has an imaginary part
\begin{equation}
    \text{Im} [\omega_n^{(j=1)}]=0,\ \text{Im} [\omega_n^{(j=2)}]=\frac{-1}{2\beta}\ln\delta,
\end{equation}
meaning that, the Matsubara energies have already \textit{damping modes}. The two standard limits, there are no such damping modes. To see this we notice that:
\begin{equation}
    F:\begin{cases}
    \text{Im} [\omega_n^{(j=1)}]=0 \\
    \text{Im} [\omega_n^{(j=2)}]=0
    \end{cases}
    ,\ B:\begin{cases}
    \text{Im} [\omega_n^{(j=1)}]=0 \\
    \text{Im} [\omega_n^{(j=2)}]=+\infty
    \end{cases}
\end{equation}
For the bosonic case, this implies that the ... expansion of the second frequencies $j=2$ is zero (since $i\omega_n$ appears in the denominator), guaranteeing that the Green's function is obtained using the first term $j=1$. The amplitude is defined as:
\begin{equation}
    \begin{split}
    \mathcal{A}_\delta(\zeta_\nu,i\omega_n^{(j)})&\equiv-\left(1+\vartheta n_\nu\right)\left(e^{\beta(i\omega_n^{(j)}-\zeta_\nu)}-1\right)\\
    &=-\left(1+\vartheta n_\nu\right)\left[\left((-1)^n_{}\sqrt{\delta}\right)^{j-1}_{}e^{-\beta\zeta_\nu^{}}_{}-1\right].
    \end{split}
\end{equation}
There is a problem with this for of the Green's function amplitude: it diverges as $\beta\to \infty$ for $\zeta_\nu<0$. Note that this divergence is not UV or IR divergence as one sees in quantum field theories since it is statistics-generated divergence for $T\to 0$. In this case, we need a regulator, that prevent this divergence:
\begin{equation}
    \begin{split}
    \mathcal{A}^{(\epsilon)}_\delta(\zeta_\nu,i\omega_n^{(j)})&\equiv \chi_\epsilon(\beta\zeta_\nu)\mathcal{A}_\delta(\zeta_\nu,i\omega_n^{(j)}).
    \end{split}
    \label{Eq:regulated}
\end{equation}
where $\epsilon$ is a small real positive quantity, and the regulator is defined as
\begin{equation}
    \chi_\epsilon(x)\equiv \exp\left[\Theta(-x)f(-\epsilon x)x\right]
    \label{Eq:Regulator}
\end{equation}
and $f(X)$ is a smooth function that is negligibly small for $X<1$, and is 1 for $X\gg1$. One choice is
\begin{equation}
    f(X)\equiv e^{-\frac{1}{X}},
\end{equation}
while one may have other choices like the stretched sigmoid. is related to the physical limitations of the system. In particular, one may choose 
\begin{equation}
    \epsilon\equiv \delta(1-\delta)\frac{k_B T_m}{\mu},    
\end{equation}
where $T_m$ denotes the minimum temperature experimentally accessible in the laboratory.
Alternatively, one may start with the equation of motion (see Eq.~\eqref{Eq:EqofMo} In Appendix~\ref{GF-EOM}):
\begin{equation}
    \frac{\partial G^{\mathcal{M}}_{\hat{q}}(\nu,\tau)}{\partial\tau}+\zeta_\nu G^{\mathcal{M}}_{\hat{q}}(\nu,\tau)=-\delta(\tau),
\end{equation}
Taking a direct Fourier transformation of this equation gives rise exactly to Eq.~\eqref{Eq:GFomega}.
\begin{figure}
	\centering
	\includegraphics[width=1\linewidth]{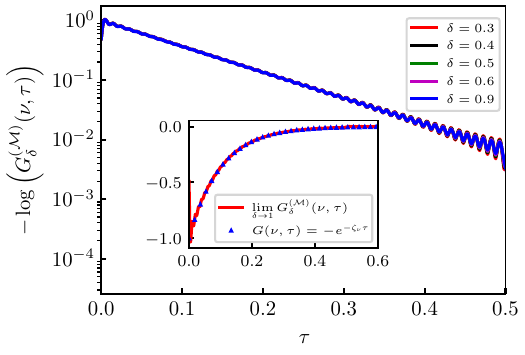}
	\caption{ The standard single-particle Green’s function, defined in Eq.~\eqref{Eq:GFomega}, is plotted as a function of $\tau$ for several values of $\delta$. The observed deviations between the corresponding Green’s functions are relatively small yet finite. The inset highlights the case of $\delta=1$, where the result is compared with the conventional expression for normal fermions, $G(\epsilon_{\nu},\tau)=-\exp(-\zeta_{\nu}\tau)$.}
	\label{Fig:GreenFunction}
\end{figure}
Fig.~\ref{Fig:GreenFunction} shows the imaginary-time Matsubara Green's function for various $\delta$ values. For different values of $\delta$, the behavior of the Green's function is the same and their values have little difference.
In the limit $\delta\to 1$, the behavior reduces to that of the fermionic Green's function.
%=========================================
\subsection{A Fermi-Liquid Description}\label{SpectralDensity}
To have a Fermi liquid description, one should find a one-to-one correspondence between the states on non-interacting and interacting systems, and importantly the Green's function. Our idea in this section is to find a correspondence between the states of the system with deformed statistics, and the interacting system with conventional statistics. We start with the Dyson's expansion theorem, stating that the single-particle Green's function in the momentum space can be written as follows:
\begin{equation}
    G^{\mathcal{M}}_{\delta}(\textbf{p},i\omega_n^{(j)})=
    \frac{\mathcal{A}_{\delta}(\zeta_{\textbf{p}},i\omega_n^{(j)})}{\,i\omega_n^{(j)} - \zeta_{\textbf{p}}^{} - \mathcal{A}_\delta(\zeta_{\textbf{p}},i\omega_n^{(j)})\Sigma_{\delta}(\textbf{p},i\omega_n^{(j)})},
    \label{eq:Dyson1}
\end{equation}
where $\zeta_{\textbf{p}}^{} = \textbf{p}^{2}/2m - \mu$ is the bare electron energy measured from the chemical potential $\mu$, and $\Sigma_{\delta}(\textbf{p},i\omega_n^{(j)})$ is the self-energy a complex, momentum and frequency-dependent function that encodes the full effect of many-body interactions. Note that, according to Eq.~\eqref{Eq:GFomega}:
\begin{equation}
G^{\mathcal{M}}_{\delta}(\textbf{p},i\omega_n^{(j)})^{-1}=G^{\mathcal{M}0}_{\delta}(\textbf{p},i\omega_n^{(j)})^{-1}-\Sigma_{\delta}(\textbf{p},i\omega_n^{(j)}).
\end{equation}
$\Sigma_{\delta}$ is directly related to the contributions of Feynman diagrams, to be explored in the following sections. \\

To find the retarded zero-temperature ($T=0$) Green's function we utilize the analytic continuation $i\omega_n^{(j)}\to \omega+i0^+$ described in SEC.~\ref{AppendixC}. Substituting this into Eq.~\eqref{Eq:GFomega} yields the retarded Green's function $G_{\delta}^{\mathcal{R}}(\textbf{p},\omega)$: 
    \begin{equation}
    G^{\mathcal{R}}_{\delta}(\textbf{p},\omega)=
    \frac{\mathcal{A}_\delta^{\mathcal{R}}(\zeta_{\textbf{p}},\omega)}{\,\omega - \zeta_{\textbf{p}}^{} - \tilde{\Sigma}_{\delta}^{\mathcal{R}}(\textbf{p},\omega)},
    \label{eq:Dyson2}
\end{equation}
where $\Sigma_{\delta}^{\mathcal{R}}(\zeta_{\textbf{p}},\omega)\equiv \Sigma_{\delta}(\zeta_{\textbf{p}},i\omega^{(j)}_n\to\omega+i0^+)$ is the retarded self energy, and the retarded amplitude is
\begin{equation}
\begin{split}
\mathcal{A}_\delta^{\mathcal{R}}(\zeta_{\textbf{p}},\omega)&\equiv \lim_{\beta\to\infty}\mathcal{A}_\delta(\zeta_{\textbf{p}},i\omega^{(j)}_n\to\omega+i0^+)\\
&=1+\vartheta n_{\textbf{p}}=\mathcal{A}_\delta^{\mathcal{R}}(\zeta_{\textbf{p}})
\end{split}
\label{Eq:AP}
\end{equation}
and $\tilde{\Sigma}_{\delta}^{\mathcal{R}}(\textbf{p},\omega)\equiv \mathcal{A}_\delta^{\mathcal{R}}(\zeta_{\textbf{p}})\Sigma_{\delta}^{\mathcal{R}}(\textbf{p},\omega)$. In the remainder of the paper, all $\mathcal{A}$ are the regulated ones according to Eq.~\eqref{Eq:regulated}, and to avoid complications in notation, we do not explicitly write its $\epsilon$ dependence. We notice that, as pointed out in the last line $\mathcal{A}_{\delta}^{\mathcal{R}}$ is independent of $\omega$. Note that the poles of Eq.~\eqref{eq:Dyson2} identifies the energy spectrum of the deformed particles:
\begin{equation}
\begin{aligned}
    \omega- \zeta_\textbf{p}^{} - \tilde{\Sigma}_{\delta}^{\mathcal{R}}(\textbf{p},\omega) &= 0.
    \label{eq:spectrum_eq1}
\end{aligned}
\end{equation}
As a normal strategy, one can consider only the real part of the self-energy, and find the solution of the above equation, and the expand the Green function around that solution. Let us call this solution as $\omega^*\equiv\tilde{\zeta}_\textbf{p}$, and:
\begin{equation}
    \begin{aligned}
        &\tilde{\Sigma}_{\delta}^{\mathcal{R}}(\textbf{p},\omega)  
    \approx\\
    &\tilde{\Sigma}_{\delta}^{\mathcal{R}}(\textbf{p},\tilde{\zeta}_\textbf{p})
    +
    \left.\frac{\partial~\tilde{\Sigma}_{\delta}^{\mathcal{R}}}{\partial \omega}\right|_{\omega=\tilde{\zeta}_\textbf{p}^{}}
    (\omega - \tilde{\zeta}_\textbf{p}^{})
    + \mathcal{O}\!\left[(\omega-\tilde{\zeta}_\textbf{p}^{})^{2}\right].
    \label{eq:Taylor1}
    \end{aligned}
\end{equation}
Substituting~\eqref{eq:Taylor1} into~\eqref{eq:Dyson2} and invoking the
on-shell condition~\eqref{eq:spectrum_eq1} ($\zeta_\textbf{p} + \text{Re}\tilde{\Sigma}^{\mathcal{R}}(\textbf{p},\tilde{\zeta}_\textbf{p}^{}) = \tilde{\zeta}_\textbf{p}$), the
Green's function denominator simplifies to
\begin{equation}
    \begin{aligned}
    &G^{\mathcal{R}}(\mathbf{p},\omega)\approx
    \frac{Z_{\textbf{p}}}{\omega - \tilde{\zeta}_{\textbf{p}}^{}-\frac{i}{\tau_{\textbf{p}}}}\,,
    \label{eq:G_QP1}
    \end{aligned}
\end{equation}
where the \emph{wave-function renormalization factor} (or quasiparticle residue), and the particle life time are defined as
\begin{equation}
    Z^{\delta}_\textbf{p} =
    \frac{\mathcal{A}_\delta(\zeta_{\textbf{p}})}{
        1 - \left.\dfrac{\partial\tilde{\Sigma}_{\delta}^{\mathcal{R}}}{\partial \omega}
            \right|_{\omega=\tilde{\zeta}_\textbf{p}^{}}}
    , \ \tau_\textbf{p}^{-1}\equiv \frac{\text{Im}\tilde{\Sigma}_{\delta}^{\mathcal{R}}(\text{p},\omega)}{1 - \left.\dfrac{\partial\tilde{\Sigma}_{\delta}^{\mathcal{R}}}{\partial \omega}
            \right|_{\omega=\tilde{\zeta}_\textbf{p}^{}}}
    \label{eq:Zp1}
\end{equation}
Now it is a good place to make a duality between an interacting fermionic system obeying standard statistics and a non-interacting system with deformed particles, both following the Fermi liquid scenario. For the former case $\mathcal{A}_\delta(\text{p})$ is unity, while the denominator is different from one, while in the dual (latter) system the denominator is unity while $\mathcal{A}_\delta(\text{p})$ is given by Eq.~\eqref{Eq:AP}. The duality then reads: 
\begin{equation}
\boxed{
\left.\mathcal{A}_{\delta}(\text{p})\right|_{\text{DNIS}}=\left(1 - \left.\dfrac{\partial\Sigma_{\delta=1}^{\mathcal{R}}}{\partial \omega}
            \right|_{\omega=\tilde{\zeta}_\textbf{p}^{}}\right)^{-1}_{\text{OIS}},}
\end{equation}
where DNIS stands for ``deformed non-interacting system" and OIS stands for ``ordinary interacting system". In the remainder of this paper, we assess this duality proposition.\\

The spectral function $ A_{\delta}(\textbf{p},\omega)$ is then derived by taking the imaginary part of the retarded Green's function, which describes the probability distribution for finding a particle or hole with energy $\omega$ and momentum $\textbf{p}$:
	\begin{equation}
		\begin{aligned}
			A_{\delta}(\textbf{p},\omega)=-\frac{1}{\pi}\text{Im}G_{\delta}^{\mathcal{R}}(\textbf{p},\omega)
		\end{aligned}
	\end{equation}
For an analytic expression, we take $\tau_\textbf{p}$ a large quantity in Eq.~\eqref{eq:G_QP1}, meaning that the quasi particles' life time is large. In the limit $\tau_\textbf{p}^{-1}\to 0^+$ we obtain
\begin{equation}
    \begin{split}
    A_{\delta}(\textbf{p},\omega)&=Z^{\delta}_\textbf{p}\delta(\omega-\tilde{\zeta}_\textbf{p}),
    \end{split}
\end{equation}
where $Z^\delta_\textbf{p}$ is given in Eq.~\eqref{eq:Zp1}, and we used the identity $\lim_{\eta\to 0}\frac{1}{x\pm i\eta}=\mathcal{P} \left(\frac{1}{x}\right)\mp i\pi\delta(x)$, in which $\mathcal{P}$ is the Cauchy principal value. This relation is in accordance with the standard Fermi liquid scenario where $Z^\delta_\textbf{p}$ shows the amount of drop of the particles number in the \textit{effective Fermi surface}. \\
\begin{figure}
	\centering
	\includegraphics[width=1\linewidth]{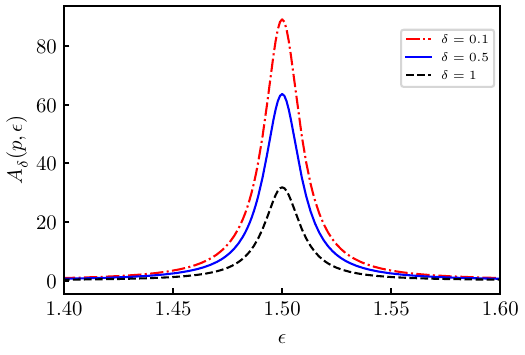}
	\caption{ This figure displays a distinct, symmetrical peak in the spectral function around $\epsilon \approx 1.49$. }
	\label{fig:SDQ}
\end{figure}
%===============================================

\begin{figure}
	\centering
	\includegraphics[width=1\linewidth]{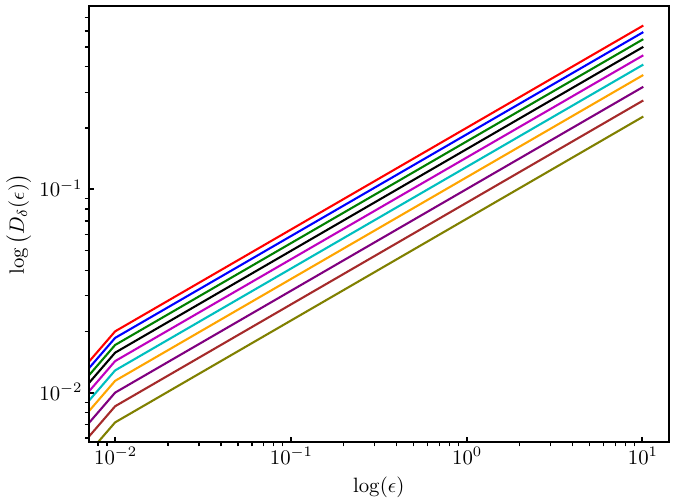}
	\caption{This figure presents the logarithm of the density of states as a function of energy $\epsilon$, 
	for different statistical parameter ($\delta$) values, ranging from 0.1 at the top to 1 at the bottom.}
	\label{fig:DOS}
\end{figure}

This form of the spectral function, allows one to calculate the density of states (DOS, shown by $\mathcal{D}_\delta(\epsilon)$), where $\epsilon$ is the single-particle energy. To avoid unnecessary complications, we consider the non-interacting case, where $\tilde{\zeta}_\textbf{p}=\zeta_\textbf{p}=\frac{\textbf{p}^2}{2m}$, and $\mu\to 0$. This leads us to conclude:
\begin{equation}\label{Eq:DOS}
    \begin{split}
     \frac{(2\pi)^3}{V}\mathcal{D}_\delta(\epsilon)&=\int A_\delta(\textbf{p},\epsilon)\text{d}\textbf{p}\\
     &=\int Z^\delta_\textbf{p}~\delta\left(\epsilon-\frac{\hbar^2\textbf{p}^2}{2m}\right) \text{d}\textbf{p}
    \end{split}
\end{equation}
where $V$ is the system volume. The results in the following simple expression, valid for free deformed particles (see Appendix~\ref{App:DOS} for the details):
\begin{equation}
	\frac{\mathcal{D}_\delta(\epsilon)}{\mathcal{D}_\text{F}(\epsilon)}=
    % \frac{1}{3}(1+2\delta),
    (3-2\delta),
\end{equation}
where $\mathcal{D}_\text{F}(\epsilon)=(V/4\pi^2)(2m/\hbar^2)^{3/2}\epsilon^{1/2}$ is the DOS for free fermionic system. This function is shown for various $\delta$ values in Fig.~\ref{fig:DOS}.
%===========================figure==============
%=========================================
\section{Diagrammatic Technique}\label{SEC:DiagrammaticParticles}
%%%%%%%%%%%%%%%%%%%%%%%%%%%%%%%%%%%%%%%%
To establish diagrammatic techniques for systems governed by unified quantum statistics, it is necessary to generalize the standard Feynman rules on the basis of the generalized Wick's theorem and the $S$-matrix expansion formulated earlier. In what follows, we outline these generalized Feynman rules for both zero and finite temperature. To avoid unnecessary complications, we adopt the finite-temperature notation throughout, since the corresponding zero-temperature limit follows trivially from it.

We consider a translationally invariant system of interacting, deformed spinless particles, described by the interaction Hamiltonian
\begin{equation}
	\hat{H}_{\text{int}} 
	=
	\frac{1}{2}\int d\textbf{x}\, d\textbf{x}' v_0(x,x') \hat{\psi}^{\dagger}(x) \hat{\psi}^{\dagger}(x') \hat{\psi}(x') \hat{\psi}(x),
    \label{Eq:Hint}
\end{equation}
where $v_0(x,x')\equiv v_0(\textbf{x}-\textbf{x}')$ denotes a bare instantaneous interaction between particles located at $\textbf{x}$ and $\textbf{x}'$, and the subscript $0$ indicates that this is the bare interaction. In this expression, the field operators are understood to be evaluated at equal imaginary time, so that an implicit factor $\delta(\tau-\tau')$ is contained within the interaction; for particles carrying spin, the corresponding spin degrees of freedom must be incorporated appropriately.

This interaction potential admits the momentum-space expansion
\begin{equation}
	\hat{H}_{\text{int}}=\sum_{pp'Q}v_0(p,p',Q)a^\dagger_{p+Q}a^\dagger_{p'-Q}a^{}_p a^{}_{p'},
\end{equation}
where $p$, $p'$, and $Q$ denote momenta, and $v_0(p,p',Q)\equiv v_0(Q)$ (the jellium model) is the interaction expressed in Fourier space, depending only on the momentum transfer $Q$.

Using this interaction, and as derived in detail in Appendix~\ref{FeynmanD}, the formal perturbative expression for the Matsubara Green's function follows from the $S$-matrix expansion technique and takes the form
\begin{equation}
	\begin{aligned}
		G^{\mathcal{M}}_{\hat{q}}(x,y) &=
		\sum_{m=0}^{\infty} \frac{(-i)^m}{m!} \int_0^{\beta} d\tau_1 \cdots \int_0^{\beta} d\tau_m 
		\\&\times
		\left\langle T_{\hat{q}}\left[\hat{H}_{\text{int}}(\tau_1) \cdots \hat{H}_{\text{int}}(\tau_m) \hat{\psi}_{}(x) \hat{\psi}^{\dagger}_{}(y)\right]\right\rangle_0,
    \end{aligned}
\end{equation}
where, as before, the subscript $0$ indicates that the thermal expectation value is to be evaluated with respect to the non-interacting Hamiltonian. This form of the expansion is precisely what allows us to apply the generalized Wick's theorem developed in Sec.~\ref{SEC:WICKsTHEOREM} to evaluate each term order by order.

To illustrate the procedure, consider the first-order term in this perturbative series, which involves an expectation value of the type (denoting space-time coordinates by $x_1$ and $x_2$, applicable to both real- and imaginary-time formalisms)
\begin{equation}
	\langle T_{\hat{q}}[\underbrace{\hat{\psi}^\dagger(x_1)\hat{\psi}^\dagger(x'_1)\hat{\psi}(x'_1)\hat{\psi}(x_1)}_{\text{Int}}\underbrace{\hat{\psi}(x)\hat{\psi}^\dagger(y)}_{\text{GF}}]\rangle.
\end{equation}
This six-operator expectation value can be evaluated systematically using the generalized Wick's theorem, according to which every operator interchange requires a multiplicative factor of $\hat{q}$. Since $\hat{q}^{m=2n}_{}=1$ and $\hat{q}^{m=2n+1}_{}=\hat{q}$, where $m$ is the number of field permutations needed to perform the contraction, the sign (or, more precisely, the $\hat{q}$-factor) associated with each contraction is fully determined by the parity of the corresponding permutation.

The resulting perturbative expansion admits a diagrammatic representation based on the following set of basic rules:
\begin{itemize}
      \item The basic diagrammatic element is a solid line, representing the contraction of two field operators at $x_1$ and $x_2$: $\langle T_{\hat{q}}\{\hat{\psi}(x_1)\hat{\psi}(x_2)\}\rangle
	=
	\overbracket{\hat{\psi}(x_{1})\hat{\psi}}(x_{2})$;
    \item Higher-order contractions, arising from the generalized statistics, are depicted using bold lines, representing the corresponding quonic propagators;
    \item A wavy line connecting two spatial points $x_1$ and $x_2$ represents the instantaneous interaction $v_0(x_1,x_2)$.
\end{itemize}
Concrete examples illustrating these rules can be found in Appendix~\ref{FeynmanD}. A particular caution is warranted when treating exchange terms carrying an extra factor of $\hat{q}$: because $\hat{q}$ is itself an operator rather than a c-number, this extra factor cannot simply be factored out of the Green's function, and must instead be retained inside the corresponding expectation value. Finally, we note that in momentum space, momentum conservation at each vertex follows directly from the structure of the formalism outlined above. Based on this, the diagrammatic representation of an interaction line is as follows:
%\begin{figure}[H]
%    \centering
%    \includegraphics[width=0.5\linewidth]{01.jpg}
%    \label{fig:placeholder}
%\end{figure}
	 \begin{equation}
	 	\begin{tikzpicture}[baseline={(current bounding box.center)}]
	 		\begin{feynman}
	 			\vertex (a1) at (-0.8,1.2) {$P$};
	 			\vertex (a2) at (-0.2,0) {};
	 			\vertex (a3) at (-0.8,-1.2) {$P+Q$};
	 			\vertex (a4) at (1.4,0) {};
	 			\vertex (a5) at (2,1.2) {$P'$};
	 			\vertex (a6) at (2,-1.2) {$P'-Q$};
	 			%===============================================
	 			\vertex (b1) at (0,-0.13) {};
	 			\vertex (b3) at (0,0.13) {};
	 			\vertex (b5) at (1.2,-0.13) {};
	 			\vertex (b6) at (1.2,0.13) {};
	 			%===============================================
	 			\diagram{(b1) -- [fermion,edge , line width=0.8pt] (a1)};
	 			\diagram{(a3) -- [fermion,edge , line width=0.8pt] (b3)};
	 			\diagram{(a2) -- [photon,edge , line width=0.8pt,edge label'=$Q$] (a4)};
	 			\diagram{(b5) -- [fermion,edge , line width=0.8pt] (a5)};
	 			\diagram{(a6) -- [fermion,edge , line width=0.8pt] (b6)};
	 		\end{feynman}
	 	\end{tikzpicture}
	 \end{equation}
	
   %================================
In evaluating the Green's function $G^{\mathcal{M}}_{\delta}(x,y)$, it is instructive to first examine how the generalized statistics manifests itself diagrammatically. One may verify by direct inspection (see Eq.~\eqref{Eq:FeynmanD}) that each closed loop appearing in a given diagram contributes an overall prefactor $\hat{q}$, reflecting the additional exchange phase accumulated whenever a fermion-like line is traced back onto itself.

In addition to loop contributions, further factors of $\hat{q}$ arise from the contraction structure of the diagram itself. Specifically, contractions connecting the external vertices $x$ and $y$ to internal vertices (e.g., $x^{}_1$, $x'_1$), as well as contractions among the internal vertices themselves (such as $x^{}_1 \rightarrow x'_1$), each contribute an additional factor of $\hat{q}$. To keep track of these two distinct sources of $\hat{q}$-factors, we introduce $\ell'$ to denote the total number of such contraction-related operations, and $\ell$ to denote the total number of closed loops in the diagram.

With these two bookkeeping quantities in hand, the diagrammatic rules for computing the $n$-th order contribution to the single-particle Green's function $G^{\mathcal{M}}_\delta(x,y)$ can now be stated systematically:

\begin{enumerate}
	\renewcommand{\labelenumi}{(\roman{enumi})}
	\item We construct all topologically distinct, connected diagrams containing $n$ interaction (wavy) lines and $2n+1$ directed solid lines, since these are precisely the diagrams contributing at $n$-th order in the interaction.
		
	\item Each vertex is assigned a four-dimensional space-time coordinate $x^{}_i$ when working in real space. Alternatively, when working in momentum space, each vertex instead carries the relevant conserved quantum numbers, and every propagator line is labeled by a corresponding momentum, consistent with momentum conservation at each vertex.
		
	\item Each solid line represents a propagator connecting $x$ to $y$, corresponding to a \textit{bare Green's function} $G_{\delta}^{\mathcal{M}0}(x,y)$-taken to be the Matsubara Green's function at finite temperature, or the ordinary (real-time) Green's function at zero temperature. Each wavy line, in turn, denotes an instantaneous interaction $v_0(x,y) = v_0(\mathbf{x}-\mathbf{y})\,\delta(t_x - t_y)$, where $v_0(\mathbf{x}-\mathbf{y})$ is the bare interaction potential between the spatial points $\mathbf{x}$ and $\mathbf{y}$, and the delta function enforces the instantaneous (equal-time) nature of the interaction.
		
	\item All internal coordinates, both spatial and temporal, associated with the internal vertices must be integrated over, since these are not fixed by the external points $x$ and $y$.

	\item Finally, to obtain the value of $G^{\mathcal{M}}_\delta(x,y)$ at a given order, each diagram is assigned an overall sign factor $(\hat{q})^{\ell+\ell'}$, determined by the loop count $\ell$ and contraction count $\ell'$ defined above, and every $n$-th order contribution is further multiplied by the combinatorial factor $(i/\hbar)^n$.
\end{enumerate}
These rules provide the complete diagrammatic prescription for constructing the perturbative expansion of the Green's function within the generalized statistics framework, and will be used extensively in the sections that follow.
%=================================================
\section{Average Energy and Correlation Energy}\label{AverageEnergy}
The average interaction energy for unified statistics is obtained from the expectation value of the interaction Hamiltonian. We start from the average of Eq.~\ref{Eq:Hint}, e.g.
\begin{equation}\label{Eq:EnergyPotantiol}
	\begin{aligned}
		&\langle\hat{H}_{\text{int}}\rangle =\frac{1}{2}\int \text{d}\textbf{x}\text{d}\textbf{x}'v_0(x,x')\langle\hat{\psi}^\dagger(x)\hat{\psi}^\dagger(x')\hat{\psi}(x')\hat{\psi}(x)\rangle.
	\end{aligned}
\end{equation}
%========================================
As derived in Appendix \ref{App:Average energy}, employing the generalized commutation relations, the four-operator product can be systematically rearranged. This algebraic manipulation, detailed in Eqs.~(\ref{App:Eq:EnergyPotantiol1},\ref{App:Eq:EnergyPotantiol2},\ref{App:Eq:EnergyPotantiol3}), reduces the interaction energy to:
\begin{equation}\label{EnergyPotantio}
	\begin{aligned}
		&\langle\hat{H}_{\text{int}}\rangle =\frac{1}{2}\int d\textbf{x} d\textbf{x}'v_0(x,x')
		\\&~~~~\times\left(\langle\hat{n}(x)\hat{n}(x')\rangle-\frac{1}{V}\sum_{\textbf{p}}e^{i\textbf{p}.(\textbf{x}-\textbf{x}')}\langle\mathbf{1}^{\hat{q}}\hat{\psi}^\dagger(x)\hat{\psi}(x')\rangle\right),
	\end{aligned}
\end{equation}
where we have used Eq.~\eqref{Eq:modeexpantion}, and the fact that the fields are equal-time. Defining the variation of the density correlation function
\begin{equation}
	\Delta D_{\hat{q}}(x,x')\equiv D_{\hat{q}}(x,x')-D_{\hat{q}}^0(x,x')
    \label{Eq:Dxxp}
\end{equation}
enables one to write the interaction term as
\begin{equation}\label{Eq:averageEnergy}
    \begin{aligned}
	\langle\hat{H}_{\text{int}}\rangle
	=\langle\hat{H}_{\text{int}}\rangle_0
	+\frac{1}{2}\int d\textbf{x} d\textbf{x}'v_0(x,x')\Big[\Delta D_{\hat{q}}(x,x')\Big],
	\end{aligned}
\end{equation}
i.e. the interaction energy to separate into a first-order Hartree-Fock contribution ($\langle\hat{H}_{\text{int}}\rangle_0$) and a correlation energy term~\cite{fetter2012quantum}. In the above relation 
\begin{equation}
   \begin{split}
   D_{\hat{q}}^{}(x-x')&=\langle{T}_{\hat{q}}^{}[\tilde{n}(x)\tilde{n}(x')]\rangle,
   \end{split}
\end{equation} 
is the density correlation (polarization) function, where the fluctuation density operator is defined as
\begin{equation}\label{average(n)}
	\tilde{n}(x)\equiv \hat{n}(x)-\langle\hat{n}(x)\rangle.
\end{equation}
The first-order interaction energy $(\langle\hat{H}_{\text{int}}\rangle_0)$ is the expectation value in the non-interacting ground state, while the second integral represents the correlation energy $E_{\text{corr}}$, sometimes represented as:
\begin{equation}
	\begin{aligned}
	E_{\text{corr}}&=\frac{1}{2}\int^1_0 \frac{d\lambda}{\lambda}\int d\textbf{x} d\textbf{x}'\lambda v_0(x,x')\Big[\Delta D_{\hat{q}}(x,x')\Big],
	\end{aligned}
\end{equation}
in which $x$ and $x'$ are assumed to be equal time. This expression leads to the following relation in the Fourier space:
\begin{equation}
	\begin{aligned}
	E_{\text{corr}}&\propto\int_0^1 \frac{d\lambda}{\lambda}\int d^4 Q \lambda v_0(Q)\Big[\Delta D_{\hat{q}}(Q)\Big].
	\end{aligned}
\end{equation}
where $\lambda$ is a coupling parameter, $Q=(\textbf{Q},\omega)$ is the four‑vector of wave‑vector and frequency. This energy arises from the dynamical screening and collective effects not captured in the non‑interacting picture. The correlation energy can also be expressed in terms of the full screened interaction $v_{\hat{q}}(x,x')$:
\begin{equation}\label{Eq:Dyson-equation}
	\begin{aligned}
	&v_{\hat{q}}(x,x')
	=v_0^{}(x,x')
	\\&+\int d^4x_1 d^4x'_1 v_0^{}(x,x_1)\Pi^*_{\hat{q}}(x_1,x'_1)v_{\hat{q}}^{}(x'_1,x'),
	\end{aligned}
\end{equation}
where $\Pi^*_{\hat{q}}(x,x')$ is irreducible polarization function, which is indeed the irreducible component of the full density-density correlation function $D_{\hat{q}}(x,x')$. A bubble expansion of irreducible polarization function gives rise to 
\begin{equation}
    v_{\hat{q}}(Q)=\frac{v_0(Q)}{1-v_0(Q)\Pi^*_{\hat{q}}(Q)}\equiv \frac{v_0(Q)}{\varepsilon_{\hat{q}}(Q)},
\end{equation}
which itself results in
\begin{equation}
	\begin{aligned}
		E^{}_{\text{corr}}&\propto i
		\int^1_0\frac{d\lambda}{\lambda}\int d^4Q\Big[\frac{1}{\varepsilon^\lambda_{\hat{q}}(Q)}-1-\lambda v_{0}(Q)\Pi^{0}_{\hat{q}}(Q)
		\Big].
	\end{aligned}
\end{equation}
where $\varepsilon^\lambda_{\hat{q}}(Q)$ is the full dielectric function for an interaction strength of $\lambda$.\\
The second term of Eq.~\eqref{Eq:Dxxp} is the irreducible polarization function in zeroth order, which reads:
\begin{equation}\label{Eq:D0&noninteractingsystem}
	\begin{aligned}
    D^0_{\hat{q}}(x-x')&=\langle{T}_{\hat{q}}^{}[\tilde{n}(x)\tilde{n}(x')]\rangle_0^{}\\
	&=\left\langle T_{\hat{q}}\big[\hat{n}(x')\hat{n}(x)\big]\right\rangle_0
	-\left\langle\hat{n}(x')\right\rangle_0\left\langle\hat{n}(x)\right\rangle_0.
	\end{aligned}
\end{equation}
Expanding the density operators in terms of field operators and performing the allowed contractions leads to the final result as a product of two zeroth‑order single‑particle Green’s functions (See Appendix \ref{App:Average energy}). As we noted above, $\hat{q}$ is inside the expectation values in the Wick's theorem: 
% ({\color{blue}What happened to the factor $i$?}):
\begin{equation}\label{Eq:D0-Noninteracting-System}
    \begin{aligned}
		D^0_{\hat{q}}(x,x')&\approx\hat{q} G_{\hat{q}}^{\mathcal{M}0}(x,x')G_{\hat{q}}^{\mathcal{M}0}(x',x),
	\end{aligned}
\end{equation}
which is diagrammatically represented by a bubble diagram in coordinate space. An approximation, which corresponds to a minimal variation of Fermionic and Bosonic situations is to replace $\hat{q}$ with its expectation value, which is explored in the next section.
	%-------------------------------------------
\section{Polarization Operator, and Dielectric Function}\label{DielectricFunction}
In this section we use the diagrammatic techniques to calculate many-body response functions and other quantities. This section is much more technical. The zeroth-order (or bare) polarization function $\Pi^0_{}$ is based on Eq.~\eqref{Eq:D0-Noninteracting-System}:
\begin{equation}
	\Pi^0_{\hat{q}}(x,x')=D^0_{\hat{q}}(x,x'),
\end{equation}
or in the momentum–frequency space:
\begin{equation}\label{LindhardF}
    \begin{aligned}
	\Pi^0_{\hat{q}}(Q,\omega^{(j)}_n)&=\frac{{\hat{q}}}{\beta}\sum_{j',m}\int \frac{d^3p}{(2\pi)^3}\\&\times G_{\hat{q}}^{\mathcal{M}0}(p,i\epsilon_m^{(j')})G_{\hat{q}}^{\mathcal{M}0}(p+Q,i\omega_n^{(j)}+i\epsilon^{(j')}).
	\end{aligned}
\end{equation}
It's represented diagrammatically as:
	 \begin{equation}
	 %\begin{center}
	         \begin{tikzpicture}
	         \begin{feynman}
	             \vertex (a) ;
	             \vertex [right=of a] (b) ;
	             \vertex [right=of b] (c) ;
	             \vertex [right=of c] (d) ;
	
	             \diagram{
	             (b) -- [fermion, quarter right,edge label'=$p~\text{,}~i\epsilon_m^{(j')}$] (c);
	             (c) -- [fermion, quarter right,edge label'=$p+Q~\text{,}~i\epsilon_m^{(j')}+i\omega_n^{(j)}$] (b);
	             (a) -- [boson] (b);
	             (c) -- [boson] (d);
	             };
	             %%%%%%%%%%%%%%%%%%%%%%%%%%
	         \node [left] at (a.south) {$Q~\text{,}~i\omega_m^{(j)}$};
	          \node [draw, circle, fill=black, inner sep=1pt] at (b) {};
	          %%%%%%%%%%%%%%%%%%%%%%%%%%
	         \node [draw, circle, fill=black, inner sep=1pt] at (c) {};
	          \node [right] at (d.south) {$Q~\text{,}~i\omega_m^{(j)}$};
	          %%%%%%%%%%%%%%%%%%%%%%%%%%%%%%
	         \end{feynman}
	     \end{tikzpicture}
	 %\end{center}
	 \end{equation}
     \vspace{0.5\baselineskip}
Starting from the definition of the finite-temperature Green’s function for unified statistics, we insert its explicit form into the bubble integral:
\vspace{0.5\baselineskip}
\begin{widetext}
\begin{equation}
	\begin{aligned}
		\Pi^0_{\hat{q}}(Q,\omega^{(j)}_n)&=\frac{{\hat{q}}}{\beta}\sum_{j',m}\int \frac{d^3p}{(2\pi)^3} 
		\left(\frac{\chi_\epsilon(\beta\zeta_{p}) B_{\hat{q}}(p)\left(e^{\beta(i\epsilon^{(j')}_m-\zeta_{p})}-1\right)}{i\epsilon^{(j')}_m-\zeta_p}\right)
		% \\&\times
        \left(\frac{\chi_\epsilon(\beta\zeta_{p+Q}) B_{\hat{q}}(p+Q)\left(e^{\beta(i\epsilon^{(j')}_m+i\omega_n^{(j)}-\zeta_{p+Q})}-1\right)}{i\epsilon^{(j')}_m+i\omega_n^{(j)}-\zeta_{p+Q}}\right),
	\end{aligned}
\end{equation}
where $B_{\hat{q}}(p)=\left(1+{\hat{q}} n_{p}\right)$, and $\chi_\epsilon(.)$ is the regulator defined in Eq.~\eqref{Eq:Regulator}.
The product of the two Green’s functions can be combined using a partial-fraction decomposition:
\begin{equation}
	\begin{aligned}
		\Pi^0_{\hat{q}}(Q,\omega^{(j)}_n)
		=&\frac{{\hat{q}}}{\beta}\sum_{j',m}\int \frac{d^3p}{(2\pi)^3}
        \frac{B_{\hat{q}}(p)B_{\hat{q}}(p+Q)\chi_\epsilon(\beta\zeta_{p})\chi_\epsilon(\beta\zeta_{p+Q})\left(e^{\beta(i\epsilon^{(j')}_m-\zeta_{p})}-1\right)\left(e^{\beta(i\epsilon^{(j')}_m+i\omega_n^{(j)}-\zeta_{p+Q})}-1\right)}{i\omega_n^{(j)}-\zeta_{p+Q}+\zeta_p}
        \\&\times
		\left(\frac{1}{i\epsilon^{(j')}_m-\zeta_p}-\frac{1}{i\epsilon^{(j')}_m+i\omega_n^{(j)}-\zeta_{p+Q}}\right).
	\end{aligned}
\end{equation}
\end{widetext}
\vspace{0.5\baselineskip}
Shifting the summation index in the second term $m\to m-n$ and $j\to j'$, allows us to exploit the Matsubara sum relations independently. The result takes a compact form:
\begin{equation}\label{Eq:polarization function}
	\begin{aligned}
			&\Pi^0_{\hat{q}}(Q,\omega)
			\\&=\frac{\hat{q}}{\beta}\int \frac{d^3p}{(2\pi)^3}\frac{B_{\hat{q}}(p)B_{\hat{q}}(p+Q)\left[S_1(\zeta_p)-S_2(A_{p+Q})\right]}{\zeta_{p+Q}-\zeta_p-\omega-i\eta},
	\end{aligned}
\end{equation}
with the auxiliary sums defined as:
\begin{equation}
	\begin{aligned}
		&S_1(\zeta_p)=\frac{1}{\beta}\sum_{j',m}
        \chi_\epsilon(\beta\zeta_{p})\chi_\epsilon(\beta\zeta_{p+Q})\\&\times\frac{\left(e^{\beta(i\epsilon^{(j')}_m-\zeta_{p})}-1\right)\left(e^{\beta(i\epsilon^{(j')}_m+i\omega_n^{(j)}-\zeta_{p+Q})}-1\right)\zeta_{p}}{\epsilon^{(j')2}_m+\zeta_{p}^2},
        \\&S_2(A_p)=\frac{1}{\beta}\sum_{j',m}
        \chi_\epsilon(\beta A_{p})\chi_\epsilon(\beta A_{p-Q})\\&\times\frac{\left(e^{\beta(i\epsilon^{(j')}_m-A^{(j')}_{p})}-1\right)\left(e^{\beta(i\epsilon^{(j')}_m-i\omega_n^{(j')}-A^{(j')}_{p-Q})}-1\right)A^{(j')}_{p}}{\epsilon^{(j')2}_m+A^{(j')2}_{p}},
	\end{aligned}
\end{equation}
where $A^{(j)}_{p}=\zeta_p-(j-1)\dfrac{\ln{\delta}}{2\beta}$. If in the second term of the above expression we let $p+Q \to -p'$, and then $p'$ be relabeled as $p$, one ends up with
\begin{equation}
    \begin{aligned}
		\Pi^0_{\hat{q}}&(Q,\omega)
		=\frac{\hat{q}}{\beta}\int \frac{d^3p}{(2\pi)^3}B_{\hat{q}}(p)B_{\hat{q}}(p+Q)
        \\&\times\left[\frac{S_1(\zeta_{p})}{\zeta_{p+Q}-\zeta_p-\omega-i\eta}+\frac{S_2(A_{p})}{\zeta_{p+Q}-\zeta_p+\omega+i\eta}\right],
    \end{aligned}
\end{equation} 
This expression, after a straightforward $\hat{q}$ algebra, and substituting $\hat{q}\to 1-2\delta$, gives the following relation: 
\begin{equation}\label{Eq:polarization function}
	\begin{aligned}
		\Pi^0_{\delta}&(Q,\omega)
		=\int \frac{d^3p}{(2\pi)^3}\mathcal{B}_{pQ}(\beta,\delta)
        \\&\times\left[\frac{S_1(\zeta_{p})}{\zeta_{p+Q}-\zeta_p-\omega-i\eta}+\frac{S_2(A_{p})}{\zeta_{p+Q}-\zeta_p+\omega+i\eta}\right],
	\end{aligned}
\end{equation}
where we have defined
\begin{equation}
    \begin{aligned}
   &\mathcal{B}_{pQ}(\beta,\delta)=\left. \hat{q}B_{\hat{q}}(p)B_{\hat{q}}(p+Q)\right|_{\hat{q}\to 1-2\delta}
   \\&=\frac{-\frac{4 \delta }{b^2-\delta }+\frac{2 \delta }{b-1}-2 \delta }{a^2-\delta }+\frac{\frac{2
   (2-a) \delta }{b^2-\delta }+\frac{a-2 \delta }{b-1}+a (1-2 \delta )+2 \delta }{a-1},
\end{aligned}
\end{equation}
where $a=e^{\beta\zeta_p}$ and $b=e^{\beta\zeta_{p+Q}}$.
Note that in the limit $\beta\to\infty$ ($\zeta_p,\zeta_{p+Q}\ne 0$), 
\begin{equation}
    \lim_{\beta\to\infty}\mathcal{B}_{pQ}(\beta,\delta)\equiv\mathcal{B}(\delta)=\begin{cases}
    \frac{-4(1-\delta)^2}{\delta} & \zeta_p<0,\zeta_{p+Q}<0 \\
     1-2\delta & \zeta_p>0,\zeta_{p+Q}>0\\
    2-2\delta & \zeta_p>0,\zeta_{p+Q}<0\\
    2-2\delta & \zeta_p<0,\zeta_{p+Q}>0\\
   \end{cases}
\end{equation}
It is worth mentioning that this formula is applicable for all $\delta$ values except $\delta=1$ were the regulator Eq.~\eqref{Eq:Regulator} should not be emploied, i.e. it generates an unnecessary prefactor, which is zero in the $\beta=\infty$ limit.
This function has also singular points at $\zeta_p=0$, and $\zeta_{p+Q}=0$, which can lead to BEC for $0\le\delta\le 0.5$, while for $0.5\le\delta\le 1$ it is of fermionic character, which causes the singularities to cancel each other out. We consider the fermionic case in the remainder of this paper to avoid complexities arisen from BEC singularities.
Also, $S_1(\zeta_{p})$ and $S_2(A_{p})$ take on the following values when $\beta\to\infty$:
\begin{equation}
    \lim_{\beta\to\infty}S_1(\zeta_{p})=\lim_{\beta\to\infty}S_2(A_{p})=\begin{cases}
    -1 & \zeta_p<0,\zeta_{p+Q}<0 \\
     3 & \zeta_p>0,\zeta_{p+Q}>0\\
    -1 & \zeta_p>0,\zeta_{p+Q}<0\\
   1 & \zeta_p<0,\zeta_{p+Q}>0\\
    \end{cases}.
\end{equation}
We will use these expressions in the following sections to calculate the thremodynamic quantities.
%==========================section====================
\subsubsection{Dielectric function for unified fermion gas}
%==========================section====================
In the fermionic limit ($\delta=1$), the unified statistics reduces to conventional Fermi-Dirac statistics. At zero temperature, the occupation number distribution simplifies to Eq.~\eqref{Eq:FermionAccuNumZero}. With this step-function occupation, the zeroth-order polarization function
$\Pi^0_{\delta}(Q,\omega^{(j)}_n)$ can be evaluated analytically. For a parabolic dispersion, the energy difference appearing in the denominator of Eq.~\eqref{Eq:polarization function} takes the form:
\begin{equation}\label{Eq:deltazeta}
	\zeta_{pQ}=\zeta_{p+Q}-\zeta_{p}=\frac{\hbar^2}{2m}(Q^2+2p.Q).
\end{equation}
In the following we do a similar calculations based on an analytic continuation to real frequencies $i\omega^{(j)}_n=\omega+i\eta$.\\
    
%===========================================
To evaluate the polarization function in Eq.~\eqref{Eq:polarization function}, we consider four possible cases arising from the integration domain. It is found that only two of these contributions are non-vanishing, while the remaining terms vanish identically.
The two vanishing contributions are associated with the cases ($\beta\to\infty,\zeta_p<0,\zeta_{p+Q}<0$) and ($\beta\to\infty,\zeta_p>0,\zeta_{p+Q}>0$). In the Appendix~\ref{polarization_function} we show explicitly that the first of these contributions ($\left.\Pi^0_{\delta}(Q,\omega)\right|_{\zeta_p,\zeta_{p+Q}<0}$) is zero. $\Pi^0$ for the case $\zeta_p<0$ and $\zeta_{p+Q}>0$ reduces to 
\begin{equation}\label{Eq:RePi}
    \begin{aligned}
        &\Pi^0_{\delta}(Q,\omega)=
        \frac{\mathcal{B}(\delta)}{(2\pi)^3}\int d^3p~\Theta\left(p_F^{(\delta)}-|p|\right)
        \\&\times \left[\frac{1}{\frac{\hbar^2}{2m}(Q^2+2p.Q)-\omega-i\eta}+\frac{1}{\frac{\hbar^2}{2m}(Q^2+2p.Q)+\omega+i\eta}\right],
    \end{aligned}
\end{equation}
where $p_F^{(\delta)}$ is the Fermi momentum given in Eq.~\eqref{Eq:FermiMomentum}. One can show that the result for the case $\zeta_p>0$ and $\zeta_{p+Q}<0$ is exactly the same, both leading to (see Eq.~\eqref{Eq:RePiAppendix}):
\begin{equation}\label{Eq:1term}
    \begin{aligned}
    \Pi^0_{\delta}(Q,\omega)
    =\frac{\mathcal{B}(\delta)m~{p_F^{(\delta)}}^2}{\hbar^2(2\pi)^2Q}
    \sum_{i=1}^2\left(Z_i+\frac{1-Z_i^2}{2}\ln\frac{Z_i+1}{Z_i-1}\right),
    \end{aligned}
\end{equation}
 where $Z^{}_{1,2}=\frac{Q}{2p_F^{(\delta)}}\mp\frac{m\omega}{\hbar^2p_F^{(\delta)}Q}\mp i\eta$.
Therefore, combining the four contributions, of which only two expressions are non-zero, we arrive at the following result:
\begin{equation}
    \begin{aligned}
    \Pi^0_{\delta}(Q,\omega)
    =\frac{2\mathcal{B}(\delta)m~{p_F^{(\delta)}}^2}{\hbar^2(2\pi)^2Q}
    \sum_{i=1}^2\left(Z_i+\frac{1-Z_i^2}{2}\ln\frac{Z_i+1}{Z_i-1}\right),
    \end{aligned}
\end{equation}
%=================================
Inserting this into the random‑phase approximation (RPA) yields the dynamical dielectric function for the unified fermion gas:
\begin{equation}
	\label{dielectric-constant0}
	\begin{aligned}
		&\varepsilon^{{\text{RPA}}}(\omega^{(j)}_n,Q)=1-v_0(Q)\left(\Pi_{\delta}^0(\omega^{(j)}_n,Q)\right)\\
		&=1-\frac{4me^2 \mathcal{B}(\delta) p^{(\delta)}_F}{\hbar^2\pi Q^2}(\frac{p^{(\delta)}_F}{2Q})\sum_{i=1}^2\Bigg(Z^{}_i+\frac{1-Z_i^2}{2}\ln{\frac{Z^{}_i+1}{Z^{}_i-1}}\Bigg).
	\end{aligned}
\end{equation}
where $v_0(Q)=\frac{4\pi e^2}{Q^2}$ is the bare Coulomb interaction in Fourier space. Separating the real and imaginary parts,
\begin{equation}
	\varepsilon^{{\text{RPA}}}=\varepsilon_1^{{\text{RPA}}}+i\varepsilon_2^{{\text{RPA}}},
\end{equation}
allows the identification of absorption (Landau damping) and dispersive properties.
Static limit ($\omega^{(j)}_n\to 0$): 
\begin{equation}\label{Eq:RealDF}
	\begin{aligned}
		\varepsilon_s^{{\text{RPA}}}(Q)\equiv \varepsilon_1^{{\text{RPA}}}(0,Q)=   1+\left(\frac{Q_{\text{TF}}^{(\delta)}}{Q^2}\right)^2F(x_\delta).
	\end{aligned}
\end{equation}    
with the Lindhard screening function
\begin{equation}
	F(x)\equiv \frac{1}{2}+\frac{1-x^2}{4x}\ln\left|\frac{x+1}{x-1}\right|,
\end{equation}
and $x_\delta\equiv \frac{Q}{2p_F^{(\delta)}}$. The generalized Thomas–Fermi wave‑vector is
\begin{equation}
	Q_{\text{TF}}^{(\delta)}\equiv \sqrt{\frac{8m e^2 \mathcal{B}(\delta) {p^{(\delta)}_F}^2}{\hbar^2\pi^2}}.
\end{equation}
We note that the overall structure of the functions remains similar to that obtained in the standard fermionic limit \cite{fetter2012quantum,mahan2013many,sadovskii2006diagrammatics}, with the corresponding parameters appropriately renormalized due to the deformation. In this sense, unified statistics can be naturally incorporated into the Fermi-liquid framework.
Noting that in small $x_\delta$ limit $F(x_q)\to 1$, i.e. one retrieves a generalized TF model:
\begin{equation}
	\begin{aligned}
	\varepsilon^{{\text{TF}}}(Q)\equiv\lim_{\text{small }Q}\varepsilon_s^{{\text{RPA}}}(Q)=1+\frac{{Q_{\text{TF}}^{(\delta)}}^2}{Q^2}.
	\end{aligned}
\end{equation}
Furthermore, we find that $Q^{(\delta)}_{\mathrm{TF}}$ can be regarded as a tuning parameter that determines the value of $\delta$ most appropriate for the system under consideration. In the large $Q$ limit, $F(x)\to \frac{1}{3x_\delta^2}$, which consequently yields
\begin{equation}
	\begin{aligned}
	\lim_{Q\gg p_F^{(\delta)}}\varepsilon_s^{{\text{RPA}}}(Q)=
	1+\frac{4}{3}\frac{{Q_{\text{TF}}^{(\delta)}}^2{p_{F}^{(\delta)}}^2}{Q^4}.
	\end{aligned}
	\label{Eq:largeQ}
\end{equation}
%====================figure=====================
\begin{figure}[h!]
	\centering
	\includegraphics[width=1\linewidth]{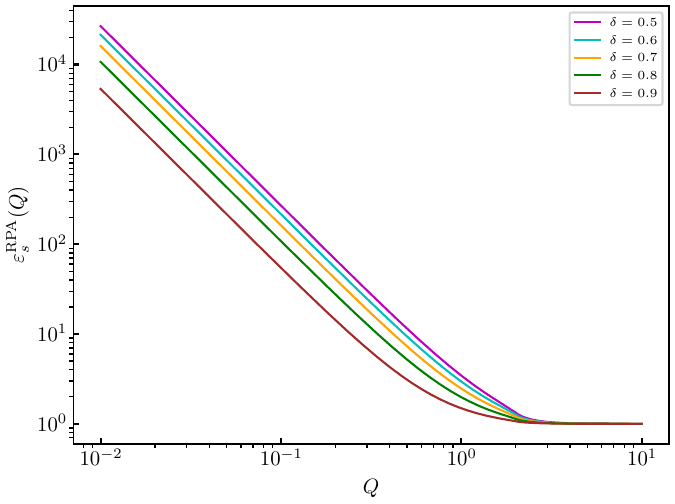}
	\caption{ The main panel displays $\varepsilon_1^{\text{RPA}}(0,Q)$ as a function of $Q/p_F^{(\delta)}$ for $\delta$ values ranging from 0.5 to 0.9.
    }
	\label{fig:placeholder}
\end{figure}
%================================================
 FIG.~\ref{fig:placeholder} displays the static real part of the RPA dielectric function,
$\varepsilon^{\text{RPA}}_s(Q)=\varepsilon^{\text{RPA}}_1(0,Q)$ of Eq.~\eqref{Eq:RealDF}, versus $Q/p_F^{(\delta)}$ for $\delta=0.5$--$0.9$.
 The curves exhibit the  characteristic Lindhard behavior, diverging as $1/Q^2$ in the  long-wavelength (Thomas--Fermi) limit and decaying as $1/Q^4$ at large  $Q$, with a kink at $Q=2p_F^{(\delta)}$ that signals the onset of Friedel oscillations. 
Since the deformation parameter renormalizes both
$p_F^{(\delta)}$ and $Q^{(\delta)}_{TF}$, varying $\delta$ rescales the envelope while preserving the functional form, smoothly interpolating the dielectric response within the fermionic regime $0.5<\delta<1$.
%===============================================
\subsubsection{Plasmons, Friedel Oscillations and Energy Loss Function}\label{Friedel-Oscillations}
The response of a unified Fermi gas to an external charge is characterized by the parameter $\delta$. This dependence gives rise to Friedel oscillations, which originate from the singularity at $Q=2p_F^{(\delta)}$ \cite{grosso2013solid}. In particular, the induced charge density $\rho_{\text{ind}}(\mathbf{r})$ surrounding a point impurity with charge $Ze$ can be expressed as \cite{mahan2013many}
\begin{equation}
	\begin{split}
	\rho_{\text{ind}} &\equiv\frac{Ze}{(2\pi)^3}\int \bigg(\frac{1}{\varepsilon_s(Q)}-1\bigg)e^{i\textbf{Q}.\textbf{r}}\text{d}\textbf{Q}\\
	&=\frac{Ze}{r^3}\int_0^{\infty}g''(Q)\sin (Qr) \text{d}Q,
    \end{split}
\end{equation}
The auxiliary function g(Q) is defined as:
\begin{equation}
g(Q)\equiv\frac{Q}{2\pi^2}\frac{\varepsilon_s(Q)-1}{\varepsilon_s(Q)},
\end{equation}
where $\varepsilon_s(Q)$ is the static dielectric function and
\begin{equation}
    g''(Q)\equiv \frac{\mathrm{d}^2g}{\mathrm{d}Q^2}
    =\frac{\alpha^{}_\delta}{Q-2p_F^{(\delta)}} ~~,~~ \alpha^{}_\delta\equiv \frac{{Q_{\text{TF}}^{(\delta)}}^2}{16{p_F^{(\delta)}}^3}.
\end{equation}
Here, we note that $g(Q)$ vanishes in both limiting cases, i.e., $\lim_{Q\to 0}g(Q)=\lim_{Q\to\infty}g(Q)=0$. Restricting the integration to a narrow region of width $\sigma$ around $Q=2p_F^{(\delta)}$, and evaluating the corresponding principal-value integral in the asymptotic limit $r\to\infty$, one obtains the Friedel oscillatory induced density:
\begin{equation}
	\begin{aligned}
	\rho_{\text{ind}}\approx & \frac{Ze\alpha^{}_\delta}{r^3}\int_{2p_F^{(\delta)}-\sigma}^{2p_F^{(\delta)}+\sigma}\frac{\sin(Qr)}{Q-2p_F^{(\delta)}}\text{d}Q\\
	&=\frac{Ze\alpha^{}_\delta\cos(2p_F^{(\delta)}r)}{r^3}\int_{-r\sigma}^{r\sigma}\frac{\sin y}{y}\text{d}y. 
	\end{aligned}
	\label{Eq:Friedel}
\end{equation}
%====================figure=====================
\begin{figure}[h!]
	\centering
	\includegraphics[width=1\linewidth]{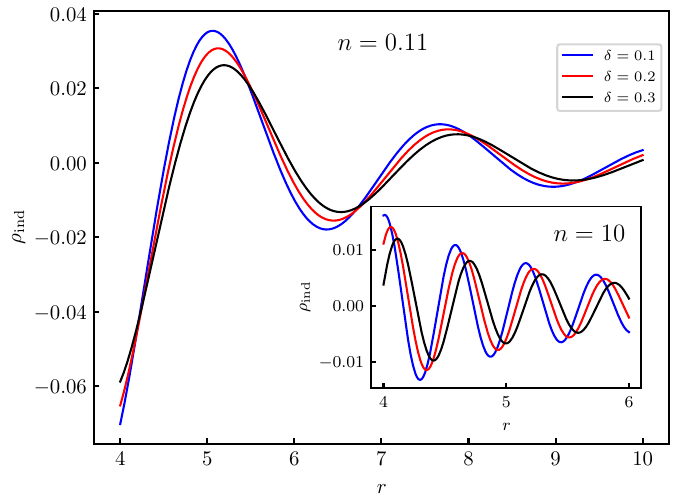}
	\caption{Visualized here are Friedel oscillations corresponding to n = 0.11 (and n = 10 in the inset) for $\delta$ equal to 0.1, 0.2, and 0.3. 
    }
	\label{fig:rho-ind}
\end{figure}
	%================================================
FIG.~\ref{fig:rho-ind} displays the Friedel oscillations of the induced charge density
$\rho_{\mathrm{ind}}(r)$ around a point impurity of charge $Ze$, which asymptotically behaves as
$\rho_{\mathrm{ind}}(r)\approx \pi Z e\,\alpha^{}_\delta\, r^{-3}\cos(2p_F^{(\delta)}r)$.
The main panel corresponds to the low-density regime $n=0.11$ and the inset to the high-density case $n=10$, for $\delta=0.1,0.2,0.3$.
Increasing $\delta$ shortens the oscillation period and suppresses the amplitude, while the oscillations are markedly more pronounced at low density.
\\
At very high frequencies, the dielectric function reduces to the Drude‑like form:
\begin{equation}
	\varepsilon(\omega^{(j)}_n,0)=1-\frac{{\Omega_P^{(\delta)}}^2}{(\omega^{(j)}_n+i\eta)^2_{}},
\end{equation}
with the generalized plasma frequency $\Omega_P^{(\delta)}=2(1-\delta)\omega_p^2$. 
Introducing a phenomenological relaxation rate $\tau^{-1}$ (where  $2\eta\to 1/\tau$) yields:
\begin{equation}
	\begin{split}
    &\varepsilon_1(\omega,0)=1-\frac{{\Omega_P^{(\delta)}}^2}{\omega^2_{}+1/\tau^2},\\
    &\varepsilon_2(\omega,0)=\frac{{\Omega_P^{(\delta)}}^2}{\tau\omega\big(\omega^2+1/\tau^2\big)}.
    \end{split}
\end{equation}
%====================figure=====================
\begin{figure}[h!]
	\centering
	\includegraphics[width=1\linewidth]{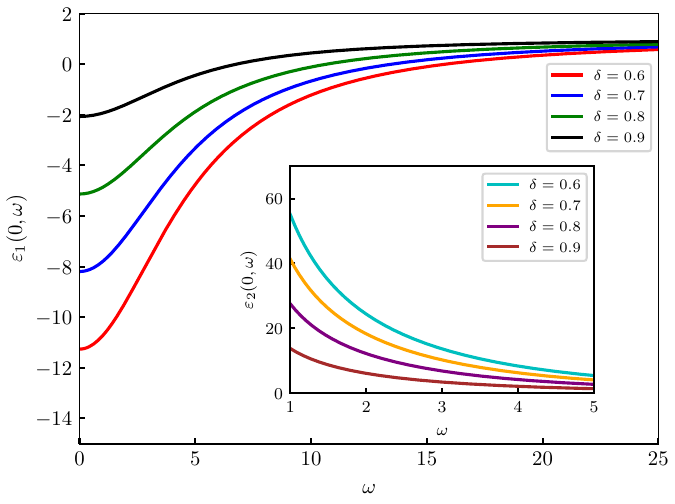}
	\caption{ The main graph illustrates the variation of $\varepsilon_1$ with frequency, while the inset plot shows the corresponding behavior of $\varepsilon_2$ for various values of $\delta$.}
	\label{fig:combined}
\end{figure}
%================================================
FIG.~\ref{fig:combined} shows the real part $\varepsilon_1(0,\omega)$ of the dielectric function versus frequency in the main panel, with the imaginary part $\varepsilon_2(0,\omega)$ shown in the inset, for $\delta=0.6,0.7,0.8,0.9$.
Within the Drude-like form $\varepsilon_1$ and
$\varepsilon_2$, both quantities vary non-monotonically with  $\delta$, reflecting the $\delta$-dependent renormalization of the plasma frequency $\Omega_P^{(\delta)}$.
\\
Expanding the full RPA dielectric function to second order in Q and solving $\epsilon(\omega,Q)=0$ gives the plasmon dispersion relation for small wavevectors:
\begin{equation}\label{Expand}
	1-\frac{{\Omega^{(\delta)}_P}^2}{\omega^2}-\frac{3}{5}\frac{{\Omega^{(\delta)}_P}^2}{\omega^4}{\mathrm{v}_F^{(\delta)}}^2 Q^2=0,
\end{equation}
where $\mathrm{v}_F^{(\delta)}=\hbar p_F^{(\delta)}/m$.
The solution is:
\begin{equation}\label{wplasmon}
	\omega_{\text{plasmon}}(Q)=\Omega^{(\delta)}_P\left[1+\frac{3}{10}\frac{{\mathrm{v}_F^{(\delta)}}^2 Q^2}{{\Omega^{(\delta)}_P}^2}+\cdots\right].
\end{equation}
%====================figure======================
\begin{figure}[h!]
	\centering
	\includegraphics[width=1\linewidth]{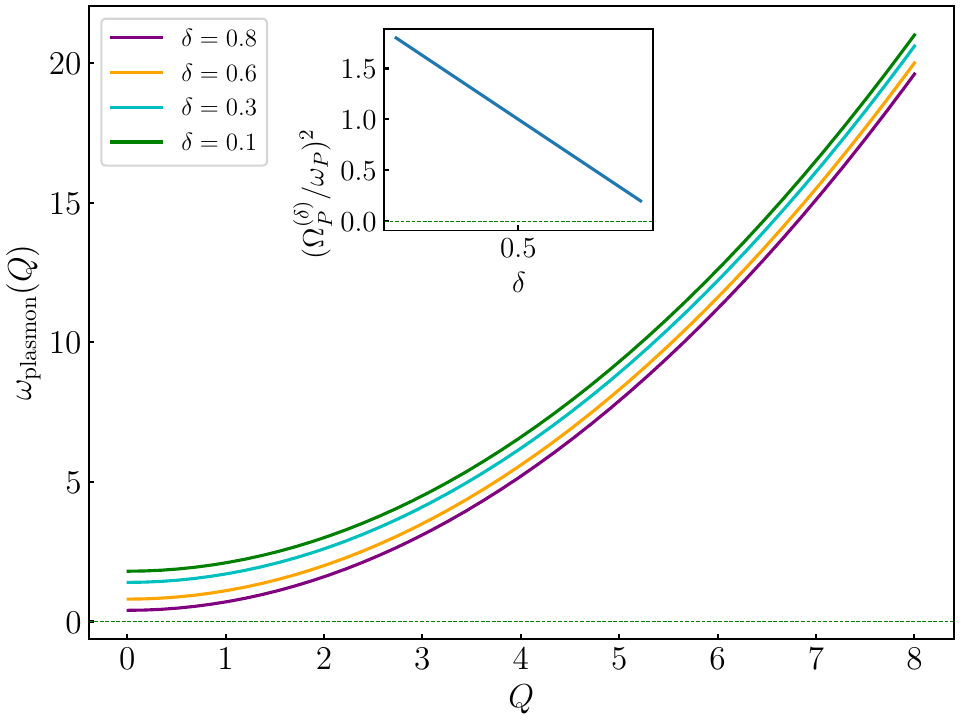}
	\caption{ The dispersion relation described by Eq.~\eqref{wplasmon} is presented in terms of $Q$ for various values of $\delta$. The inset illustrates the behavior of the function $\Omega_P^{(\delta)}/\omega_P$ as a function of $\delta$
    .
	}
	\label{fig:DispersionRelation}
\end{figure}
 This expression defines the plasmon dispersion relation. 
Fig.~\ref{fig:DispersionRelation} presents the plasmon dispersion relation of Eq.~\eqref{wplasmon},
versus $Q$ for $\delta=0.1,0.3,0.6,0.8$. 
Each branch originates from a finite gap $\Omega_P^{(\delta)}$ at $Q=0$ and rises quadratically with a stiffness set by the $\delta$-dependent Fermi velocity $\mathrm{v}_F^{(\delta)}$, so that decreasing $\delta$ shifts the entire plasmon
branch upward.
The inset plots $(\Omega_P^{(\delta)}/\omega_P)^2$ against $\delta$, isolating the purely statistical renormalization of the bulk plasma frequency, which reduces to the ordinary Fermi value
$\omega^{}_P=\sqrt{4\pi n e^2/m}$ at $\delta\to 1$.
%================================================

\begin{figure}
	\centering
	\includegraphics[width=1\linewidth]{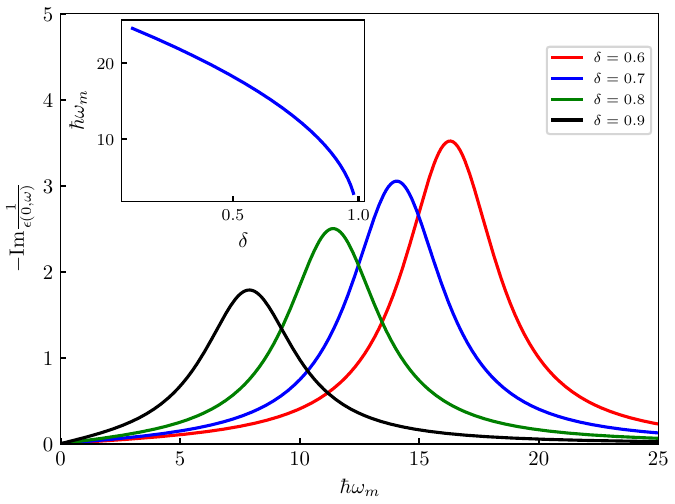}
	\caption{ The main panel shows the Energy-loss function versus $\hbar\omega_m$ for $\delta$ values from 0.6 to 0.9.
    The inset displays the $\hbar\omega_m$ behavior as a function of $\delta$ within the energy loss function.}
	\label{fig:PELF}
\end{figure}

The imaginary part of the inverse dielectric function,$-\text{Im}(1/\varepsilon)$ quantifies the energy dissipation experienced by a fast charged particle traversing the medium. It exhibits a sharp peak at the plasmon frequency, providing a direct spectroscopic signature of collective excitations in the unified Fermi gas.
\begin{equation}\label{Eq:ELF}
	-\text{Im}\frac{1}{\varepsilon(\omega,0)}=\frac{\tau^{-1}\omega{\Omega^{(\delta)}_P}^2}{\bigg(\omega^2_{}-{\Omega^{(\delta)}_P}^2\bigg)^2_{}+\omega^2_{}/\tau^2_{}}.
\end{equation}
FIG.~\ref{fig:PELF} plots the energy-loss function $-\mathrm{Im}\,\varepsilon^{-1}(\omega,0)$
of Eq.~\eqref{Eq:ELF} versus $\hbar\omega_m$ for $\delta=0.6$--$0.9$, with the inset
showing the $\delta$-dependence of the peak position $\hbar\omega_m$.
The loss function exhibits a sharp resonance peak at the plasmon frequency, whose location and height shift with $\delta$ through the renormalized plasma frequency
$\Omega_P^{(\delta)}$, providing a direct spectroscopic signature of the collective excitations.
%\newpage
%====================figure======================
%================================================
\section{Conclusion}

In this work, we have developed a comprehensive many-body theoretical framework for particles obeying unified quantum statistics, a formalism that interpolates continuously between Bose-Einstein and Fermi-Dirac statistics through a single deformation parameter $\delta$. Starting from the underlying algebraic structure defined by the exchange-statistics factor operator $\hat{q}$, we constructed the necessary field-theoretic machinery-namely, a generalized $S$-matrix expansion, an extended Wick's theorem incorporating the operator $\hat{q}$, and a corresponding set of generalized Feynman diagrammatic rules-thereby placing systems of deformed particles on the same systematic footing as conventional bosonic and fermionic many-body theory.

Using this formalism, we derived the finite-temperature single-particle Green's function for the unified-statistics gas, both in real space and in momentum space, and obtained the associated spectral density and density of states. We argue that there is a correspondence between the non-interacting Green's function, and the interacting Green's function for the standard particles in low-interaction regime. Building on the single-particle results, we computed the density-density correlation function and the dielectric response function within the random phase approximation for the interacting deformed fermion gas, and used these results to analyze collective phenomena such as Friedel oscillations, plasmonic excitations, and the dynamical energy loss function.

Taken together, these results demonstrate that the deformation parameter $\delta$ does not merely interpolate between the two conventional quantum statistics at the level of occupation numbers, but also directly controls the strength and character of the effective interactions and collective response of the many-body system. 
It enters the Green's function through the factor $(1+\vartheta\, n_\nu)$ with $\vartheta = 1-2\delta$, introduces $\delta$-dependent generalized Matsubara frequencies, and renormalizes the density of states by the factor $(3-2\delta)$.
In the fermion-like regime ($0.5<\delta\le 1$), it also rescales the Fermi momentum $p_F^{(\delta)}$ and the generalized Thomas--Fermi wave vector $Q_{\mathrm{TF}}^{(\delta)}$.
Within the random phase approximation, the dielectric function retains the Lindhard form, while its characteristic scales and the plasma frequency $\Omega_P^{(\delta)}$ acquire an explicit dependence on $\delta$. 
As a result, the Friedel oscillations (amplitude and period), the plasmon dispersion, and the position and height of the energy-loss peak all shift systematically with $\delta$, and they recover the conventional Fermi-gas results in the limit $\delta\to 1$.
In this way, the formalism links the algebraic structure of unified quantum statistics to computable and, in principle, measurable collective responses, and it supplies an operational diagrammatic many-body framework that had been missing for infinite and generalized statistics. \\

A natural extension would be to go beyond the random phase approximation and incorporate higher-order correlation effects, such as vertex corrections or self-energy renormalization beyond the quasiparticle approximation, in order to assess the robustness of the results presented here. It would also be of interest to investigate whether the effective attractive or repulsive interactions induced by tuning $\delta$ can give rise to genuinely new collective instabilities, such as an emergent pairing or condensation mechanism, in analogy with more conventional strongly correlated systems. Finally, extending this formalism to systems with internal degrees of freedom (such as spin) or to low-dimensional and lattice geometries could provide a valuable bridge between the abstract framework of unified quantum statistics and physically realizable platforms, such as cold-atom systems or strongly correlated condensed matter systems exhibiting fractional or intermediate statistics.
%@@@@@@@@@@@@@@@@@@@@@@@@@@@@@@@@@@@@@@@@@@@@@@@@@@@@@@
% \newpage
\bibliography{refpaper}
%@@@@@@@@@@@@@@@@@@@@@@@@@@@@@@@@@@@@@@@@@@@@@@@@@@@@@@
%================================================
% \newpage
% \thispagestyle{empty}
% \newpage
% %%%%%%%%%%%%% Acknowledgments %%%%%%%%%%%%%%%%
% %\section*{Acknowledgments}
% %=================appendix=====================
% \clearpage
% \newpage
\appendix
%=============================================
\begin{widetext}
\section{Fock space construction}~\label{SEC:FockSpaceApp}
In this section we construct the Fock space for deformed particles, adapting the notation of Ref.~\cite{yan2021statistical}. A Fock space for a many body system is constructed using the following representation 
\begin{equation}
\left| \nu_1,\nu_2,...,\nu_N\right)\equiv \left| \nu_1\right\rangle \otimes \left| \nu_2\right\rangle\otimes...\otimes\left| \nu_N\right\rangle,
\end{equation}
where $\left\lbrace  \left| \nu_i\right\rangle\right\rbrace$ is an orthonormal basis of the Hilbert space. Note that the many body state is shown by a \textit{parasynthesis-like} ending symbol to be distinguishable from the final state to be shown in the end. The completeness relation is the Fock space reads
\begin{equation}
    \sum_{\nu_1,...,\nu_N}\left| \nu_1,...,\nu_N\right) \left(\nu_1,...,\nu_N \right| =1.
    \label{Eq:orth}
\end{equation}
Here we suppose that the particles are identical (are in the same $\hat{q}$ eigenstate). To investigate the statistical properties of a quantum system using the occupation number representation, we define a projection operator that helps us describe the occupancy of quantum states.
The (Hermitian and unitary) symmetrization projection operator $\hat{Q}$ which enforces a $\hat{q}$-exchange symmetry on an arbitrary $N$-particle wave function is defined as follows:
\begin{equation}
	\hat{Q}=\frac{1}{N!}\sum_P\hat{q}^{[P]}_{}P,
    \label{Eq:Exchange}
\end{equation}
where $P$ is the permutation operator and the summation is over the $N!$ permutations of the $N$ particles, and $\hat{q}^{[P]}=\textbf{1}$ ($\hat{q}^{[P]}=\hat{q}$) for an even (odd) permutation. One can simply show that $\hat{Q}^2=\hat{Q}$, i.e. $\hat{Q}$ is a projection operator. One then can construct the $\hat{q}$-symmetrized many-body states as follows
\begin{equation}
    \begin{split}
    \left| \nu_1,\nu_2,...,\nu_N\right\rbrace^{\hat{q}}&\equiv \sqrt{N!}\hat{Q} \left| \nu_1,\nu_2,...,\nu_N\right)=\frac{1}{\sqrt{N!}}\sum_P\hat{q}^{[P]}\left| \nu_{P_1},\nu_{P_2},...,\nu_{P_N}\right),
    \end{split}
\end{equation}
where $P_j$ is a permutation group member for the $j$th particle, and the superscript $\hat{q}$ denotes explicitly that the state is $\hat{q}$-symmetrized.  Note that this state is shown by a "right brace" ending symbol to be distinguishable from the final state to be shown in the end. Here it would be nice to separate the permutations to two categories. In the particle number representation, suppose that the number of particles in each $\nu$ state is $n_\nu$. Note that $\sum_\nu n_\nu=N$. Now we decompose $\hat{Q}$ to two \textit{internal} and \textit{external} components $\hat{Q}=\hat{Q}_I\hat{Q}_E$, where the internal permutations corresponds to the exchanges of two particles within a single state, i.e. two exchanging particles are within a same quantum number, while the exchanging particles in the $\hat{Q}_E$ part belong to two different states. The internal exchange whithin e.g. the state $\nu$ has $\nu!$ components half of which have even permutation (with a prefactor $\hat{q}^0=1$), and the remaining half contain odd permutations with a prefactor of $\hat{q}$, so that the total wave function pick a factor $\frac{n_\nu!}{2}\left(1+\hat{q}\right)$. Therefore, the state is expanded as follows:
\begin{equation}
\begin{split}
    \left| \nu_1,\nu_2,...,\nu_N\right\rbrace^{\hat{q}}&=\frac{1}{\sqrt{N!}}\left(\prod_\nu n_\nu!\right)\left(\frac{1+\hat{q}}{2}\right)^M\sum_{P_E}\hat{q}^{[P_E]}\left| \nu_{P^E_1},\nu_{P^E_2},...,\nu_{P^E_N}\right),
\end{split}
\end{equation}
where $P_E$ stands for the external permutations, and $M$ is the number of states with $n_\nu>0$. Using the orthonormality property of single states, one may readily show that 
\begin{equation}
    \begin{split}
    {}^{\hat{q}}\left\lbrace \nu'_1,...,\nu'_N\right.\left| \nu_1,...,\nu_N\right\rbrace^{\hat{q}}&=N!\left(\nu'_1,...,\nu'_N \right|\hat{Q}^2 \left| \nu_1,...,\nu_N\right) =N!\left(\nu'_1,...,\nu'_N \right|\hat{Q} \left| \nu_1,...,\nu_N\right)\\
    &=\left(\prod_\nu n_\nu!\right)\left(\frac{1+\hat{q}}{2}\right)^M\sum_{P_E}\hat{q}^{[P_E]}\left\langle \nu'_1\right| \nu_{P^E_1}\left. \right\rangle ... \left\langle \nu'_1\right| \nu_{P^E_N}\left. \right\rangle.
    \end{split}
\end{equation}
The last line is always zero except for the case $P_E$ is identity, leading one to get:
\begin{equation}
    \begin{split}
     {}^{\hat{q}}\left\lbrace \nu'_1,...,\nu'_N\right.\left| \nu_1,...,\nu_N\right\rbrace^{\hat{q}}&=\left(\prod_\nu n_\nu!\right)\left(\frac{1+\hat{q}}{2}\right)^M=\left(\prod_\nu n_\nu!\right)\textbf{1}^{\hat{q}}
    \end{split}
\end{equation}
where for the last line we used the fact that $\left(\frac{1+\hat{q}}{2}\right)^M=\frac{1+\hat{q}}{2}$ if $M\ge 1$, and
\begin{equation}
    \mathbf{1}^{\hat{q}}_{} =
    \begin{cases}
        \mathbf{1}, & \text{if all } n_i = 0 \text{ or } 1, \\
        \frac{1}{2}(\mathbf{1} + \hat{q}), & \text{otherwise},
    \end{cases}
\end{equation}
This is recognized as the modified identity operator in the \textit{quon Fock space}. Now one can define 
\begin{equation}
	\begin{split}
    |\hat{\Psi}^{\hat{q}}_{}\rangle\equiv |\nu_{P_1}&,\nu_{P_2},...,\nu_{P_N}\rangle^{\hat{q}} \equiv \frac{1}{\sqrt{\prod_\nu n_\nu !}}\left| \nu_{P_1},\nu_{P_2},...,\nu_{P_N}\right\rbrace ^{\hat{q}}\\
    &=\frac{1}{\sqrt{N!\prod_\nu n_\nu !}}\sum_P\hat{q}^{[P]}\left| \nu_{P_1},\nu_{P_2},...,\nu_{P_N}\right)\\
    &=\sqrt{\frac{\prod_\nu n_\nu !}{N!}}\textbf{1}^{\hat{q}}\sum_{P_E}\hat{q}^{[P_E]}\left| \nu_{P^E_1},\nu_{P^E_2},...,\nu_{P^E_N}\right),
	\end{split}
\end{equation}
so that $\langle\hat{\Psi}^{\hat{q}}_{}|\hat{\Psi}^{\hat{q}}_{}\rangle= \textbf{1}^{\hat{q}}$. This shows explicitly that these states form a modified complete orthonormal basis set. These states are employed as the $\hat{q}$-symmetrized basis for the Fock space of deformed particles. 
%====================================================
\section{Unified statistics and Partition function calculations}\label{AppendixUSpf}
Within the framework of unified quantum statistics, the underlying algebraic structure is characterized by the generalized commutation relation introduced in the main text (Eqs.~(\ref{Eq:commutators1}, \ref{Eq:commutators2})). Using this same generalized commutation relation, together with the $\hat{q}$-commutation property $\hat{a}_\mu\hat{a}_\nu=\hat{q}\hat{a}_{\nu}\hat{a}_{\mu}$, we derive the commutators of the number operator $\hat{n}^{}_{\mu}=\hat{a}^\dagger_{\mu} \hat{a}^{}_{\mu}$ with the annihilation and creation operators.
\begin{equation}\label{Eq:commutator1}
	\begin{aligned}
		\left[\hat{n}^{}_{\mu},\hat{a}^{}_\nu\right]=
		\left[\hat{a}^\dagger_\mu \hat{a}^{}_\mu,\hat{a}^{}_\nu\right]&=
		\hat{a}^\dagger_\mu \hat{a}^{}_\mu\hat{a}^{}_\nu-\hat{a}^{}_\nu\hat{a}^\dagger_\mu \hat{a}^{}_\mu =\hat{a}^\dagger_\mu\hat{q}\hat{a}^{}_\nu\hat{a}^{}_\mu-\hat{a}^{}_\nu\hat{a}^\dagger_\mu \hat{a}^{}_\mu=
		\left(\hat{q}\hat{a}^\dagger_\mu\hat{a}^{}_\nu-\hat{a}^{}_\nu\hat{a}^\dagger_\mu \right)\hat{a}^{}_\mu
		\\&=-
		\left(\hat{a}^{}_\nu\hat{a}^\dagger_\mu-\hat{q}\hat{a}^\dagger_\mu\hat{a}^{}_\nu \right)\hat{a}^{}_\mu=-
		[\hat{a}_\nu,\hat{a}^{\dagger}_{\mu}]^{}_{\hat{q}}\;\hat{a}_\mu
		= -\mathbf{1}^{\hat{q}}\delta^{}_{\nu\mu}\hat{a}^{}_\mu,
	\end{aligned}
\end{equation}
And,
\begin{equation}\label{Eq:commutator2}
	\begin{aligned}
		\left[\hat{n}^{}_{\mu},\hat{a}^{\dagger}_\nu\right]&=
		\left[\hat{a}^\dagger_\mu \hat{a}^{}_\mu,\hat{a}^{\dagger}_\nu\right]=
		\hat{a}^\dagger_\mu \hat{a}^{}_\mu\hat{a}^{\dagger}_\nu-\hat{a}^{\dagger}_\nu\hat{a}^\dagger_\mu \hat{a}^{}_\mu =\hat{a}^\dagger_\mu\hat{a}^{}_\mu\hat{a}^{\dagger}_\nu-\hat{q}\hat{a}^{\dagger}_\mu\hat{a}^\dagger_\nu \hat{a}^{}_\mu\\
		&=
		\hat{a}_\mu^{\dagger}\left(\hat{a}_\mu\hat{a}^{\dagger}_\nu-\hat{q}\hat{a}^{\dagger}_\nu\hat{a}_\mu \right)
		=\hat{a}_\mu^{\dagger}\left[\hat{a}_\mu^{},\hat{a}_\nu^{\dagger}\right]_{\hat{q}}
		=\mathbf{1}^{\hat{q}}\delta^{}_{\nu\mu}\hat{a}^{\dagger}_\mu.
	\end{aligned}
\end{equation}
%============================================
In the grand canonical ensemble, the grand partition function for a system of non-interacting particles obeying unified quantum statistics (quons) is constructed as follows (Eq.~\eqref{Eq:Partition-Function1}):
\begin{equation}
	\mathcal{Z}(\hat{q}) = \mathrm{Tr}_{\hat{q}} \left[ e^{-\beta [H(\hat{q}) - \mu N(\hat{q})]} \right],   
\end{equation}
where $\beta=\frac{1}{k_{\beta} T}$, $\mu$ is the chemical potential, and $\text{Tr}_{\hat{q}}$ denotes the trace taken in the $\hat{q}$-symmetrized Fock space.
For non-interacting particles, the Hamiltonian and number operator are diagonal in the occupation number basis:
\begin{equation}
	H(\hat{q}) = \sum_\nu \varepsilon_\nu \hat{n}_\nu, \quad N(\hat{q}) = \sum_\nu \hat{n}_\nu. 
\end{equation}
In the occupation number representation, the trace becomes a sum over all possible occupation configurations of the system. Since the Fock space is spanned by the basis states $|n_1, n_2, \dots, n_\nu, \dots\rangle$, where $n_\nu$ denotes the occupation number of the single-particle state $\nu$, the trace of any operator $\hat{O}$ can be written as
\begin{equation}
\mathrm{Tr}[\hat{O}] = \sum_{\{n_\nu\}} \langle n_1, n_2, \dots | \hat{O} | n_1, n_2, \dots \rangle,
\end{equation}
where the sum runs independently over all allowed values of each occupation number $n_\nu$, consistent with the underlying quantum statistics. This representation allows the trace, originally defined over the full many-body Hilbert space, to be reduced to a set of sums (or, in the continuum limit, integrals) over the occupation numbers of the individual single-particle states, thereby considerably simplifying the evaluation of thermodynamic quantities such as the partition function.
\begin{equation}
	\mathcal{Z}(\hat{q}) =\sum_{\{n_\nu\}} {}^{\hat{q}}\langle \{n_\nu\} | e^{-\beta\sum^{}_{\nu} (\epsilon_\nu - \mu) \hat{n}_\nu} | \{n_\nu\} \rangle^{\hat{q}},
\end{equation}
Since the exponential operator factorizes over the independent single-particle modes and the Fock-space basis states are simple product states over these modes, the trace naturally separates into a product of independent single-mode contributions:
\begin{equation}
	\mathcal{Z}(\hat{q}) =\prod_\nu\sum_{\{n_\nu\}} {}^{\hat{q}}\langle \{n_\nu\} | e^{-\beta (\epsilon_\nu - \mu) \hat{n}_\nu} | \{n_\nu\} \rangle^{\hat{q}}.
\end{equation}
%================================
The evaluation of the trace relies on the completeness and orthogonality of the occupation-number basis in the $\hat{q}$-symmetric Fock space:
\begin{align}\label{Eq:RelationOrthogonal}
	\sum_{\{n_\nu\}}
	|\{n_\nu\}\rangle^{\hat{q}}\;{}^{\hat{q}}\langle \{n_\nu\}|
	&=\mathbf{1}^{\hat{q}},
	\\
	{}^{\hat{q}}\langle \{n_\nu\}|\{n'_\nu\}\rangle^{\hat{q}}
	=
	\prod_\nu&\mathbf{1}^{\hat{q}}\delta^{}_{n^{}_\nu n'_\nu}.
\end{align}
With these relations, the norm of a basis state is found to be:
\begin{equation}\label{Eq:Base}
    \begin{aligned}
		&{}^{\hat{q}}\langle \{n_\nu\}|\mathbf{1}^{\hat{q}}|\{n_\nu\}\rangle^{\hat{q}}
		=
    	\sum_{\{n'_\nu\}}
		{}^{\hat{q}}\langle \{n_\nu\}|\{n'_\nu\}\rangle^{\hat{q}}\;{}^{\hat{q}}\langle \{n'_\nu\}|\{n_\nu\}\rangle^{\hat{q}}
    	% \\&
        =\sum_{\{n'_\nu\}}
		(\prod_\nu\mathbf{1}^{\hat{q}}\delta_{n_\nu n'_\nu})
		(\prod_\nu\mathbf{1}^{\hat{q}}\delta_{n_\nu n'_\nu})
		=
		(\mathbf{1}^{\hat{q}})^2
		\equiv\mathbf{1}^{\hat{q}},
	\end{aligned}
\end{equation}
%================================
The sum over all many-body configurations ${n_i}$ reduces, for each mode $i$, to a sum over the possible occupation numbers $n_i$ of that mode. The inner product introduces the $\hat{q}$ -dependent identity operator $\mathbf{1}^{\hat{q}}$ as defined in the Fock space construction. 
\begin{equation}
	\mathcal{Z}(\hat{q}) =\prod_\nu \sum_{n_\nu} e^{-\beta (\epsilon_\nu - \mu) n_\nu} \mathbf{1}^{\hat{q}},
\end{equation}
Thus, for each single-particle state i, we obtain the single-mode partition function:
\begin{equation}
	\mathcal{Z}_{\hat{q}}^{(\nu)} =\sum_{n_\nu} e^{-\beta (\epsilon_\nu - \mu) n_\nu} \mathbf{1}^{\hat{q}},
\end{equation}
The allowed values of $n_\nu$ and the explicit form of $1^{\hat{q}}$ depend on the exchange statistics factor operator $\hat{q}$ :
\begin{itemize}
	\item For $n_\nu = 0$, the contribution is $1$,
	\item For $n_\nu = 1$, the contribution is $e^{-\beta(\epsilon_\nu-\mu)}$,
	\item For $n_\nu = 2$, the contribution is $\frac{1}{2}(1+\hat{q})e^{-2\beta(\epsilon_\nu-\mu)}$.
\end{itemize}
Therefore, we can write:
\begin{equation}
	\mathcal{Z}_{\hat{q}}^{(\nu)}= 1 + e^{-\beta (\epsilon_\nu - \mu)} + \frac{1}{2}(1 + \hat{q}) \Big\{e^{-2\beta (\epsilon_\nu - \mu)}+...\Big\},
\end{equation}
Introducing the variable $x_\nu = e^{\beta(\mu - \epsilon_\nu)}$, which represents the fugacity factor associated with mode $\nu$, we note that this can equivalently be written as $x_\nu = e^{-\beta(\epsilon_\nu - \mu)}$, with $x_\nu^2 = e^{-2\beta(\epsilon_\nu - \mu)}$, and so on for higher powers. Substituting this expression, we obtain:
\begin{equation}\label{PF:SingleParticle}
	\mathcal{Z}_{\hat{q}}^{(\nu)} = 1 + x_\nu + \frac{1}{2}(1 + \hat{q}) \Big\{x_\nu^2+x_\nu^3+...\Big\}=\frac{1 - \frac{1 - \hat{q}}{2} x_\nu^2}{1 - x_\nu},
\end{equation}
Since the modes are independent, the total grand partition function is the product over all single-particle states:
\begin{equation}\label{Equation:gpfunction}
	\mathcal{Z}(\hat{q})=\prod_\nu \frac{1 - \frac{1 - \hat{q}}{2} x^2_\nu}{1-{x_\nu}} = \prod_\nu \frac{1 - \frac{1 - \hat{q}}{2} e^{2\beta(\mu - \epsilon_\nu)}}{1 - e^{\beta(\mu - \epsilon_\nu)}}.
\end{equation}
In the limiting cases $\hat{q} = 1$ and $\hat{q} = -1$, the grand partition function reduces to the familiar results for bosonic and fermionic statistics, respectively.
\begin{equation}
	\begin{cases}
	    \dfrac{1 - \hat{q}}{2} = 0 \quad \Rightarrow \quad \mathcal{Z}(+1) = \prod^{}_\nu \dfrac{1}{1 - e^{\beta(\mu - \varepsilon_\nu)}}, \,\ & 
        \text{ Bose–Einstein grand partition function},
        \\
        \dfrac{1 - \hat{q}}{2} = 1 \quad \Rightarrow \quad \mathcal{Z}(-1) = 
		\prod_\nu (1 + e^{\beta(\mu - \varepsilon_\nu)}), \,\ & 
        \text{ Fermi–Dirac grand partition function}.
	\end{cases}
\end{equation}
For intermediate values of the deformation parameter $\hat{q}$, the particles exhibit statistical behavior that interpolates continuously between bosonic and fermionic character. Physically, the resulting $\delta$-dependent grand partition function can be understood as arising from a statistical admixture of intrinsically bosonic and fermionic contributions, rather than corresponding to either pure statistics alone.
\begin{equation}\label{Eq:pf}
	\mathcal{Z}(\delta) = \prod_i \frac{1 - \delta z^2 e^{-2\beta \epsilon_\nu}}{1 - z e^{-\beta \epsilon_\nu}},
\end{equation}
		where $z = e^{\beta \mu}$ is the fugacity and $\delta =\frac{1-\hat{q}}{2}$, $0\eqslantless	
		\delta\eqslantless	
		1$. This smoothly interpolates between bosons ($\delta = 0$) and fermions ($\delta = 1$).
		%--------------------------------------
		To obtain the mean occupation number \( \langle n_\nu \rangle \) of a single-particle state in energy level $\nu$
		in the grand canonical ensemble ,we apply the standard thermodynamic relation from the grand canonical ensemble:
		\begin{equation}\label{Equation:OCCnumber}
			\langle n_\nu \rangle = \frac{1}{\beta} \frac{\partial \ln \mathcal{Z}_{\hat{q}}^{(\nu)}}{\partial \mu}.
		\end{equation}
where $\mathcal{Z}_{\hat{q}}^{(\nu)}$ is the single-mode partition function corresponding to the $\nu$-th energy level.
From Eq.~\eqref{Eq:pf}, the single-state partition function is:
\begin{equation}
	\mathcal{Z}_{\hat{q}}^{(\nu)} = \frac{1 - \delta z^2 x^2}{1 - z x},
\end{equation}
where $x = e^{-\beta \epsilon_\nu}$. To compute the mean occupation number  $\langle n_\nu \rangle$ via the thermodynamic identity Eq.~\eqref{Equation:OCCnumber}, we first take the logarithm of $\mathcal{Z}_{\hat{q}}^{(\nu)}$:
\begin{equation}
	\ln \mathcal{Z}_{\hat{q}}^{(\nu)} = \ln(1 - \delta z^2 x^2) - \ln(1 - z x),
\end{equation}
We then differentiate this expression with respect to the chemical potential $\mu$:
\begin{equation}
	\frac{\partial \ln \mathcal{Z}_{\hat{q}}^{(\nu)}}{\partial \mu}
	= \left[ \frac{-2\delta z x^2}{1 - \delta z^2 x^2} + \frac{x}{1 - z x} \right] \cdot \frac{\partial z}{\partial \mu}
    = \beta z \left[ \frac{x}{1 - z x} - \frac{2\delta z x^2}{1 - \delta z^2 x^2} \right],
\end{equation}
Thus, using Eq.~\eqref{Equation:OCCnumber} the mean occupation number becomes:
\begin{equation}
	\langle n_\nu
    \rangle = z \left[ \frac{x}{1 - z x} - \frac{2\delta z x^2}{1 - \delta z^2 x^2} \right],
\end{equation}
Substituting back $x = e^{-\beta \epsilon_\nu}$ and $z = e^{\beta \mu}$, and simplifying the resulting expression, we obtain:
\begin{equation}
	\langle n_\nu \rangle = \frac{1}{e^{\beta(\epsilon_\nu - \mu)} - 1}
	- \frac{2\delta}{e^{2\beta(\epsilon_\nu - \mu)} - \delta}.
\end{equation}
This distribution function is the central result of the unified quantum statistics framework. It smoothly interpolates between the Bose–Einstein distribution (when $\delta=0$) and the Fermi–Dirac distribution (when $\delta=1$). 
%==================================
\section{GENERALIZED WICK'S THEOREM}\label{wick-zero}
In order to treat systems governed by unified quantum statistics, the conventional Wick's theorem must be generalized to account for the exchange-statistics operator $\hat{q}$. Such a generalization is essential for evaluating many-body response functions, including density-density correlations and dielectric properties. \\
Within this formalism, a generic field operator is expressed as a linear combination of its creation and annihilation components, as given in Eq.~\eqref{Eq:field_operator}, with the explicit plane-wave expansion provided in Eq.~\eqref{Eq:psi_plus_minus}. The generalized time-ordering operator $T_{\hat{q}}$, defined in Eq.~\eqref{Eq:Tq_definition}, orders these operators according to their time arguments, introducing a factor of $\hat{q}$ whenever the order of two operators is interchanged.
Correspondingly, the generalized normal-ordering operator $N_{\hat{q}}$, defined in Eq.~\eqref{Eq:Nq_definition}, arranges all creation operators to the left of all annihilation operators, attaching a factor of $\hat{q}$ for each transposition required to achieve this ordering.
With these definitions, the time‑ordered product of two generic fields in unifed statistics for $t_1>t_2$ can be decomposed as:
\begin{equation}
	\begin{aligned}
		&T_{\hat{q}}\left\lbrace\hat{\Psi}(x_1)\hat{\Psi}(x_2)\right\rbrace=\hat{\Psi}(x_1)\hat{\Psi}(x_2)=\left(\hat{\psi}^+(x_1)+\hat{\psi}^-(x_1)\right)\left(\hat{\psi}^+(x_2)+\hat{\psi}^-(x_2)\right)
		\\&
		=\hat{\psi}^{(+)}(x_1)\hat{\psi}^{(+)}(x_2)+\hat{q}\hat{\psi}^{(-)}(x_2)\hat{\psi}^{(+)}(x_1)+\left[\hat{\psi}^{(+)}(x_1),\hat{\psi}^{(-)}(x_2)\right]_{\hat{q}}+\hat{\psi}^{(-)}(x_1)\hat{\psi}^{(+)}(x_2)+\hat{\psi}^{(-)}(x_1)\hat{\psi}^{(-)}(x_2)\\
		&=N_{\hat{q}}\{\hat{\Psi}_1(x_1)\hat{\Psi}_2(x_2)\}+\overparent{\hat{\Psi}_1(x_1)\hat{\Psi}_2}(x_2)
	\end{aligned}
\end{equation}
The second term of this expression is the contraction of two field operators $\hat{\Psi}_1(x_1),\hat{\Psi}_2(x_2)$ in unified statistic.
For three fields, considering all the contractions between the fields, we will have:
\begin{equation}
	\begin{aligned}
	&T_{\hat{q}}\left\lbrace\hat{\Psi}(x_1)\hat{\Psi}(x_2)\hat{\Psi}(x_3)\right\rbrace
    \\&
    =N_{\hat{q}}\left\lbrace\hat{\Psi}_1(x_1)\hat{\Psi}_2(x_2)\hat{\psi}(x_3)+\overparent{\hat{\Psi}_1(x_1)\hat{\Psi}_2}(x_2)\hat{\psi}(x_3)+\hat{q}\overparent{\hat{\Psi}_1(x_1)\hat{\Psi}_3}(x_3)\hat{\Psi}_2(x_2)+\hat{\Psi}_1(x_1)\overparent{\hat{\Psi}_2(x_2)\hat{\Psi}_3}(x_3)\right\rbrace.
\end{aligned}
\end{equation}
This result can be systematically generalized to an arbitrary number of fields:
\begin{equation}\label{WickTheorem}
	\begin{aligned}
    T_{\hat{q}}&\left\lbrace\hat{\Psi}_1(x_1)\hat{\Psi}_2(x_2)...\hat{\Psi}_n(x_n)\right\rbrace
	=N_{\hat{q}}\Big\{\hat{\Psi}_1(x_1)\hat{\Psi}_2(x_2)...\hat{\Psi}_n(x_n)\Big\}+\overparent{\hat{\Psi}_1(x_1)\hat{\Psi}_2}(x_2)N_{\hat{q}}\Big\{\hat{\Psi}_3(x_3)...\hat{\Psi}_n(x_{n})\Big\}\\&
	+\hat{q}\overparent{\hat{\Psi}_1(x_1)\hat{\Psi}_3}(x_3)
	N_{\hat{q}}\Big\{\hat{\Psi}_2(x_2)...\hat{\Psi}_n(x_{n})\Big\}+...+(\hat{q})^{n-2}\overparent{\hat{\Psi}_1(x_1)\hat{\Psi}_n}(x_n)N_{\hat{q}}\Big\{\hat{\Psi}_3(x_3)...\hat{\Psi}_n(x_{n-1})\Big\}+\text{all pair contractions}\\
	&+...+(\hat{q})^{\mathcal{P}_{i,j;i',j'}}\overparent{\hat{\Psi}_i(x_i)\hat{\Psi}_j}(x_j)\overparent{\hat{\Psi}_{i'}(x_{i'})\hat{\Psi}_{j'}}(x_{j'})N_{\hat{q}}\Big\{\hat{\Psi}_1(x_1)...\hat{\Psi}_n(x_{n})\Big\}_{i,j,i',j'}+\text{all possible higher order contractions},
	\end{aligned}
\end{equation}
The full Wick expansion for a product of $n$ operators includes not only all possible pairwise contractions, but also accounts for cases where multiple contractions and remaining normal-ordered factors appear simultaneously.
%======================================================
\section{WICK'S THEOREM FOR NONZERO TEMPERATURE}\label{SEC:wick-nonzero}
We provide a self-contained derivation of the generalized Wick’s theorem in the imaginary-time (Matsubara) formalism for particles obeying unified quantum statistics.
The proof establishes the diagrammatic rules required for systematic perturbative calculations at finite temperature.
We define the thermal contraction of two arbitrary operators $\hat{A}$ and $\hat{B}$ in the interaction picture as the thermal average of their $\hat{q}$ -time-ordered product with respect to the non-interacting grand-canonical ensemble:
\begin{equation}\label{Eq:grand-CA}
	\overparent{\hat{A}\hat{B}}
    =\text{Tr}_{\hat{q}}\left\{\hat{\rho}_{\hat{q}}T_{\hat{q}}\left[\hat{A}\hat{B}\right]\right\},
\end{equation}
A fundamental example is the contraction of an annihilation and a creation operator, which yields the free finite-temperature Green’s function:
\begin{equation}
	\overparent{\hat{\psi}_I^{}(\text{x}\tau)\hat{\psi}_I^{\dagger}}(\text{x}'\tau')=-G^{\mathcal{M}}_{\hat{q}}(\text{x}\tau,\text{x}'\tau').
\end{equation}
Consider the thermal average of a $\hat{q}$ -time ordered product of four operators $\hat{A},\hat{B},\hat{C},\hat{D}$:
\begin{equation}\label{Eq:wicknonzero}
	\left\langle T_{\hat{q}}\left[\hat{A}\hat{B}\hat{C}\hat{D}\right]\right\rangle_0=\left\langle \hat{A}\hat{B}\hat{C}\hat{D}\right\rangle_0
	=\left\langle\overparent{\hat{A}\hat{B}}\overparent{\hat{C}\hat{D}}\right\rangle_0
	+\hat{q}\left\langle\overparent{\hat{A}\hat{C}}\overparent{\hat{B}\hat{D}}\right\rangle_0
	+\left\langle\overparent{\hat{A}\hat{D}}\overparent{\hat{B}\hat{C}}\right\rangle_0.
\end{equation}
In the interaction picture, each operator can be expanded in a single-particle basis:
\begin{equation}\label{Eq:state-SP}
	\hat{\Psi}_I~~\text{or}~~ \hat{\Psi}_I^\dagger=\sum_{j}\chi_j(\text{x}\tau)\alpha_j,
\end{equation}
where $\alpha_j$, denotes $a_j$, or $a^\dagger_j$ and $\chi_j(\text{x}\tau)$ denotes $\phi^0_j(\text{x}) e^{-\tau\zeta_j}$ or $\phi^0_j(\text{x})^\dagger e^{\tau\zeta_j}$. So the thermal average then becomes a sum over mode indices:
\begin{equation}
	\left\langle \hat{A}\hat{B}\hat{C}\hat{D}\right\rangle_0=\sum_{a}\sum_{b}\sum_{c}\sum_{d}\chi_a\chi_b\chi_c\chi_d\text{Tr}_{\hat{q}}\left(\hat{\rho}_{\hat{q}}\alpha_a(\tau)\alpha_b\alpha_c\alpha_d\right).
\end{equation}
Using the generalized commutation relation  $\hat{\rho}_{\hat{q}}=e^{-\beta K_0}/\mathcal{Z}_{\hat{q}}$ and $\alpha_a\alpha_b=\left[\alpha_a,\alpha_b\right]_{\hat{q}}+\hat{q}\alpha_b\alpha_a$
and the cyclic property of the trace, we commute $\alpha_a(\tau)$ successively to the right:
\begin{equation}\label{Eq:alpha}
	\begin{aligned}
		\text{Tr}_{\hat{q}}&\left(\hat{\rho}_{\hat{q}}\alpha_a(\tau)\alpha_b\alpha_c\alpha_d\right)=
		\text{Tr}_{\hat{q}}\left(\hat{\rho}_{\hat{q}}\left[\alpha_a(\tau),\alpha_b\right]_{\hat{q}}\alpha_c\alpha_d\right)
		+\hat{q}\text{Tr}_{\hat{q}}\left(\hat{\rho}_{\hat{q}}\alpha_b\alpha_a(\tau)\alpha_c\alpha_d\right)
		\\&=\text{Tr}_{\hat{q}}\left(\hat{\rho}_{\hat{q}}\left[\alpha_a(\tau),\alpha_b\right]_{\hat{q}}\alpha_c\alpha_d\right)
		+\hat{q}\text{Tr}_{\hat{q}}\left(\hat{\rho}_{\hat{q}}\alpha_b\left[\alpha_a(\tau),\alpha_c\right]_{\hat{q}}\alpha_d\right)
		+\text{Tr}_{\hat{q}}\left(\hat{\rho}_{\hat{q}}\alpha_b\alpha_c\alpha_a(\tau)\alpha_d\right)
		\\&=\text{Tr}_{\hat{q}}\left(\hat{\rho}_{\hat{q}}\left[\alpha_a(\tau),\alpha_b\right]_{\hat{q}}\alpha_c\alpha_d\right)
		+\hat{q}\text{Tr}_{\hat{q}}\left(\hat{\rho}_{\hat{q}}\alpha_b\left[\alpha_a(\tau),\alpha_c\right]_{\hat{q}}\alpha_d\right)
		+\text{Tr}_{\hat{q}}\left(\hat{\rho}_{\hat{q}}\alpha_b\alpha_c\left[\alpha_a(\tau),\alpha_d\right]_{\hat{q}}\right)
		+\hat{q}\text{Tr}_{\hat{q}}\left(\hat{\rho}_{\hat{q}}\alpha_b\alpha_c\alpha_d\alpha_a(\tau)\right).
	\end{aligned}
\end{equation}
The last term can be rewritten using the periodicity property of the imaginary-time operators,
the time evolution of the creation and annihilation operators in the interaction picture is governed by the generator $\hat{K}_0 $. According to the Heisenberg equation of motion in imaginary time, these operators $(\hat{a}_\nu(\tau),\hat{a}^{\dagger}_\nu(\tau))$ evolve as follows:
\begin{equation}
	\begin{aligned}
		\hat{a}_\nu(\tau)&=e^{\tau\hat{K}_0} \hat{a}^{}_\nu e^{-\tau\hat{K}_0}=
		e^{-\tau\zeta^{}_\nu} \mathbf{1}^{\hat{q}}
		\hat{a}_\nu,
		\\
		\hat{a}^{\dagger}_\nu(\tau)&=e^{\tau\hat{K}_0} \hat{a}^{}_\nu e^{-\tau\hat{K}_0}=
		e^{\tau\zeta^{}_\nu} \mathbf{1}^{\hat{q}}
		\hat{a}^{\dagger}_\nu,
	\end{aligned}   
\end{equation}
Furthermore, a straightforward generalization leads to the transformation of the operator $\alpha_\nu$:
\begin{equation}
	\begin{aligned}
		\alpha_\nu(\tau)&=e^{\tau\hat{K}_0}\alpha^{}_\nu e^{-\tau\hat{K}_0}=
		e^{\lambda_\nu\tau\zeta^{}_\nu} \mathbf{1}^{\hat{q}}
		\alpha_\nu,
	\end{aligned}   
\end{equation}
Where $\lambda_\nu=1$ if $\alpha_\nu$ is a creation operator, and $\lambda_\nu=-1$ if $\alpha_\nu$ is a annihilation operator. This relation is equivalent to the equation 
\begin{equation}
	\alpha_\nu\hat{\rho}_{\hat{q}}^{}=\hat{\rho}_{\hat{q}}^{}\alpha_\nu e^{\lambda_\nu\tau\zeta^{}_\nu} \mathbf{1}^{\hat{q}},
\end{equation}
where $\hat{\rho}_{\hat{q}}^{}$ denotes the density matrix.
Similarly, we can write the imaginary-time evolution with a shift $\beta$ as:
\begin{equation}
	\begin{aligned}
		\alpha_\nu(\tau+\beta)&=e^{\beta\hat{K}_0}\alpha^{}_\nu(\tau) e^{-\beta\hat{K}_0}=
		e^{\lambda_\nu\beta\zeta^{}_\nu} \mathbf{1}^{\hat{q}}
    	\alpha_\nu(\tau),
	\end{aligned}   
\end{equation}
where $\beta$ denotes the inverse temperature.
\begin{equation}
	\alpha_\nu(\tau)\hat{\rho}_{\hat{q}}^{}=\hat{\rho}_{\hat{q}}^{}\alpha_\nu(\tau) e^{\lambda_\nu\beta\zeta^{}_\nu} \mathbf{1}^{\hat{q}}
\end{equation}
Finally, applying the cyclic property of the trace, the last term in Eq.~\eqref{Eq:alpha} can be rewritten as:
\begin{equation}
	\hat{q}\text{Tr}_{\hat{q}}\left(\alpha_a(\tau)\hat{\rho}_{\hat{q}}\alpha_b\alpha_c\alpha_d\right)
	=\hat{q}e^{\lambda_\nu\beta\zeta^{}_\nu}\text{Tr}_{\hat{q}}\left(\hat{\rho}_{\hat{q}}\alpha_a(\tau)\alpha_b\alpha_c\alpha_d\right),
\end{equation}
Thus, we obtain the following trace identity:
\begin{equation}\label{fourfield}
	\begin{aligned}
	\text{Tr}_{\hat{q}}\left(\hat{\rho}_{\hat{q}}\alpha_a(\tau)\alpha_b\alpha_c\alpha_d\right)
	&=\frac{\left[\alpha_a(\tau),\alpha_b\right]_{\hat{q}}}{1-\hat{q}e^{\lambda_\nu\beta\zeta^{}_\nu}}\text{Tr}_{\hat{q}}\left(\hat{\rho}_{\hat{q}}\alpha_c\alpha_d\right)
	+\hat{q}\frac{\left[\alpha_a(\tau),\alpha_c\right]_{\hat{q}}}{1-\hat{q}e^{\lambda_\nu\beta\zeta^{}_\nu}}\text{Tr}_{\hat{q}}\left(\hat{\rho}_{\hat{q}}\alpha_b\alpha_d\right)
	+\frac{\left[\alpha_a(\tau),\alpha_d\right]_{\hat{q}}}{1-\hat{q}e^{\lambda_\nu\beta\zeta^{}_\nu}}\text{Tr}_{\hat{q}}\left(\hat{\rho}_{\hat{q}}\alpha_b\alpha_c\right)
	\\&
	=\overparent{\alpha_a(\tau)\alpha_b}\text{Tr}_{\hat{q}}\left(\hat{\rho}_{\hat{q}}\alpha_c\alpha_d\right)
	+\hat{q}\overparent{\alpha_a(\tau)\alpha_c}\text{Tr}_{\hat{q}}\left(\hat{\rho}_{\hat{q}}\alpha_b\alpha_d\right)
	+\overparent{\alpha_a(\tau)\alpha_d}\text{Tr}_{\hat{q}}\left(\hat{\rho}_{\hat{q}}\alpha_b\alpha_c\right).
	\end{aligned}
\end{equation}
Now, to show that the expression $\overparent{\alpha_a(\tau)\alpha_b}$ is equal to the Green's function, we can write Wick's theorem on two fields, which will be equal to the Green's function. so we have ($\tau_A>\tau_B$):
\begin{equation}\label{Eq:TO_TwoField}
	\left\langle T_{\hat{q}}\left[\hat{A}\hat{B}\right]\right\rangle_0
	=\left\langle\hat{A}\hat{B}\right\rangle_0
	=\left\langle\overparent{\hat{A}\hat{B}}\right\rangle_0,
\end{equation}
Using Eq.~\eqref{Eq:grand-CA} and Eq.~\eqref{Eq:state-SP},we can write
\begin{equation}
	\left\langle\hat{A}\hat{B}\right\rangle_0=\sum_{a}\sum_{b}\chi_a\chi_b\text{Tr}_{\hat{q}}\left(\hat{\rho}_{\hat{q}}\alpha_a(\tau)\alpha_b\right),
\end{equation}
The trace can be decomposed using the unified statistics commutator:
\begin{equation}\label{Eq:TR-FT}
	\text{Tr}_{\hat{q}}\left(\hat{\rho}_{\hat{q}}\alpha_a(\tau)\alpha_b\right)
	=\text{Tr}_{\hat{q}}\left(\hat{\rho}_{\hat{q}}\left[\alpha_a(\tau),\alpha_b\right]_{\hat{q}}\right)
	+\hat{q}\text{Tr}_{\hat{q}}\left(\hat{\rho}_{\hat{q}}\alpha_b\alpha_a(\tau)\right).
\end{equation}
Given that $\alpha_a(\tau)\hat{\rho}_{\hat{q}}=\hat{\rho}_{\hat{q}}\alpha_a(\tau)e^{\lambda_a\beta\zeta_a}\mathbf{1}^{\hat{q}}$, we can substitute into Eq.~\eqref{Eq:TR-FT} and use the cyclic property of the trace to obtain:
\begin{equation}\label{Eq:Tr.t1}
	\text{Tr}_{\hat{q}}\left(\hat{\rho}_{\hat{q}}\alpha_a(\tau)\alpha_b\right)
	=\text{Tr}_{\hat{q}}\left(\hat{\rho}_{\hat{q}}\left[\alpha_a(\tau),\alpha_b\right]_{\hat{q}}\right)
	+\hat{q}e^{\lambda_a\beta\zeta_a}\text{Tr}_{\hat{q}}\left(\hat{\rho}_{\hat{q}}\alpha_a(\tau)\alpha_b\right),
\end{equation}
Solving Eq.~\eqref{Eq:Tr.t1} for the trace yields the explicit form:
\begin{equation}
	\overparent{\alpha_a(\tau)\alpha_b}=-G^{\mathcal{M}}_{\hat{q}}(\tau,0)=\text{Tr}_{\hat{q}}\left(\hat{\rho}_{\hat{q}}\alpha_a(\tau)\alpha_b\right)
   =\frac{\text{Tr}_{\hat{q}}\left(\hat{\rho}_{\hat{q}}\left[\alpha_a(\tau),\alpha_b\right]_{\hat{q}}\right)}{1-\hat{q}e^{\lambda_a\beta\zeta_a}}.
 % =\frac{\left[\alpha_a(\tau),\alpha_b\right]_{\hat{q}}}{1-\hat{q}e^{\lambda_a\beta\zeta_a}}.
    \label{Eq:contr}
\end{equation}
Therefore, the time-ordered two-point function in the non-interacting ensemble reduces to Eq.~\eqref{Eq:TO_TwoField} and applying Wick’s theorem for the four-operator trace in the presence of $\hat{q}$-statistics yields:
\begin{equation}\label{fourfield}
	\begin{aligned}
	\text{Tr}_{\hat{q}}\left(\hat{\rho}_{\hat{q}}\alpha_a(\tau)\alpha_b\alpha_c\alpha_d\right)
	&=\overparent{\alpha_a(\tau)\alpha_b}\overparent{\alpha_c\alpha_d}
	+\hat{q}\overparent{\alpha_a(\tau)\alpha_c}\overparent{\alpha_b\alpha_d}
	+\overparent{\alpha_a(\tau)\alpha_d}\overparent{\alpha_b\alpha_c},
	\end{aligned}
\end{equation}
Consequently, the time-ordered four-point function in the non-interacting ensemble decomposes as:
\begin{equation}\label{Eq:wicknonzero}
	\left\langle T_{\hat{q}}\left[\hat{A}\hat{B}\hat{C}\hat{D}\right]\right\rangle_0=\left\langle \hat{A}\hat{B}\hat{C}\hat{D}\right\rangle_0
	=\left\langle\overparent{\hat{A}\hat{B}}\overparent{\hat{C}\hat{D}}\right\rangle_0
	+\hat{q}\left\langle\overparent{\hat{A}\hat{C}}\overparent{\hat{B}\hat{D}}\right\rangle_0
	+\left\langle\overparent{\hat{A}\hat{D}}\overparent{\hat{B}\hat{C}}\right\rangle_0.
\end{equation}
which reflects the sum over all possible pairwise contractions, with the factor $\hat{q}$ accounting for the exchange statistics ($+1$ for bosons, $-1$ for fermions) when permuting operators under the generalized time ordering $T_{\hat{q}}$.
%%%%%%%%%%%%%%%%%%%%%%%%%%%%%%%%%%%%%
\section{GREEN'S FUNCTION CALCULATIONS }\label{AppendixB}
In this section, we examine the single-particle Green's function within the framework of unified quantum statistics. The system is described by a non-interacting Hamiltonian
\begin{equation}
	\begin{aligned}
	\hat{H}_{\hat{q}}=\sum_\nu\varepsilon_\nu\hat{n}_\nu=\sum_{\nu}\varepsilon_\nu \hat{a}^{\dagger}_{\nu}\hat{a}^{}_{\nu},
	\end{aligned}
\end{equation}
Within the grand-canonical ensemble, we introduce the grand-canonical Hamiltonian $\hat{K}$ by incorporating the chemical potential $\mu$ into the Hamiltonian system.  In the single-particle state representation, it can be written as
\begin{equation}
	\begin{aligned} 
	\hat{K}=\hat{H}_{\hat{q}}-\mu\hat{N}     &=\sum_{\nu}\zeta_\nu \hat{a}^{\dagger}_{\nu}\hat{a}^{}_{\nu} 
	~~~,~~~~~~
    \zeta_\nu=\varepsilon_\nu-\mu,
	\end{aligned}
\end{equation}
The time evolution of an operator follows from the Baker–Hausdorff expansion:
\begin{equation}
	e^A C e^{-A}
	=
	C+[A,C]+\frac{1}{2!}[A,[A,C]]+\ldots,
\end{equation}
This expansion provides an exact expression when the series converges, and it terminates identically when the nested commutators vanish beyond a finite order
In our context, we set $A=\tau\hat{K}$
and $C=\hat{a}_\nu$, enabling us to compute the Heisenberg-picture operator $e^{\tau\hat{K}} \hat{a}^{}_\nu e^{-\tau\hat{K}}$ order by order in $\tau$.
Applying the expansion and using the commutation relation $[\hat{K},\hat{a}_\nu]=-\zeta_\nu 1^{\hat{q}}$ $\hat{a}_\nu$, we obtain the closed form:
\begin{equation}
	\begin{aligned}
	\hat{a}_\nu(\tau)=e^{\tau\hat{K}} \hat{a}^{}_\nu e^{-\tau\hat{K}}
	&=\hat{a}_\nu+[\hat{K},\hat{a}_\nu]+\ldots
	=
	\hat{a}_\nu+(-\tau\zeta_\nu \mathbf{1}^{\hat{q}}) \hat{a}_\nu+\frac{1}{2!}(-\tau\zeta_\nu \mathbf{1}^{\hat{q}})^2  
	\hat{a}_\nu+\frac{1}{3!}(-\tau\zeta_\nu \mathbf{1}^{\hat{q}})^3  
	\hat{a}_\nu+...
	\\&=
	\Big(\sum_{n=0}^\infty\frac{(-\tau\zeta_\nu \mathbf{1}^{\hat{q}})^n}{n!}\Big)
	\hat{a}_\nu
	=
    \Big(\sum_{n=0}^\infty\frac{(-\tau\zeta_\nu )^n}{n!}\Big)\mathbf{1}^{\hat{q}}
	\hat{a}_\nu
	=
	e^{-\tau\zeta^{}_\nu} \mathbf{1}^{\hat{q}}
	\hat{a}_\nu,
	\end{aligned}  
    \label{Eq:anonInt}
\end{equation}
and similarly for the creation operator $\hat{a}^{\dagger}_{\nu}(\tau)$: 
\begin{equation}
	\begin{aligned}
		\hat{a}^{\dagger}_\nu(\tau)&=
		e^{\tau\zeta^{}_\nu} \mathbf{1}^{\hat{q}}
		\hat{a}^{\dagger}_\nu,
	\end{aligned}
    \label{Eq:adagger}
\end{equation}
The standard single particle finite-temperature Green’s function for unified quantum statistics is given by
\begin{equation}\label{Eq:SingleParticle1}
	G^{\mathcal{M}0}_{\hat{q}}(\nu;\tau,\tau')
	=-\left\langle T_{\hat{q}}(\hat{a}_\nu(\tau)\hat{a}^{\dagger}_{\nu}(\tau'))\right\rangle,
\end{equation}
We assume that $\tau'=0$ and $\tau>0$, so the $\tau$ dependence of Green's function becomes:
\begin{equation}\label{Eq:SingleParticle2}
	\begin{aligned}
	G^{\mathcal{M}0}_{\hat{q}}(\nu;\tau)
	&=
	-\left\langle \hat{a}_\nu(\tau)\hat{a}^{\dagger}_{\nu}(0)\right\rangle=
	-\frac{1}{\mathcal{Z}_{\hat{q}}}\text{Tr}\left[e^{-\beta(\hat{H}-\mu \hat{N})}\hat{a}_\nu(\tau)\hat{a}^{\dagger}_{\nu}(0)\right],
	\end{aligned}
\end{equation}
Substituting $\hat{a}_\nu(\tau)=e^{-\zeta_\nu\tau}\mathbf{1}^{\hat{q}}\hat{a}_\nu$ 
and using the cyclic property of the trace, the expression becomes:
\begin{equation}\label{equ3}
    \begin{aligned}
		G^{\mathcal{M}0}_{\hat{q}}(\nu;\tau)
		&=-\frac{e^{-\zeta_\nu\tau}}{\mathcal{Z}_{\hat{q}}}
		\sum_{\{n_i\}}\langle\{n_i\}|e^{-\beta(\hat{H}-\mu \hat{N})}\mathbf{1}^{\hat{q}}\hat{a}_\nu \hat{a}^{\dagger}_{\nu})|\{n_i\}\rangle
		-\frac{e^{-\zeta_\nu\tau}}{\mathcal{Z}_{\hat{q}}}
		\prod_i\sum_{\{n_i\}}e^{-\beta(\varepsilon_i-\mu){n}_i} \left\langle\{n_i\}|
        \mathbf{1}^{\hat{q}}\hat{a}_{\nu}\hat{a}^{\dagger}_\nu|\{n_i\}\right\rangle
		\\&=
		-\frac{e^{-\zeta_\nu\tau}}{\mathcal{Z}_{\hat{q}}}
		\prod_i\sum_{\{n_i\}}e^{-\beta(\varepsilon_i-\mu )n_i}\langle\{n_i\}|\mathbf{1}^{\hat{q}}(\mathbf{1}^{\hat{q}}+\hat{q}\hat{a}^{\dagger}_\nu\hat{a}_{\nu})|\{n_i\}\rangle
		\\&=
		-e^{-\zeta_\nu\tau}\Big\{
		\frac{1}{\mathcal{Z}_{\hat{q}}}
		\prod_i\sum_{\{n_i\}}e^{-\beta(\varepsilon_i-\mu )n_i}\langle\{n_i\}|\mathbf{1}^{\hat{q}}|\{n_i\}\rangle
		+\hat{q}\frac{1}{\mathcal{Z}_{\hat{q}}}
		\prod_i\sum_{\{n_i\}}e^{-\beta(\varepsilon_i-\mu )n_i}\langle\{n_i\}| \mathbf{1}^{\hat{q}}\hat{a}^{\dagger}_\nu\hat{a}_{\nu}|\{n_i\}\rangle\Big\},
	\end{aligned}
\end{equation}
Carrying out the summations, the first contribution from the identity part yields  (See Eq.~\eqref{Eq:RelationOrthogonal} and Eq.~\eqref{Eq:Base}):
\begin{equation}
	\begin{aligned}
	\frac{1}{\mathcal{Z}_{\hat{q}}}
	\prod_i\sum_{\{n_i\}}
	e^{-\beta(\varepsilon_i-\mu )n_i}\;
	{}^{\hat{q}}\langle\{n_i\}|\mathbf{1}^{\hat{q}}|\{n_i\}\rangle^{\hat{q}}
	=
	\frac{
	\prod_i\sum_{\{n_i\}}e^{-\beta(\varepsilon_i-\mu )n_i}(\mathbf{1}^{\hat{q}})^2}{\prod_i\sum_{\{n_i\}}e^{-\beta(\varepsilon_i-\mu )n_i}\mathbf{1}^{\hat{q}}}
	=
	\mathbf{1}^{\hat{q}},
	\end{aligned}
\end{equation}
And the second term becomes,
\begin{equation}
	\begin{aligned}
	\frac{1}{\mathcal{Z}_{\hat{q}}}
	\prod_i\sum_{\{n_i\}}
	e^{-\beta(\varepsilon_i-\mu )n_i}\;
	{}^{\hat{q}}\langle\{n_i\}|\mathbf{1}^{\hat{q}}\hat{a}^{\dagger}_{\nu}\hat{a}_\nu|\{n_i\}\rangle^{\hat{q}}
	=
	\frac{1}{\mathcal{Z}_{\hat{q}}}
	\prod_i\sum_{\{n_i\}}\sum_{\{n'_i\}}
	e^{-\beta(\varepsilon_i-\mu )n_i}\;
	{}^{\hat{q}}\langle\{n_i\}|\{n'_i\}\rangle^{\hat{q}}\;
	{}^{\hat{q}}\langle\{n'_i\}|\hat{a}^{\dagger}_{\nu}\hat{a}_\nu|\{n_i\}\rangle^{\hat{q}}
	&\\=
	\frac{1}{\mathcal{Z}_{\hat{q}}}
	\prod_i\sum_{\{n_i\}}
	e^{-\beta(\varepsilon_i-\mu )n_i}\;
	n_\nu(\mathbf{1}^{\hat{q}})^2
	=
	\frac{
	\prod_i\sum_{\{n_i\}}e^{-\beta(\varepsilon_i-\mu )n_i}n_\nu(\mathbf{1}^{\hat{q}})^2}{\prod_i\sum_{\{n_i\}}e^{-\beta(\varepsilon_i-\mu )n_i}\mathbf{1}^{\hat{q}}}
	=
	\mathbf{1}^{\hat{q}}n_\nu&,
	\end{aligned}
\end{equation}
where $n_\nu$ is the number operator: 
$n_\nu={}^{\hat{q}}\langle\hat{a}^{\dagger}_\nu\hat{a}_\nu\rangle^{\hat{q}}$, and $^{\hat{q}}\langle\mathbf{1}^{\hat{q}}\rangle^{\hat{q}}=\mathbf{1}^{\hat{q}}$.
%%%===========================================
The single-particle finite-temperature Green’s function in unified quantum statistics is defined as Eq.~\eqref{equ3}
\begin{equation}
	\begin{aligned}
	&G^{\mathcal{M}0}_{\hat{q}}(\nu;\tau)
	=-\frac{1}{\mathcal{Z}_{\hat{q}}}e^{-\tau\zeta_\nu}\sum_{\{n_i\}}e^{-\beta\sum_{i}\zeta_i n_i}\Bigg(\left\langle n_1,n_2,\ldots|\mathbf{1}^{\hat{q}}|n_1,n_2,\ldots\right\rangle
	+\left\langle n_1,n_2,\ldots|\mathbf{1}^{\hat{q}}\hat{q}a^\dagger_\nu a_\nu|n_1,n_2,\ldots\right\rangle\Bigg),
	\end{aligned}
\end{equation}
which is written in the occupation-number basis. Using the definition of the partition function, we obtain 
\begin{equation}
	\begin{aligned}
	\mathcal{Z}_{\hat{q}}=\sum_{\{n_i\}}e^{-\beta\sum_{i}\zeta_i n_i}\left\langle n_1,n_2,\ldots|\mathbf{1}^{\hat{q}}|n_1,n_2,\ldots\right\rangle,
	\end{aligned}
\end{equation}
Then
\begin{equation}\label{Eq:BNV}
	\begin{aligned}
	G^{\mathcal{M}0}_{\hat{q}}(\nu;\tau)=-\frac{1}{\mathcal{Z}_{\hat{q}}}e^{-\tau\zeta_\nu}\Bigg(\mathcal{Z}_{\hat{q}}+\hat{q}\sum_{\{n_i\}}e^{-\beta\sum_{i}\zeta_i n_i}\mathbf{1}^{\hat{q}}n_\nu\Bigg).
	\end{aligned}
\end{equation}
Since the total partition function factorizes into contributions from each single-particle state, this sum can be separated into a product over all modes, with the mode $\nu$ treated explicitly. 
\begin{equation}
	\begin{aligned}
	&\sum_{\{n_i\}}e^{-\beta\sum_{i}\zeta_i n_i}\mathbf{1}^{\hat{q}}n_\nu
	=\sum'_{\{n_i\}}e^{-\beta\sum'_{i}\zeta_i n_i}\mathbf{1}^{\hat{q}}\left(\sum_{n_\nu}e^{-\beta\zeta_\nu n_\nu}\mathbf{1}^{\hat{q}}n_\nu\right)\\
	&=\left[\frac{\sum_{n_\nu}e^{-\beta\zeta_\nu n_\nu}\mathbf{1}^{\hat{q}}}{\sum_{n_\nu}e^{-\beta\zeta_\nu n_\nu}\mathbf{1}^{\hat{q}}}\right]\sum'_{\{n_i\}}e^{-\beta\sum'_{i}\zeta_i n_i}\mathbf{1}^{\hat{q}}\left(\sum_{n_\nu}e^{-\beta\zeta_\nu n_\nu}\mathbf{1}^{\hat{q}}n_\nu\right)
	=\mathcal{Z}_{\hat{q}}\frac{\sum_{n_\nu}e^{-\beta\zeta_\nu n_\nu}\mathbf{1}^{\hat{q}}n_\nu}{\sum_{n_\nu}e^{-\beta\zeta_\nu n_\nu}\mathbf{1}^{\hat{q}}}=-\mathcal{Z}_{\hat{q}}\frac{1}{\zeta_\nu}\frac{\partial}{\partial\beta}\ln{\mathcal{Z}^\nu_{\hat{q}}},
	\end{aligned}
\end{equation}
Therefore, the equation ~\eqref{Eq:BNV} becomes as follows:
\begin{equation}
	\begin{aligned}
	G^{\mathcal{M}0}_{\hat{q}}(\nu;\tau)&=-e^{-\tau\zeta_\nu}\Bigg(1-\hat{q}\frac{1}{\zeta_\nu}\frac{\partial}{\partial\beta}\ln{\mathcal{Z}^\nu_{\hat{q}}}\Bigg),
	\end{aligned}
\end{equation}
By substituting Eq.~\eqref{PF:SingleParticle} and taking the derivative and simplifying, we will have:
\begin{equation}\label{Eq:GF-M0}
	G^{\mathcal{M}0}_{\hat{q}}(\nu;\tau)=e^{- \tau\zeta_\nu  } \left(-\frac{\hat{q}}{e^{\beta  \zeta_\nu }-1}-\frac{2
	(\hat{q}-1) \hat{q}}{2 e^{2 \beta  \zeta_\nu }+\hat{q}-1}-1\right).
\end{equation}
The expectation value of $\hat{q}$ is $1-2\delta$. Replacing the operator $\hat{q}$ by its expectation value turns the expression into a function based on the numeric parameter $\delta$, that $\langle\hat{q}\rangle=\vartheta=1-2\delta$ and we obtain
\begin{equation}\label{Eq:GFf}
	\begin{aligned}
	G^{\mathcal{M}0}_{\delta}(\nu,\tau)
	=-e^{-\zeta_\nu  \tau } \left(1+(1-2 \delta ) \left(\frac{1}{e^{\beta\zeta_\nu }-1}-\frac{2 \delta }{e^{2 \beta  \zeta_\nu }-\delta}\right)\right)
	=-e^{-\zeta_\nu  \tau } \left(1+\vartheta n_\nu\right),
	\end{aligned}
\end{equation}
Its Fourier transform is the Green's function of frequency and defined as
\begin{equation}
	\begin{aligned}
	G^{\mathcal{M}0}_{\delta}(\nu,i\omega_n^{(j)})&=\int_0^\beta d\tau e^{i\omega_n^{(j)}\tau} G^{\mathcal{M}0}_{\delta}(\nu,\tau)
	=-\left(1+\vartheta n_\nu\right)\int_0^\beta d\tau e^{\tau(i\omega_n^{(j)}-\zeta_\nu)}
	=-\frac{\left(1+\vartheta n_\nu\right)(e^{\beta(i\omega_n^{(j)}-\zeta_\nu)}-1)}{i\omega_n^{(j)}-\zeta_\nu}.
	\end{aligned}
\end{equation}
The generalized Matsubara frequencies arise from the poles of the unified distribution function: 
\begin{equation}
		i\beta\omega_n^{(j)}=-(j-2)i2n\pi+(j-1)\frac{i(2n\pi)+\ln{\delta}}{2}.
\end{equation}
So that $j=1,2$ give boson like and fermion like (for $\delta=0$) feriquencies.
The single-particle finite-temperature Green’s function for negative times $(\tau<0)$ is as follows:
\begin{equation}
	\begin{aligned}
		G^{\mathcal{M}0}_{\hat{q}}(\nu;\tau)
		&=-\left\langle T_{\hat{q}}a_\nu(\tau)a^{\dagger}_{\nu}(0)\right\rangle
		\\
		&=-\frac{1}{\mathcal{Z}_{\hat{q}}}\text{Tr}\left(e^{-\beta(H-\mu N)} T_{\hat{q}}a_\nu(\tau)a^{\dagger}_{\nu}(0)\right)
    	=-\frac{1}{\mathcal{Z}_{\hat{q}}}e^{-\tau\zeta_\nu}\text{Tr}\left(e^{-\beta(H-\mu N)}\mathbf{1}^{\hat{q}}\hat{q} a^{\dagger}_{\nu}a_\nu\right)\\&
		=-\frac{1}{\mathcal{Z}_{\hat{q}}}e^{-\tau\zeta_\nu}\sum_{\{n_i\}}e^{-\beta\sum_{i}\zeta_i n_i}\left\langle n_1,n_2,\ldots|\mathbf{1}^{\hat{q}}\hat{q}a^\dagger_\nu a_\nu|n_1,n_2,\ldots\right\rangle,
	\end{aligned}
\end{equation}
As a result, it can be written
\begin{equation}
	\begin{aligned}
	&G^{\mathcal{M}0}_{\hat{q}}(\nu;\tau)=-\frac{1}{\mathcal{Z}_{\hat{q}}}e^{-\tau\zeta_\nu}\hat{q}\sum_{\{n_i\}}e^{-\beta\sum_{i}\zeta_i n_i}\mathbf{1}^{\hat{q}}n_\nu
    =-\frac{1}{\mathcal{Z}_{\hat{q}}}e^{-\tau\zeta_\nu}\Bigg(\hat{q}\mathcal{Z}_{\hat{q}}\frac{1}{\zeta_\nu}\frac{\partial}{\partial\beta}\ln{\mathcal{Z}^\nu_{\hat{q}}}\Bigg)
	=-e^{-\tau\zeta_\nu}\hat{q}\frac{1}{\zeta_\nu}\frac{\partial}{\partial\beta}\ln{\mathcal{Z}^\nu_{\hat{q}}}.
	\end{aligned}
\end{equation}

An alternative strategy consists of eliminating the operator $\hat{q}$ from the denominator of the expression, as illustrated in the following derivation. This is achieved by suitably rearranging the terms so that $\hat{q}$ appears only in the numerator, thereby avoiding the need to invert the operator $\hat{q}$ and simplifying the subsequent algebraic manipulations.
\begin{equation}
	G^{\mathcal{M}0}_{\hat{q}}(\nu,\tau)=
	e^{-\zeta_\nu\tau}_{}
	\Big(-\frac{\hat{q}}{e^{\beta\zeta_\nu}_{}-1}
	+\frac{2(1-\hat{q})}{(1-2e^{2\beta\zeta_\nu}_{})}\hat{f}-1\Big),
\end{equation}
We have used the following relation
\begin{equation}\label{Eq:hatf}
	\begin{aligned}
	&\frac{1}{(1-2e^{2\beta\zeta_\nu}_{})-\hat{q}}=
	\frac{\hat{f}}{(1-2e^{2\beta\zeta_\nu}_{})},
	\end{aligned}
\end{equation}
As a result $\hat{f}=\dfrac{1}{1-\dfrac{\hat{q}}{(1-2e^{2\beta\zeta_\nu}_{})}}$
Defining $\hat{\epsilon}=\dfrac{\hat{q}}{(1-2e^{2\beta\zeta_\nu}_{})}$ and solving Eq.~\eqref{Eq:hatf} for $\hat{f}$ gives:
\begin{equation}
	\begin{aligned}
	\hat{f}&=
	\frac{1}{1-\hat{\epsilon}}=
	\sum^{\infty}_{n=0}\hat{\epsilon}^n\to
	\sum^{\infty}_{n=0}
	\left(\frac{1}{1-2e^{2\beta\zeta_\nu}_{}}\right)^n\hat{q}^n,
	\end{aligned}
\end{equation}
We split the sum into even and odd powers to obtain:
\begin{equation}
	\begin{aligned}
		\hat{f}
		&=\sum^{\infty}_{m=0}
		\left(\frac{1}{1-2e^{2\beta\zeta_\nu}_{}}\right)^{2m}
		+\hat{q}
		\sum^{\infty}_{m=0}
		\left(\frac{1}{1-2e^{2\beta\zeta_\nu}_{}}\right)^{2m+1}  
		=\sum^{\infty}_{m=0}
		\left(\frac{1}{(1-2e^{2\beta\zeta_\nu}_{})^2}\right)^{m}
		\left(
		1+\frac{\hat{q}}{(1-2e^{2\beta\zeta_\nu}_{})}\right)
		\\&=
		\frac{1}{1-
		\frac{1}{(1-2e^{2\beta\zeta_\nu}_{})^2}}
		\left(
		1+\frac{\hat{q}}{(1-2e^{2\beta\zeta_\nu}_{})}
		\right)=
		\frac{(1-2e^{2\beta\zeta}_{})^2}{(1-
		2e^{2\beta\zeta}_{})^2-1}
		\left(
		\frac{1-2e^{2\beta\zeta_\nu}_{}+\hat{q}}{(1-2e^{2\beta\zeta_\nu}_{})}
		\right)=
		\frac{(1-2e^{2\beta\zeta_\nu}_{})(1-2e^{2\beta\zeta_\nu}_{}+\hat{q})}{(1-
		2e^{2\beta\zeta_\nu}_{})^2-1},
	\end{aligned}
\end{equation}
Using this new definition, the corresponding Green's function can be expressed in the following form:	
\begin{equation}
	\begin{aligned}
	&G_{\hat{q}}(\nu,\tau)
	\\&=e^{-\zeta_\nu\tau}_{}
	\Big\{-\frac{\hat{q}}{e^{\beta\zeta_\nu}_{}-1}
	+\frac{2(1-\hat{q})}{(1-2e^{2\beta\zeta_\nu}_{})}\hat{f}
	-1\Big\}
	=e^{-\zeta_\nu\tau}_{}
	\Big\{-\frac{\hat{q}}{e^{\beta\zeta_\nu}_{}-1}
	+\frac{2(1-\hat{q})(1-2e^{2\beta\zeta_\nu}_{}+\hat{q})}{(1-2e^{2\beta\zeta_\nu}_{})^2-1}-1\Big\}
	\\&=
	e^{-\zeta_\nu\tau}_{}
	\Big\{-\frac{\hat{q}}{e^{\beta\zeta_\nu}_{}-1}
	+2\frac{1-2e^{2\beta\zeta_\nu}_{}+\hat{q}-\hat{q}+2\hat{q}e^{2\beta\zeta_\nu}_{}-1}{(1-2e^{2\beta\zeta_\nu}_{})^2-1}-1\Big\}
	=
	e^{-\zeta_\nu\tau}_{}
	\Big\{-\frac{\hat{q}}{e^{\beta\zeta_\nu}_{}-1}
	+\frac{(-4e^{2\beta\zeta}_{})(1-\hat{q})}{1-4e^{2\beta\zeta}_{}+4e^{4\beta\zeta_\nu}_{}-1}
	-1\Big\}
	\\&=
	e^{-\zeta_\nu\tau}_{}
	\Big\{
	-\frac{\hat{q}}{e^{\beta\zeta_\nu}_{}-1}+
	\frac{(-4e^{2\beta\zeta_\nu}_{})(1-\hat{q})}{(-4e^{2\beta\zeta}_{})(1-2e^{2\beta\zeta_\nu}_{})}
	-1\Big\}=
	e^{-\zeta_\nu\tau}_{}
	\Big\{-\frac{\hat{q}}{e^{\beta\zeta_\nu}_{}-1}
	+\frac{(1-\hat{q})}{(1-2e^{2\beta\zeta_\nu}_{})}
	-1\Big\}.
	\end{aligned}
\end{equation}
We now demonstrate that the expression given in Eq.~\eqref{Eq:contr} is in fact equivalent to the Green's function defined in Eq.~\eqref{Eq:GF-M0}. Establishing this equivalence confirms that the two formulations describe the same underlying quantity, allowing us to write:
\begin{equation}
    \begin{aligned}
         G^{\mathcal{M}}_{\hat{q}}(\tau,0)=-\frac{\text{Tr}_{\hat{q}}\left(\hat{\rho}_{\hat{q}}\left[\alpha_a(\tau),\alpha_b\right]_{\hat{q}}\right)}{1-\hat{q}e^{\lambda_a\beta\zeta_a}},
    \end{aligned}
\end{equation}
% By choosing $\lambda_\alpha = -1$, we can write
Setting $\lambda_\alpha = -1$, the expression can be written as:
\begin{equation}\label{Eq:F01}
   \begin{aligned}
        G^{\mathcal{M}}_{\hat{q}}(\nu;\tau)=-\frac{\text{Tr}_{\hat{q}}\left(\hat{\rho}_{\hat{q}}\left[a_\nu(\tau),a^\dagger_\nu\right]_{\hat{q}}\right)}{1-\hat{q}e^{-\beta\zeta_\nu}}
        =-\frac{\text{Tr}_{\hat{q}}\left(\hat{\rho}_{\hat{q}}e^{-\tau\zeta_\nu}_{}\mathbf{1}^{\hat{q}}\right)}{1-\hat{q}e^{-\beta\zeta_\nu}}
        =-\frac{e^{-\tau\zeta_\nu}}{1-\hat{q}e^{-\beta\zeta_\nu}}=-e^{-\tau\zeta_\nu}\left(1+\hat{q}\frac{1}{e^{\beta\zeta_\nu}-\hat{q}}\right),
   \end{aligned}
\end{equation}
On the other hand, Eq.~\eqref{Eq:GF-M0} can be further simplified as follows:
\begin{equation}\label{Eq:F2}
	-e^{-\tau\zeta\nu } \left(1+\frac{\hat{q}}{e^{x}-1}+\frac{2
	(\hat{q}-1) \hat{q}}{2 e^{2 x}+\hat{q}-1}\right)
    =-e^{-\tau\zeta\nu } \left(1+\hat{q}\left(\frac{1}{e^{x}-1}+\frac{2
	(\hat{q}-1)}{2 e^{2 x}+\hat{q}-1}\right)\right),
\end{equation}
where $x=\beta\zeta_\nu$.
For Eq.~\eqref{Eq:F01} and Eq.~\eqref{Eq:F2} to be equivalent, the following equality must hold:
\begin{equation}
   \begin{aligned}
       \frac{1}{e^{x}-\hat{q}}=\frac{1}{e^{x}-1}+\frac{2
	(\hat{q}-1)}{2 e^{2 x}+\hat{q}-1},
   \end{aligned}
\end{equation}
Therefore, it can be easily shown that this equality holds.
\begin{equation}
   \begin{aligned}
   \left(e^{x}-1\right)\left(2 e^{2 x}+\hat{q}-1\right)&=\left(e^{x}-\hat{q}\right)\left(2 e^{2 x}+\hat{q}-1+(e^{x}-1)2(\hat{q}-1)\right)
   \\q e^x-q-e^x-2 e^{2 x}+2 e^{3 x}+1&=-2 q^2 e^x+q^2+q e^x-q+e^x-2 e^{2 x}+2 e^{3 x},
   \end{aligned}
\end{equation}
Using the property $\hat{q}^2 = 1$, which holds for the exchange-statistics operator in this context, the expression simplifies to:
\begin{equation}
   \begin{aligned}
    q e^x-q-e^x-2 e^{2 x}+2 e^{3 x}+1=q e^x-q-e^x-2 e^{2 x}+2 e^{3 x}+1.
    \end{aligned}
\end{equation}
Since both sides of the equation reduce to identical expressions, the equality is verified, thereby confirming the consistency of Eq.~\eqref{Eq:contr} with Eq.~\eqref{Eq:GF-M0} and completing the proof of their equivalence.
%================================================
\section{Equation of Motion approach}\label{GF-EOM}
An alternative route to the Green's function proceeds via the equation of motion. Starting from the definition
\begin{equation}
    G^{\mathcal{M}}_{\hat{q}}(\nu,\tau)=-\langle T_{\hat{q}}\psi^{}_\nu(\tau)\psi^\dagger_\nu(0)\rangle=
    -\langle \Theta(\tau)\psi^{}_\nu(\tau)\psi^\dagger_\nu(0)+\hat{q}\Theta(-\tau)\psi^\dagger_\nu(0)\psi^{}_\nu(\tau)\rangle,
\end{equation}
we differentiate with respect to $\tau$. Since the time derivative of the step function produces a delta function acting on the generalized commutator of the field operators, we obtain
\begin{equation}
    \begin{aligned}
        \frac{\partial G^{\mathcal{M}}_{\hat{q}}(\nu,\tau)}{\partial\tau}&=-\langle\delta(\tau)\left[\psi^{}_\nu(\tau),\psi^\dagger_\nu(0)\right]_{\hat{q}}\rangle
        -\langle T_{\hat{q}}\frac{\partial\psi^{}_\nu(\tau)}{\partial\tau}\psi^{\dagger}_\nu(0)\rangle
        \\&
        =-\delta(\tau)e^{-\zeta_\nu\tau}_{}-\left(\frac{\nabla^2_{}}{2m}+\mu\right)\langle T_{\hat{q}}\psi^{}_\nu(\tau)\psi^\dagger_\nu(0)\rangle
        % \\&
        =-\delta(\tau)e^{-\zeta_\nu\tau}_{}-\zeta_\nu G^{\mathcal{M}}_{\hat{q}}(\nu,\tau),
    \end{aligned}
\end{equation}
which can be rearranged into the equation of motion for the Green's function:
\begin{equation}
    \frac{\partial G^{\mathcal{M}}_{\hat{q}}(\nu,\tau)}{\partial\tau}+\zeta_\nu G^{\mathcal{M}}_{\hat{q}}(\nu,\tau)=-\delta(\tau)e^{-\zeta_\nu\tau}_{}.
    \label{Eq:EqofMo}
\end{equation}

To transform this equation into frequency space, we apply the time integral $\int_0^\beta d\tau\, e^{i\omega_n \tau}$. It is worth noting that the lower limit of integration may be taken as either $0^+$ or $0^-$, and, as shown below, both choices lead to the same final result. Performing the Fourier transform of Eq.~\eqref{Eq:EqofMo} gives
\begin{equation}
    \begin{aligned}
        \int_{0^\pm}^{\beta}e^{i\omega_n\tau}_{}\frac{\partial G^{\mathcal{M}}_{\hat{q}}(\nu,\tau)}{\partial\tau}d\tau+
        \int_{0^\pm}^{\beta}e^{i\omega_n\tau}_{}\zeta_\nu G^{\mathcal{M}}_{\hat{q}}(\nu,\tau)d\tau 
        =-\int_{0^\pm}^{\beta}e^{i\omega_n\tau}_{}\delta(\tau)e^{-\zeta_\nu\tau}_{}.
    \end{aligned}
\end{equation}
Integrating the first term by parts and evaluating the boundary contributions at $\tau=0^\pm$ and $\tau=\beta$, we find
\begin{equation}
    \begin{aligned}
        \left[e^{i\omega_n\beta}_{}G^{\mathcal{M}}_{\hat{q}}(\nu,\beta)-G^{\mathcal{M}}_{\hat{q}}(\nu,0^\pm)-\int_{0^\pm}^{\beta}e^{i\omega_n\tau}_{}(i\omega_n) G^{\mathcal{M}}_{\hat{q}}(\nu,\tau)d\tau\right]
        +\zeta_{\nu}^{}G^{\mathcal{M}}_{\hat{q}}(\nu,i\omega_n)=&\left(\begin{matrix}
            0\\
            -1
        \end{matrix}\right),
        \\
        -e^{i\omega_n\beta}_{}e^{\beta\zeta_\nu}_{}\langle\psi_\nu\psi_\nu^\dagger\rangle+\left(\begin{matrix}
            \langle\psi_\nu\psi_\nu^\dagger\rangle\\
            \langle\hat{q}\psi_\nu^\dagger\psi_\nu\rangle
        \end{matrix}\right)-(i\omega_n)G^{\mathcal{M}}_{\hat{q}}(\nu,i\omega_n)
        +\zeta_{\nu}^{}G^{\mathcal{M}}_{\hat{q}}(\nu,i\omega_n)=&\left(\begin{matrix}
            0\\
            -1
        \end{matrix}\right),
        \\
        -e^{\beta(i\omega_n-\zeta_\nu)}_{}(1+\hat{q}n_\nu)+\left(\begin{matrix}
            1+\hat{q}n_\nu\\
            \hat{q}n_\nu
        \end{matrix}\right)-(i\omega_n)G^{\mathcal{M}}_{\hat{q}}(\nu,i\omega_n)
        +\zeta_{\nu}^{}G^{\mathcal{M}}_{\hat{q}}(\nu,i\omega_n)=&\left(\begin{matrix}
            0\\
            -1
        \end{matrix}\right),
    \end{aligned}
\end{equation}
where, in each vector on the right-hand side, the upper entry corresponds to the choice $0^+$ and the lower entry to the choice $0^-$. Remarkably, both choices of the lower limit yield the identical algebraic equation:
\begin{equation}
    \begin{aligned}
        &\left(1+\hat{q}n_\nu\right)\left(1-e^{\beta(i\omega_n-\zeta_\nu)}_{}\right)=\left(i\omega_n-\zeta_\nu\right)G^{\mathcal{M}}_{\hat{q}}(\nu,i\omega_n),
    \end{aligned}
\end{equation}
which can be solved directly for the Green's function in frequency space:
\begin{equation}
    \boxed{G^{\mathcal{M}}_{\hat{q}}(\nu,i\omega_n)=\frac{\left(1+\hat{q}n_\nu\right)\left(1-e^{\beta(i\omega_n-\zeta_\nu)}_{}\right)}{\left(i\omega_n-\zeta_\nu\right)}}.
    \label{Eq:GreenIntegral}
\end{equation}
Let us examine the behavior of the Green's function for $\tau < 0$ in order to determine what form it takes as $\tau$ is shifted to $\tau + \beta$. For negative $\tau$, the Green's function is written as
\begin{equation}\label{f-tau}
     G^{\mathcal{M}}_{\hat{q}}(\nu,\tau)=-\langle T_{\hat{q}}^{}\psi_\nu(\tau)\psi_\nu^\dagger(0)\rangle=-\langle \hat{q}\psi_\nu^\dagger(0)\psi_\nu(\tau)\rangle,
\end{equation}
while for $\tau+\beta>0$ it takes the form
\begin{equation}
      G^{\mathcal{M}}_{\hat{q}}(\nu,\tau+\beta)=-\langle T_{\hat{q}}^{}\psi_\nu(\tau+\beta)\psi_\nu^\dagger(0)\rangle=-\langle \psi_\nu(\tau+\beta)\psi_\nu^\dagger(0)\rangle=-\frac{1}{\mathcal{Z}}\text{Tr}\left[e^{-\beta H}_{}\psi_\nu(\tau+\beta)\psi_\nu^\dagger(0)\right].
\end{equation}
To proceed, we make use of the identity
\begin{equation}
    e^{-\beta H}_{}\psi_\nu(\tau+\beta)=e^{-\beta H}_{}e^{(\tau+\beta) H}_{}\psi_\nu e^{-(\tau+\beta) H}_{}
    =e^{\tau H}_{}\psi_\nu e^{-\tau H}_{}e^{-\beta H}_{}=\psi_\nu(\tau)e^{-\beta H}_{},
\end{equation}
which allows the trace to be rewritten, and by the cyclic property of the trace we obtain
\begin{equation}\label{f-tau-beta}
     G^{\mathcal{M}}_{\hat{q}}(\nu,\tau+\beta)=-\frac{1}{\mathcal{Z}}\text{Tr}\left[\psi_\nu(\tau)e^{-\beta H}_{}\psi_\nu^\dagger(0)\right]
     =-\frac{1}{\mathcal{Z}}\text{Tr}\left[e^{-\beta H}_{}\psi_\nu^\dagger(0)\psi_\nu(\tau)\right]
     =-\langle\psi_\nu^\dagger(0)\psi_\nu(\tau)\rangle.
\end{equation}
Comparing Eqs.~\eqref{f-tau} and~\eqref{f-tau-beta}, we arrive at the generalized (anti-)periodicity relation
\begin{equation}
   \boxed{G^{\mathcal{M}}_{\hat{q}}(\nu,\tau+\beta)=\hat{q}G^{\mathcal{M}}_{\hat{q}}(\nu,\tau)}.
   \label{Eq:period}
\end{equation}
which reduces to the familiar periodic (bosonic) and antiperiodic (fermionic) boundary conditions in the limits $\hat{q}=1$ and $\hat{q}=-1$, respectively.
%=====================================================
\section{Analytic Continuation}\label{AppendixC}
Equation~\eqref{Eq:GF-main} provides the finite-temperature imaginary-time Green's function. It is instructive to examine the zero-temperature limit of this expression and to compare it with the corresponding real-time Green's function for $\tau>0$, which reduces to the simple exponential form
\begin{equation}
     G^{(0)}_{\delta}(\nu,\omega)= -i\int_0^\infty e^{i(\omega-\zeta_\nu+i0^+)t} \left(1+\vartheta n^{}_{T\to 0}(\epsilon_\nu)\right)dt=\frac{1+\vartheta n^{}_{T\to 0}(\epsilon_\nu)}{\omega-\zeta_\nu+i0^+}.
\end{equation}
In parallel, taking the zero-temperature limit of the finite-temperature (Matsubara) Green's function yields
\begin{equation}
	\begin{aligned}
		&G^{\mathcal{M}0}_{\delta}(\nu,i\omega_n)=\int_0^\beta d\tau e^{i\omega_n \tau } G(\nu,\tau),\\
        &\lim_{\beta\to \infty} G^{\mathcal{M}0}_{\delta}(\nu,i\omega^{(j)}_n)
        =\lim_{\beta\to \infty} \int_0^\beta d\tau\left[-e^{(i\omega_n^{(j)}-\zeta_\nu)  \tau } \left(1+\vartheta n^{}_{T\to 0}(\epsilon_\nu)\right)\right]
		=\lim_{\beta\to \infty}-\frac{1+\vartheta n^{}_{T\to 0}(\epsilon_\nu)}{i\omega_n^{(j)}-\zeta_\nu}\left(e^{(i\omega_n^{(j)}-\zeta_\nu)\beta}-1\right),
        \\
		&\lim_{\beta\to \infty} G^{\mathcal{M}0}_{\delta}(\nu,i\omega^{(j)}_n)\xlongequal{\zeta_\nu>0}\frac{1+\vartheta n^{}_{T\to 0}(\epsilon_\nu)}{i\omega_n^{(j)}-\zeta_\nu}.
	\end{aligned}
\end{equation}
In order to connect this Matsubara-frequency result to its real-time counterpart, an analytic continuation must be performed, replacing the discrete Matsubara frequency according to $i\omega_n^{(j)}\to i\omega+0^+$. Before carrying out this continuation, however, it is necessary to specify the precise form of the zero-temperature occupation function $n_{T\to 0}(\epsilon_\nu)$, which is piecewise-defined in terms of a general step-function representation:
\begin{equation}
    f(x,x_1,x_2)=f(x)\bigg(\Theta(x_1-x)+\Theta(x-x_2)\bigg),
\end{equation}
where the arguments are identified as $x=\zeta_\nu=(\epsilon_\nu-\mu)$, $x_1=\frac{1}{2\beta}\ln\delta$, and $x_2=0$. With this identification, the zero-temperature occupation function takes the explicit form
\begin{equation}
	\lim_{T\to 0}n^{}_{T\to 0}(\epsilon_\nu)=n(\zeta_\nu,T=0,\delta) = \begin{cases}
	1 &,~~(\epsilon_\nu-\mu)<\frac{1}{2\beta}\ln{\delta} \\
	0 &,~~ (\epsilon_\nu-\mu)>0\\
    \text{forbidden area} &,~~ \frac{1}{2\beta}\ln{\delta}<(\epsilon_\nu-\mu)<0.
	\end{cases}
\end{equation}
It can further be shown, by means of a Lehmann representation, that the analytic continuation $i\omega_n^{(j)}\to i\omega+0^+$ remains valid to all orders in perturbation theory, thereby justifying the identification of the Matsubara and real-time results derived above.

We now consider the strict zero-temperature limit, in which the inverse temperature satisfies $\beta\rightarrow\infty$. Under this condition, the two inequalities defining the allowed regions of the generalized statistics,
\begin{equation}
    \beta(\epsilon_\nu-\mu)>0,
\qquad \text{or} \qquad
\beta(\epsilon_\nu-\mu)<\frac{1}{2}\ln\delta,
\end{equation}
simplify considerably, since the term $\frac{1}{2\beta}\ln\delta$ vanishes as $\beta\to\infty$. The two conditions therefore reduce to the simple energy domains
\begin{equation}
\epsilon_\nu>\mu,
 \qquad \text{and} \qquad
\epsilon_\nu<\mu.
\end{equation}
Consequently, the intermediate "forbidden" region collapses, and the generalized occupation function acquires a discontinuous, step-like form: the occupation number equals unity for energies below the chemical potential and vanishes for energies above it,
\begin{equation}
    \boxed{
\langle n(\epsilon_\nu)\rangle_{T=0}
=
\Theta(\mu-\epsilon_\nu)
},
\end{equation}
where $\Theta(x)$ denotes the Heaviside step function,
\begin{equation}
    \Theta(x)=
\begin{cases}
1, & x>0,\\
0, & x<0.
\end{cases}
\end{equation}
Thus, despite the presence of generalized intermediate statistics at finite temperature, the system recovers an effective Fermi-like step distribution in the zero-temperature limit, independent of the deformation parameter $\delta$.
%=====================================================
\section{Density of States}\label{App:DOS}
According to Eq.~\eqref{Eq:DOS}, the generalized density of states is given by
\begin{equation}
		\mathcal{D}_\delta(\epsilon)=\frac{V}{(2\pi)^3}\int Z^\delta_\textbf{p}~\delta\left(\epsilon-\frac{\hbar^2\textbf{p}^2}{2m}\right) \text{d}\textbf{p}.
\end{equation}
To evaluate this integral explicitly, we make use of the quasiparticle weight expression of Eq.~\eqref{eq:Zp1} together with the zero-temperature distribution function
\begin{equation}
    \lim_{T\to 0}n^{}_{T\to 0}(\epsilon)=n(\zeta_\textbf{p},T=0,\delta) = \begin{cases}
	1 &,~~\epsilon<\mu+\frac{1}{2\beta}\ln{\delta} \\
	0 &,~~ \epsilon>\mu\\
    \text{forbidden area} &,~~ \mu+\frac{\ln{\delta}}{2\beta}<\epsilon<\mu.
	\end{cases}
\end{equation}
Identifying the single-particle energy as $\epsilon\equiv \hbar^2 \textbf{p}^2/2m$, the two conditions defining the allowed regions translate into corresponding bounds on the momentum $\textbf{p}$:
\begin{equation}
    \begin{aligned}
    \begin{cases}
    \epsilon<\mu+\dfrac{1}{2\beta}\ln{\delta} &\to~~~ \textbf{p}<\sqrt{\dfrac{2m\mu}{\hbar^2}+\dfrac{m}{\hbar^2\beta}\ln{\delta}},
    \\
    \epsilon>\mu &\to~~~ \textbf{p}>\sqrt{\dfrac{2m\mu}{\hbar^2}}.
    \end{cases}
    \end{aligned}
\end{equation}
Consequently, the momentum-space integral for $\mathcal{D}_\delta(\epsilon)$ splits naturally into two separate contributions, one from each allowed region, so that we may write
\begin{equation}
    \begin{aligned}
    \mathcal{D}_\delta(\epsilon)=\frac{V}{(2\pi)^3}\int_{-\infty}^{\sqrt{\frac{2m\mu}{\hbar^2}+\frac{m}{\hbar^2\beta}\ln{\delta}}} Z^\delta_\textbf{p}~\delta\left(\epsilon-\frac{\hbar^2\textbf{p}^2}{2m}\right) \text{d}\textbf{p}
    +
    \frac{V}{(2\pi)^3}\int^{+\infty}_{\sqrt{\frac{2m\mu}{\hbar^2}}} Z^\delta_\textbf{p}~\delta\left(\epsilon-\frac{\hbar^2\textbf{p}^2}{2m}\right) \text{d}\textbf{p}=\text{I}+\text{II}.
    \end{aligned}
\end{equation}
We now evaluate each of these two contributions in turn. The first term, associated with the region below the effective Fermi momentum, is
\begin{equation}
    \begin{aligned}
    \text{I}=&\frac{V}{(2\pi)^3}\int_{-\infty}^{\sqrt{\frac{2m\mu}{\hbar^2}+\frac{m}{\hbar^2\beta}\ln{\delta}}} Z^\delta_\textbf{p}~\delta\left(\epsilon-\frac{\hbar^2\textbf{p}^2}{2m}\right) \text{d}\textbf{p}
    =
    4\pi\frac{V}{(2\pi)^3}(2-2\delta)\int_{-\infty}^{\sqrt{\frac{2m\mu}{\hbar^2}+\frac{m}{\hbar^2\beta}\ln{\delta}}} \textbf{p}^{2}_{}~\delta\left(\epsilon-\frac{\hbar^2\textbf{p}^2}{2m}\right) d\textbf{p}.
    \end{aligned}
\end{equation}
Introducing the substitution $u=\hbar^2\textbf{p}^2/2m$ to change variables from momentum to energy, this expression can be rewritten as
\begin{equation}
    \begin{aligned}
      \text{I}&=\frac{V}{2\pi^2}(2-2\delta)\int_{-\infty}^{\mu-\frac{\ln\delta}{2\beta}} (\frac{2mu}{\hbar^2})~\delta\left(\epsilon-u\right)\left(\frac{1}{2}(\frac{2m}{\hbar^2})^{1/2}_{}u^{-1/2}_{}\right)du,
    \end{aligned}
\end{equation}
and, carrying out the integral over the delta function, reduces to
\begin{equation}
    \begin{aligned}
      \text{I}=
      \frac{V}{2\pi^2}(2-2\delta)\left(\frac{1}{2}(\frac{2m}{\hbar^2})^{3/2}_{}\right)
      \int_{-\infty}^{\mu-\frac{\ln\delta}{2\beta}} \delta\left(\epsilon-u\right)u^{1/2}_{}du
      =\frac{V}{2\pi^2}(2-2\delta)\left(\frac{1}{2}(\frac{2m}{\hbar^2})^{3/2}_{}\right)\epsilon^{1/2}.
    \end{aligned}
\end{equation}
Following an analogous procedure for the second region, the result of the second term is obtained as
\begin{equation}
    \begin{aligned}
        \text{II}=
      \frac{V}{2\pi^2}\left(\frac{1}{2}(\frac{2m}{\hbar^2})^{3/2}_{}\right)
      \int_{-\infty}^{\mu-\frac{\ln\delta}{2\beta}} \delta\left(\epsilon-u\right)u^{1/2}_{}du
      =\frac{V}{2\pi^2}\left(\frac{1}{2}(\frac{2m}{\hbar^2})^{3/2}_{}\right)\epsilon^{1/2}.
    \end{aligned}
\end{equation}
Adding the two contributions together ($\text{I}+\text{II}$), the generalized density of states takes the compact closed form
\begin{equation}
    \boxed{\mathcal{D}_\delta(\epsilon)=
    \frac{V}{4\pi^2}(3-2\delta)\left(\frac{2m}{\hbar^2}\right)^{3/2}_{}\epsilon^{1/2}.}
\end{equation}
It is instructive to compare this result with the corresponding free-fermion density of states, 
% $\mathcal{D}_\text{F}(\epsilon)=\frac{V}{4\pi^2}(\frac{2m}{\hbar^2})^{3/2}\epsilon^{1/2}$
$\mathcal{D}_\text{F}(\epsilon)=(V/4\pi^2)(2m/\hbar^2)^{3/2}\epsilon^{1/2}$
.
Written in terms of $\mathcal{D}_\text{F}(\epsilon)$, the generalized result becomes
\begin{equation}
    \begin{aligned}
    \mathcal{D}_\delta(\epsilon)=&
    \frac{V}{4\pi^2}(3-2\delta)\left(\frac{2m}{\hbar^2}\right)^{3/2}_{}\epsilon^{1/2},
    \\
    \frac{\mathcal{D}_\delta(\epsilon)}{\mathcal{D}_{\text{F}}(\epsilon)}=&(3-2\delta),
    \end{aligned}
\end{equation}
so that the generalized density of states is simply a constant multiple of the free-fermion result, with the deformation parameter $\delta$ controlling the overall enhancement or suppression factor $(3-2\delta)$.
%===================================================@@@@@@@@@@@@@@@@@@@@@@@@@@@@@@@@@@@@@@@@@@@@@@@@@@@
\section{FEYNMAN DIAGRAMS AND PERTURBATION THEORY}\label{FeynmanD}
An illustrative example that makes the diagrammatic rules developed above concrete is the perturbative expansion of the exact Green's function, which can be formulated as
\begin{equation}
	\begin{aligned}
	G^{\mathcal{M}}_{\hat{q}}(x,y) &=
	\sum_{m=0}^{\infty} \frac{(-i)^m}{m!} \int_0^{\beta} d\tau_1 \cdots \int_0^{\beta} d\tau_m 
	\left\langle T_{\hat{q}}\left[\hat{H}_{\text{int}}(\tau_1) \cdots \hat{H}_{\text{int}}(\tau_m) \hat{\psi}_{}(x) \hat{\psi}^{\dagger}_{}(y)\right]\right\rangle.
	\end{aligned}
\end{equation}
Since the interaction is instantaneous, it is convenient to express the interparticle potential appearing in $\hat{H}_{\rm int}$ as $u(x_1,x_2) \equiv v(\mathbf{x}_1,\mathbf{x}_2)\delta(\tau_1-\tau_2)$; this notation allows the time integrations to be written in a symmetric form. Restricting attention to the first-order term in this expansion, we may write
\begin{equation}
	\tilde{G}^{\mathcal{M}}_{\hat{q}}(x, y) \equiv
	G^{\mathcal{M}}_{\hat{q}}(x, y) + \tilde{G}^{\mathcal{M}}_{(1)\hat{q}}(x, y),
\end{equation}
where the zeroth-order (free) and first-order contributions are given, respectively, by
\begin{equation}
	\begin{aligned}
		G^{\mathcal{M}}_{\hat{q}}(x, y) &=
		\langle T_{\hat{q}}[\hat{\psi}(x)\hat{\psi}^{\dagger}(y)] \rangle,
		\\
		\tilde{G}_{(1)\hat{q}}^{\mathcal{M}}(x, y) &=
		(-i)\frac{1}{2}\int dx_1 dx_1' u(x_1, x_1') 
		\langle \mathcal{T}_{\hat{q}}[\hat{\psi}^{\dagger}(x_1)\hat{\psi}^{\dagger}(x_1')\hat{\psi}(x_1')\hat{\psi}^{}(x_1)\hat{\psi}(x)\hat{\psi}^{\dagger}(y)] \rangle.
	\end{aligned}
\end{equation}
To evaluate the first-order term $\tilde{G}_{(1)\hat{q}}^{\mathcal{M}}(x, y)$, we must apply the generalized Wick's theorem to the six-operator expectation value appearing above. Enumerating all possible ways of pairing the six field operators, and keeping track of the number of operator interchanges required for each pairing, we obtain
\begin{equation}
	\begin{aligned}
	&\left\langle \mathcal{T}_{\hat{q}}\{\hat{\psi}^\dagger(x_1)\hat{\psi}^\dagger(x_1')\hat{\psi}(x_1')\hat{\psi}(x_1)\hat{\psi}(x)\hat{\psi}^\dagger(y))\}\right\rangle 
	\\&~~~~~~=\left\langle(\hat{q})^2 
	\overbracket{\hat{\psi}(x_1')\hat{\psi}^{\dagger}(x_1')}\overbracket{\hat{\psi}(x_1)\hat{\psi}^{\dagger}(x_1)}\overbracket{\hat{\psi}(x)\hat{\psi}^{\dagger}(y)}\right\rangle
	+\left\langle(\hat{q})^3 
	\overbracket{\hat{\psi}(x_1')\hat{\psi}^{\dagger}(x_1)}\overbracket{\hat{\psi}(x_1)\hat{\psi}^{\dagger}(x_1')}\overbracket{\hat{\psi}(x)\hat{\psi}^{\dagger}(y)}\right\rangle
	\\&~~~~~~+\left\langle(\hat{q})^6
	\overbracket{\hat{\psi}(x)\hat{\psi}^\dagger(x_1)}\overbracket{\hat{\psi}(x_1)\hat{\psi}^{\dagger}(x_1')}\overbracket{\hat{\psi}(x_1')\hat{\psi}^{\dagger}(y)}\right\rangle
	+\left\langle(\hat{q})^5
	\overbracket{\hat{\psi}(x)\hat{\psi}^\dagger(x_1)}\overbracket{\hat{\psi}(x_1)\hat{\psi}^{\dagger}(y)}\overbracket{\hat{\psi}(x_1')\hat{\psi}^{\dagger}(x_1')}\right\rangle
	\\&~~~~~~+\left\langle(\hat{q})^4
	\overbracket{\hat{\psi}(x)\hat{\psi}^\dagger(x_1')}\overbracket{\hat{\psi}(x_1')\hat{\psi}^{\dagger}(x_1)}\overbracket{\hat{\psi}^\dagger(x_1)\hat{\psi}^{\dagger}(y)}\right\rangle
	+\left\langle(\hat{q})^5 \overbracket{\hat{\psi}(x)\hat{\psi}^\dagger(x_1')}\overbracket{\hat{\psi}(x_1')\hat{\psi}^{\dagger}(y)}\overbracket{\hat{\psi}(x_1)\hat{\psi}^{\dagger}(x_1)}\right\rangle.
	\end{aligned}
\end{equation}
Making use of the property $\hat{q}^{2n}=1$ and $\hat{q}^{2n+1}=\hat{q}$, each power of $\hat{q}$ above reduces to either the identity or a single factor of $\hat{q}$, so that the expression simplifies to
\begin{equation}\label{GFDF}
    \begin{aligned}
	&\left\langle T_{\hat{q}}\{\hat{\psi}^\dagger(x_1)\hat{\psi}^\dagger(x_1')\hat{\psi}(x_1')\hat{\psi}(x_1)\hat{\psi}(x)\hat{\psi}^\dagger(y))\}\right\rangle 
	\\&~~~~~~=
	\left\langle\overbracket{\hat{\psi}(x_1')\hat{\psi}^{\dagger}(x_1')}\overbracket{\hat{\psi}(x_1)\hat{\psi}^{\dagger}(x_1)}\overbracket{\hat{\psi}(x)\hat{\psi}^{\dagger}(y)}\right\rangle
	+\left\langle(\hat{q})\overbracket{\hat{\psi}(x_1')\hat{\psi}^{\dagger}(x_1)}\overbracket{\hat{\psi}(x_1)\hat{\psi}^{\dagger}(x_1')}\overbracket{\hat{\psi}(x)\hat{\psi}^{\dagger}(y)}\right\rangle
	\\&~~~~~~+
	\left\langle\overbracket{\hat{\psi}(x)\hat{\psi}^\dagger(x_1)}\overbracket{\hat{\psi}(x_1)\hat{\psi}^{\dagger}(x_1')}\overbracket{\hat{\psi}(x_1')\hat{\psi}^{\dagger}(y)}\right\rangle
	+\left\langle(\hat{q})
	\overbracket{\hat{\psi}(x)\hat{\psi}^\dagger(x_1)}\overbracket{\hat{\psi}(x_1)\hat{\psi}^{\dagger}(y)}\overbracket{\hat{\psi}(x_1')\hat{\psi}^{\dagger}(x_1')}\right\rangle
	\\&~~~~~~+\left\langle
	\overbracket{\hat{\psi}(x)\hat{\psi}^\dagger(x_1')}\overbracket{\hat{\psi}(x_1')\hat{\psi}^{\dagger}(x_1)}\overbracket{\hat{\psi}^\dagger(x_1)\hat{\psi}^{\dagger}(y)}\right\rangle
	+\left\langle(\hat{q}) 
	\overbracket{\hat{\psi}(x)\hat{\psi}^\dagger(x_1')}\overbracket{\hat{\psi}(x_1')\hat{\psi}^{\dagger}(y)}\overbracket{\hat{\psi}(x_1)\hat{\psi}^{\dagger}(x_1)}\right\rangle.    
	\end{aligned}
\end{equation}
Each of the contracted pairs $\overbracket{\hat{\psi}(x_i)\hat{\psi}^\dagger(x_j)}$ appearing in Eq.~\eqref{GFDF} is, by definition, directly proportional to the free single-particle Green's function connecting the corresponding coordinates. For practical calculations, it is therefore convenient to convert these abstract operator expressions into their final numerical form: this is done by replacing each contraction with the corresponding free Green's function $G^{\mathcal{M}}_{\delta}$, and by replacing the operator $\hat{q}$ with its expectation value $\vartheta=(1-2\delta)$. Applying this substitution to every term in Eq.~\eqref{GFDF} yields
\begin{equation}\label{Eq:FeynmanD}
	\begin{aligned}
	&\left\langle \mathcal{T}_{\hat{q}}
	\left\{\hat{\psi}^{\dagger}(x_1)\hat{\psi}^\dagger(x_1')\hat{\psi}(x_1')\hat{\psi}^\dagger(x_1)\hat{\psi}(x)\hat{\psi}^\dagger(y)\right\}\right\rangle 
	\\&~~~~~~=
	\underbrace{G^{\mathcal{M}}_{\delta}(x_1', x_1')G^{\mathcal{M}}_{\delta}(x_1, x_1)G^{\mathcal{M}}_{\delta}(x, y)}_{\text{(a)}} 
	+\underbrace{\vartheta G^{\mathcal{M}}_{\delta}(x_1', x_1)G^{\mathcal{M}}_{\delta}(x_1, x_1')G^{\mathcal{M}}_{\delta}(x, y)}_{\text{(b)}} 
	\\&~~~~~~+
	\underbrace{G^{\mathcal{M}}_{\delta}(x, x_1)G^{\mathcal{M}}_{\delta}(x_1, x_1')G^{\mathcal{M}}_{\delta}(x_1',y)}_{\text{(c)}} 
	+\underbrace{\vartheta G^{\mathcal{M}}_{\delta}(x, x_1)G^{\mathcal{M}}_{\delta}(x_1, y)G^{\mathcal{M}}_{\delta}(x_1', x_1')}_{\text{(d)}} 
	\\&~~~~~~+
	\underbrace{G^{\mathcal{M}}_{\delta}(x, x_1')G^{\mathcal{M}}_{\delta}(x_1', x_1)G^{\mathcal{M}}_{\delta}(x_1, y)}_{\text{(e)}} 
	+\underbrace{\vartheta G^{\mathcal{M}}_{\delta}(x, x_1')G^{\mathcal{M}}_{\delta}(x_1', y)G^{\mathcal{M}}_{\delta}(x_1, x_1)}_{\text{(f)}}.
	\end{aligned}
\end{equation}
Each of the six terms labeled (a)-(f) corresponds to a distinct topologically inequivalent contraction pattern, and hence to a distinct Feynman diagram at first order; the terms carrying an explicit factor of $\vartheta$ are precisely those in which the generalized exchange statistics introduces a nontrivial modification relative to the terms without such a factor. Equivalently, this same first-order contribution can be represented and organized diagrammatically, as shown below
 %%%%%%%%%%%%%%%%%%%%%%%%%%%%%%%%%%%%%%%%%%%%%%%%
		  \\
		  % \newpage
		      \begin{tikzpicture}
		          \begin{feynman}
		              \vertex (a) {$x$};
		              %\node [draw, circle, fill=black, inner sep=1pt, label=below:{$x$}] (a) {};
		              \vertex [right=of a] (b);
		              \vertex [right=of b] (c);
		              \vertex [right=of c] (d) {$y$};
		              %\node [draw, circle, fill=black, inner sep=1pt, label=below:{$y$}] at(d) {};
		              \vertex [above right=of a] (e) ;
		              \vertex [above right=of a] (ee) {$x_1~~~~$};
		              \vertex [above left=of d] (f) ;
		              \vertex [above left=of d] (ff) {$~~~~~x'_1$};
		              \vertex [left=of e] (g);
		              \vertex [right=of f] (h);
		
		              \diagram{
		              (a) -- [fermion] (d);
		              (e) -- [boson] (f);
		              (e) -- [fermion, half right] (g);
		              (g) -- [half right] (e);
		              (f) -- [fermion, half left] (h);
		              (h) -- [half left] (f);
		              };
		          \end{feynman}
		          \node [below] at (current bounding box.south) {(A)};
		      \end{tikzpicture}~~~~~~
		      \begin{tikzpicture}
		          \begin{feynman}
		              \vertex (a) {$x$};
		              \vertex [right=of a] (b);
		              \vertex [right=of b] (c);
		              \vertex [right=of c] (d) {$y$};
		              \vertex [above right=of a] (e) ;
		              \vertex [above right=of a] (ee) {$x_1~~~~$};
		              \vertex [above left=of d] (f) ;
		              \vertex [above left=of d] (ff) {$~~~~~x'_1$};
		              \vertex [left=of e] (g);
		              \vertex [right=of f] (h);
		
		              \diagram{
		              (a) -- [fermion] (d);
		              (e) -- [boson] (f);
		              (e) -- [fermion, quarter right] (f);
		              (f) -- [fermion, quarter right] (e);
		              };
		          \end{feynman}
		          \node [below] at (current bounding box.south) {(B)};
		      \end{tikzpicture}
		      ~~~~~~
		      \begin{tikzpicture}
		          \begin{feynman}
		              \vertex (a) {$x$};
		              \vertex [right=of a] (b);
		              \vertex [right=of b] (c);
		              \vertex [right=of c] (d) {$y$};
		              \vertex [above right=of a] (e) ;
		              \vertex [above right=of a] (ee) {$x_1~~~~$};
		              \vertex [above left=of d] (f) ;
		              \vertex [above left=of d] (ff) {$~~~~~x'_1$};
		              \vertex [left=of e] (g);
		              \vertex [right=of f] (h);
		
		              \diagram{
		              (a) -- [fermion] (f);
		              (e) -- [fermion] (d);
		              (e) -- [boson] (f);
		              (f) -- [fermion, quarter right] (e);
		              };
		          \end{feynman}
		          \node [below] at (current bounding box.south) {(C)};
		      \end{tikzpicture}
		      \\
		
		      \begin{tikzpicture}
		          \begin{feynman}
		              \vertex (a) {$x$};
		              \vertex [right=of a] (b);
		              \vertex [right=of b] (c);
		              \vertex [right=of c] (d) {$y$};
		              \vertex [above right=of a] (e) ;
		              \vertex [above right=of a] (ee) {$x_1~~~~~$};
		              \vertex [above left=of d] (f) ;
		              \vertex [above left=of d] (ff) {$~~~~~x'_1$};
		              \vertex [left=of e] (g);
		              \vertex [right=of f] (h);
		
		              \diagram{
		              (a) -- [fermion] (e);
		              (e) -- [fermion] (d);
		              (e) -- [boson] (f);
		              (f) -- [fermion, half left] (h);
		              (h) -- [half left] (f);
		              };
		          \end{feynman}
		          \node [below] at (current bounding box.south) {(D)};
		      \end{tikzpicture}~~~~~~
		      \begin{tikzpicture}
		          \begin{feynman}
		              \vertex (a) {$x$};
		              \vertex [right=of a] (b);
		              \vertex [right=of b] (c);
		              \vertex [right=of c] (d) {$y$};
		              \vertex [above right=of a] (e) ;
		              \vertex [above right=of a] (ee) {$x_1~~~~~$};
		              \vertex [above left=of d] (f) ;
		              \vertex [above left=of d] (ff) {$~~~~~~x'_1$};
		              \vertex [left=of e] (g);
		              \vertex [right=of f] (h);
		
		              \diagram{
		              (a) -- [fermion] (e);
		              (f) -- [fermion] (d);
		              (e) -- [boson] (f);
		              (e) -- [fermion, quarter right] (f);
		              };
		          \end{feynman}
		          \node [below] at (current bounding box.south) {(E)};
		      \end{tikzpicture}~~~~~~
		      \begin{tikzpicture}
		          \begin{feynman}
		              \vertex (a) {$x$};
		              \vertex [right=of a] (b);
		              \vertex [right=of b] (c);
		              \vertex [right=of c] (d) {$y$};
		              \vertex [above right=of a] (e) ;
		              \vertex [above right=of a] (ee) {$x_1~~~~$};
		              \vertex [above left=of d] (f) ;
		              \vertex [above left=of d] (ff) {$~~~~~~~x'_1$};
		              \vertex [left=of e] (g);
		              \vertex [right=of f] (h);
		
		              \diagram{
		              (a) -- [fermion] (f);
		              (f) -- [fermion] (d);
		              (e) -- [boson] (f);
		              (e) -- [fermion, half right] (g);
		              (g) -- [half right] (e);
		              };
		          \end{feynman}
		          \node [below] at (current bounding box.south) {(F)};
		      \end{tikzpicture}
%===================================================
\section{Average energy}\label{App:Average energy}
The average interaction energy in the system is expressed as
\begin{equation}\label{App:Eq:EnergyPotantiol1}
	\begin{aligned}
		&\langle\hat{H}_{\text{int}}\rangle =\frac{1}{2}\int \text{d}\textbf{x}\text{d}\textbf{x}'v(\textbf{x},\textbf{x}')\langle\hat{\psi}^\dagger(x)\hat{\psi}^\dagger(x')\hat{\psi}(x')\hat{\psi}(x)\rangle_{\hat{q}}.\\
	\end{aligned}
\end{equation}
In order to evaluate this expression, we first bring the four-operator expectation value into a more convenient ordering by making use of the generalized exchange relation
\begin{equation}\label{App:Eq:EnergyPotantiol2}
    \langle\hat{\psi}^\dagger(x)\hat{\psi}^\dagger(x')\hat{\psi}(x')\hat{\psi}(x)\rangle_{\hat{q}}=\langle\hat{q}\rangle\langle\hat{\psi}^\dagger(x)\hat{\psi}^\dagger(x')\hat{\psi}(x)\hat{\psi}(x')\rangle,
\end{equation}
which allows the interaction energy to be rewritten as
\begin{equation}\label{App:Eq:EnergyPotantiol3}
    \begin{aligned}
		\langle\hat{H}_{\text{int}}\rangle &= \frac{\langle\hat{q}\rangle}{2}\int \text{d}\textbf{x}\text{d}\textbf{x}'v(\textbf{x},\textbf{x}')
		\left\langle\hat{\psi}^\dagger(x)\Bigg( \hat{q}\hat{\psi}(x)\hat{\psi}^\dagger(x')-\frac{\hat{q}}{V}\sum_{p^{}_1}e^{ip^{}_1(\textbf{x}-\textbf{x}')}\mathbf{1}^{\hat{q}}\Bigg)\hat{\psi}(x')\right\rangle
		\\&=\frac{1}{2}\int d\textbf{x} d\textbf{x}'v(\textbf{x},\textbf{x}')
    	\left[\langle\hat{n}(x)\hat{n}(x')\rangle-\frac{1}{V}\sum_{\textbf{p}}e^{i\textbf{p}.(\textbf{x}-\textbf{x}')}\langle\hat{\psi}^\dagger(x)\mathbf{1}^{\hat{q}}\hat{\psi}(x')\rangle\right].
	\end{aligned}
\end{equation} 
In order to isolate the physically relevant density fluctuations from the trivial mean-density contribution, we introduce the fluctuation of the density operator about its expectation value,
\begin{equation}
		\tilde{n}(\textbf{x})\equiv \hat{n}(\textbf{x})-\langle\hat{n}(\textbf{x})\rangle,
\end{equation} 
together with the auxiliary quantity
\begin{equation}
	\varLambda^{(\hat{q})}(\textbf{x})\equiv \frac{1}{V}\sum_{\textbf{p}}e^{i\textbf{p}.(\textbf{x})}\langle\hat{\psi}^\dagger(\textbf{x})\mathbf{1}^{\hat{q}}\hat{\psi}(\textbf{x}')\rangle.
\end{equation}
Substituting these definitions back into Eq.~\eqref{App:Eq:EnergyPotantiol3}, the interaction energy takes the form
\begin{equation}
	\begin{aligned}
	\langle\hat{H}_{\text{int}}\rangle&=\frac{1}{2}\int d\textbf{x} d\textbf{x}'v(\textbf{x}-\textbf{x}')
	\Big[\langle\tilde{n}(\textbf{x})\tilde{n}(\textbf{x}')\rangle+\langle\hat{n}(\textbf{x})\rangle\langle\hat{n}(\textbf{x}')\rangle-\varLambda^{(\hat{q})}(\textbf{x}-\textbf{x}')\Big].
\end{aligned}
\end{equation}
The physically nontrivial contribution to this expression is the density correlation function $\langle\tilde{n}(x)\tilde{n}(x')\rangle$; we therefore focus on this term and introduce its time-ordered generalization, known as the polarization function,
\begin{equation}
    D_{\hat{q}}(x,x')=\langle T_{\hat{q}}[\tilde{n}^{}_H(x)\tilde{n}^{}_H(x')]\rangle.
\end{equation}
In terms of this polarization function, the average interaction energy becomes
\begin{equation}
	\begin{aligned}
		\langle\hat{H}_{\text{int}}\rangle
		=&\frac{1}{2}\int d\textbf{x} d\textbf{x}'v(\textbf{x}-\textbf{x}')
		\Big[D_{\hat{q}}(\textbf{x},\textbf{x}')+n^2
		-\varLambda^{(\hat{q})}(\textbf{x}-\textbf{x}')\Big]. 
    \end{aligned}
\end{equation}
To separate this energy into a first-order (mean-field) piece and a correlation correction, we denote the polarization function of the corresponding noninteracting system by $D^0_{\hat{q}}(x',x)$, and add and subtract this quantity in the expression above:
\begin{equation}
	\begin{aligned}
	\langle\hat{H}_{\text{int}}\rangle=&\frac{1}{2}\int d\textbf{x} d\textbf{x}'v(\textbf{x}-\textbf{x}')
	\Big[D^0_{\hat{q}}(\textbf{x},\textbf{x}')+n^2_{}
	-\varLambda^{(\hat{q})}(\textbf{x}-\textbf{x}')\Big]
	+\frac{1}{2}\int d\textbf{x} d\textbf{x}'v(\textbf{x}-\textbf{x}')\Big[\Delta D_{\hat{q}}(\textbf{x},\textbf{x}')\Big],
	\end{aligned}
\end{equation}
where $\Delta D_{\hat{q}}(\textbf{x},\textbf{x}')= D_{\hat{q}}(\textbf{x},\textbf{x}')-D^0_{\hat{q}}(\textbf{x},\textbf{x}')$ denotes the deviation of the interacting polarization function from its noninteracting counterpart. Identifying the first bracket as the Hartree-Fock contribution evaluated in the noninteracting ground state, $\langle\hat{H}_{\text{int}}\rangle_0 = \langle\Phi_0|\hat{H}_{\text{int}}|\Phi_0\rangle$, we may write
\begin{equation}
	\begin{aligned}
	\langle\hat{H}_{\text{int}}\rangle
	=\langle\hat{H}_{\text{int}}\rangle_0
	+\frac{1}{2}\int d\textbf{x} d\textbf{x}'v(\textbf{x}-\textbf{x}')\Big[\Delta D_{\hat{q}}(\textbf{x},\textbf{x}')\Big],
\end{aligned}
\end{equation}
or, more compactly,
\begin{equation}
	\langle\hat{H}_{\text{int}}\rangle = \langle\hat{H}_{\text{int}}\rangle_0 + E^{}_{\text{corr}}.
\end{equation}
The second term thus defines the correlation energy $E_{\text{corr}}$, which, upon introducing a coupling-constant integration and passing to momentum space, can be written as
\begin{equation}
	\begin{aligned}
		E_{\text{corr}} &=\frac{1}{2}\int^1_0 \frac{d\lambda}{\lambda}\int d\textbf{x}\, d\textbf{x}'\,\lambda v(\textbf{x}-\textbf{x}')\Big[\Delta D_{\hat{q}}(\textbf{x},\textbf{x}')\Big]
		=\frac{V}{2}(2\pi)^{-4}\int_0^1 \frac{d\lambda}{\lambda}\int d^4 Q\, \lambda v(Q)\Big[\Delta D_{\hat{q}}(\textbf{Q},\omega)\Big].
		\end{aligned}
\end{equation}
It remains to evaluate the noninteracting polarization function $D^0_{\hat{q}}(x,x')=\langle T_{\hat{q}}[\tilde{n}^{}(x)\tilde{n}^{}(x')]\rangle_0$ explicitly. This is accomplished by means of the generalized Wick's theorem, which allows this correlator to be reduced to a product of single-particle Green's functions:
\begin{equation}
	\begin{aligned}
		D^0_{\hat{q}}(x,x')&=\left\langle T_{\hat{q}}\big[\hat{n}(x')\hat{n}(x)\big]\right\rangle_0
		-\left\langle\hat{n}(x')\right\rangle_0\left\langle\hat{n}(x)\right\rangle_0
		=\left\langle T_{\hat{q}}\big[\hat{\psi}^\dagger(x)\hat{\psi}(x)\hat{\psi}^\dagger(x')\hat{\psi}(x')\big]\right\rangle_0 - n^{2}.
	\end{aligned}
\end{equation}
Performing the Wick contractions and discarding the disconnected term, which exactly cancels the subtracted $n^2$ contribution, leads to the final result
\begin{equation}
	\begin{aligned}
		D^{0}_{\hat{q}}(x,x')&= G_{\hat{q}}^{\mathcal{M}0}(x,x)G_{\hat{q}}^{\mathcal{M}0}(x',x')+\hat{q} G_{\hat{q}}^{\mathcal{M}0}(x,x')G_{\hat{q}}^{\mathcal{M}0}(x',x)-n^2
		=\hat{q} G_{\hat{q}}^{\mathcal{M}0}(x,x')G_{\hat{q}}^{\mathcal{M}0}(x',x),
	\end{aligned}
\end{equation}
where $G_{\hat{q}}^{\mathcal{M}0}$ denotes the free single-particle Green's function.
%================================
\section{Calculation of polarization function}\label{polarization_function}
Here we show that two contributions of $\Pi^0$ become zero for cases ($\beta\to\infty,\zeta_p<0,\zeta_{p+Q}<0$) and ($\beta\to\infty,\zeta_p>0,\zeta_{p+Q}>0$).
In what follows, we show explicitly why the first of these contributions vanishes
\begin{equation}\label{Eq:polarization functionR1}
	\begin{split}
		\left.\Pi^0_{\delta}(Q,\omega)\right|_{\zeta_p,\zeta_{p+Q}<0}
		=&-\frac{\mathcal{B}(\delta)}{(2\pi)^3} \int d^3p\ \Theta\left(p_F^{(\delta)}-|p|\right)\Theta\left(p_F^{(\delta)}-|Q+p|\right) \left[\frac{1}{\zeta_{p+Q}-\zeta_p-\omega-i\eta}+\frac{1}{\zeta_{p+Q}-\zeta_p+\omega+i\eta}\right],\\
		=&-\frac{\mathcal{B}(\delta)}{(2\pi)^3}\int d^3p\ \Theta\left(p_F^{(\delta)}-|p|\right)\Theta\left(p_F^{(\delta)}-|Q+p|\right)\frac{2\zeta_{pQ}}{(i\omega^{(j)}_n)^2_{}-(\zeta_{pQ})^2}.
	\end{split}
	\end{equation}
The above equation vanishes identically, because the product of step functions is even under the interchange $k\leftrightarrows k+q$, while $\zeta_{pQ}$ is odd. 

 The remaining two contributions are associated with the cases ($\beta\to\infty,\zeta_p<0,\zeta_{p+Q}>0$) and ($\beta\to\infty,\zeta_p>0,\zeta_{p+Q}<0$), which give non-zero contributions.
The detailed calculations for these two cases are provided below in the same order.
For the case ($\beta\to\infty,\zeta_p<0,\zeta_{p+Q}>0$), we obtain:
\begin{equation}\label{Eq:fistcaseDF}
	\begin{aligned}
		\Pi^0_{\delta}(Q&,\omega)
		=\frac{\mathcal{B}(\delta)}{(2\pi)^3}\int d^3p\Theta\left(|Q+p|-p_F^{(\delta)}\right)\Theta\left(p_F^{(\delta)}-|p|\right)
	 \left[\frac{1}{\zeta_{p+Q}-\zeta_p-\omega-i\eta}+\frac{1}{\zeta_{p+Q}-\zeta_p+\omega+i\eta}\right],
	\end{aligned}
\end{equation}
where
\begin{equation}\label{Eq:zetaPmPQp}
	\begin{aligned}
		&\Pi^0_{\delta}(Q,\omega)=
		\frac{\mathcal{B}(\delta)}{(2\pi)^3}\int d^3p~
		\left[1-\Theta\left(p_F^{(\delta)}-|Q+p|\right)\right] \Theta\left(p_F^{(\delta)}-|p|\right)\frac{2\zeta_{pQ}}{(i\omega^{(j)}_n)^2_{}-(\zeta_{pQ})^2},
	\end{aligned}
\end{equation}
where the first step function has been rewritten with the relation 
\begin{equation}
	\Theta(x)=1-\Theta(-x)
\end{equation}
The second term of Eq.~\eqref{Eq:zetaPmPQp} vanishes identically.

Considering the case ($\beta\to\infty,\zeta_p>0,\zeta_{p+Q}<0$), the corresponding contribution is given by:
\begin{equation}
	\begin{aligned}
		\Pi^0_{\delta}(Q&,\omega)
		=-\frac{\mathcal{B}(\delta)}{(2\pi)^3}\int d^3p\Theta\left(|p|-p_F^{(\delta)}\right)\Theta\left(p_F^{(\delta)}-|Q+p|\right) \left[\frac{1}{\zeta_{p+Q}-\zeta_p-\omega-i\eta}+\frac{1}{\zeta_{p+Q}-\zeta_p+\omega+i\eta}\right],
	\end{aligned}
\end{equation}
The expression can be rewritten with the change of variables $p'=-p-Q$; this transformation leads to 
\begin{equation}
	\begin{aligned}
		\Pi^0_{\delta}(Q&,\omega)
		=\frac{\mathcal{B}(\delta)}{(2\pi)^3}\int d^3p\Theta\left(|p+Q|-p_F^{(\delta)}\right)\Theta\left(p_F^{(\delta)}-|p|\right) \left[\frac{1}{\zeta_{p+Q}-\zeta_p+\omega+i\eta}+\frac{1}{\zeta_{p+Q}-\zeta_p-\omega-i\eta}\right],
	\end{aligned}
\end{equation}
The above expression is exactly equivalent to Eq.~\eqref{Eq:fistcaseDF}, and consequently yields the same result.

We now turn to the explicit evaluation of Eq.~\eqref{Eq:RePi},
% the real part of 
the noninteracting polarization function,
\begin{equation}\label{Eq:RePiAppendix}
   \begin{aligned}
   &\Pi^0_{\delta}(Q,\omega)=
   \frac{\mathcal{B}(\delta)}{(2\pi)^3}\int d^3p~\Theta\left(p_F^{(\delta)}-|p|\right) 
   \left[\frac{1}{\frac{\hbar^2}{2m}(Q^2+2p.Q))-\omega-i\eta}+\frac{1}{\frac{\hbar^2}{2m}(Q^2+2p.Q)+\omega+i\eta}\right],
   \end{aligned}
\end{equation}
where the step function $\Theta(p_F^{(\delta)}-|p|)$ restricts the integration to the interior of the generalized Fermi sphere of radius $p_F^{(\delta)}$.
Since the integrand depends on the momentum $\textbf{p}$ only through its magnitude $p$ and its angle $\theta$ relative to $\textbf{Q}$, it is natural to switch to spherical coordinates, $d^3p = p^2 \sin{\theta} \, dp \, d\theta \, d\phi$. In this coordinate system, each of the two denominators appearing in Eq.~\eqref{Eq:RePiAppendix} can be brought into the compact linear form $(\hbar^2Q^2/2m)+(\hbar^2p.Q/m)\mp\omega\mp i\eta=(\hbar^2Q/m)(Z^{}_{1,2}p_F^{(\delta)}+p\cos{\theta})$, where the dimensionless variables $Z_{1,2}$ are defined as $Z^{}_{1,2}=(Q/2p_F^{(\delta)})\mp(m\omega/\hbar^2p_F^{(\delta)}Q)\mp i\eta$. This substitution isolates the angular dependence in a single linear factor, which greatly simplifies the subsequent angular integration. With these definitions, Eq.~\eqref{Eq:RePiAppendix} becomes
\begin{equation}
    \begin{aligned}
    \Pi^0_{\delta}(Q,\omega)
    &=\frac{\mathcal{B}(\delta)}{(2\pi)^3}\int_0^{2\pi}\int_0^{p_F^{(\delta)}}\int_0^{\pi}
    \left[\frac{1}{\frac{\hbar^2 Q}{m}(Z_1^{}   p_F^{(\delta)}+p\cos\theta)}+\frac{1}{\frac{\hbar^2 Q}{m}(Z_2^{}   p_F^{(\delta)}+p\cos\theta)}\right]
    p^2\sin{\theta}~dp~d\theta~d\phi.
    \end{aligned}
\end{equation}
The integrand has no dependence on the azimuthal angle $\phi$, so this integration is performed trivially, contributing an overall factor of $2\pi$ which combines with the prefactor to give
\begin{equation}
    \begin{aligned}
    \Pi^0_{\delta}(Q,\omega)
    &=\frac{\mathcal{B}(\delta)m}{\hbar^2(2\pi)^2Q}\int_0^{p_F^{(\delta)}}p^2~dp\int_0^{\pi}
    \left[\frac{\sin{\theta}}{Z_1^{}   p_F^{(\delta)}+p\cos\theta}+\frac{\sin{\theta}}{Z_2^{}p_F^{(\delta)}+p\cos\theta}\right]
    ~d\theta.
    \end{aligned}
\end{equation}
The remaining angular integral is now elementary: substituting $u=\cos\theta$ converts it into a standard logarithmic integral, yielding
\begin{equation}
    \begin{aligned}
    \Pi^0_{\delta}(Q,\omega)
    &=\frac{\mathcal{B}(\delta)m}{\hbar^2(2\pi)^2Q}\int_0^{p_F^{(\delta)}}p^2\left[\left(\frac{1}{p}\ln\frac{Z_1p_F^{(\delta)}+p}{Z_1p_F^{(\delta)}-p}\right)+\left(\frac{1}{p}\ln\frac{Z_2p_F^{(\delta)}+p}{Z_2p_F^{(\delta)}-p}\right)\right]dp.
    \end{aligned}
\end{equation}
Finally, performing the remaining radial integral over $p$ from $0$ to $p_F^{(\delta)}$—which involves only elementary logarithmic and rational functions once expressed in terms of the dimensionless variables $Z_i$—leads to the closed-form result
\begin{equation}
    \begin{aligned}
    \Pi^0_{\delta}(Q,\omega)
    &=\frac{\mathcal{B}(\delta)m~{p_F^{(\delta)}}^2}{\hbar^2(2\pi)^2Q}
    \left[\left(Z_1+\frac{1-Z_1^2}{2}\ln\frac{Z_1+1}{Z_1-1}\right)+\left(Z_2+\frac{1-Z_2^2}{2}\ln\frac{Z_2+1}{Z_2-1}\right)\right].
    \end{aligned}
\end{equation}
Since the two terms in the brackets share an identical functional form, differing only in the index of $Z_i$, the result can be written compactly as a sum over $i=1,2$:

\begin{equation}
    \begin{aligned}
    \Pi^0_{\delta}(Q,\omega)
    &=\frac{\mathcal{B}(\delta)m~{p_F^{(\delta)}}^2}{\hbar^2(2\pi)^2Q}
    \sum_{i=1}^2\left(Z_i+\frac{1-Z_i^2}{2}\ln\frac{Z_i+1}{Z_i-1}\right),
    \end{aligned}
\end{equation}
which is the final closed-form expression for the generalized noninteracting polarization function.
\end{widetext}
%===========================================%%=======================================================
	%%Search  (OPEN HERE)
	%%%%%%%%%%%%%%%%%%%%%%%%%%%%%%%%%%
	%\bibliography{refpaper}
	%%%%%%%%%%%%%%%%%%%%%
\end{document}